\documentclass[twocolumn,traditabstract]{aa}
\usepackage{caption}
\usepackage{titlesec}
\titlespacing{\section}{0pt}{*3}{*2}  
\titleformat{\subparagraph}[runin]{\normalfont\normalsize\bfseries}{\thesubparagraph}{4em}{}
\usepackage{xspace}
\usepackage[switch]{lineno} 
\usepackage{amssymb}
\usepackage{placeins}
\usepackage{float}
\usepackage{pgfplots}
\usepackage{mathptmx}
\usepackage{multirow}
\usepackage{tablefootnote}
\usepackage{natbib}
\bibpunct{(}{)}{;}{a}{}{,}
\usepackage[]{graphicx}
\usepackage{booktabs}
\usepackage{longtable}
\usepackage{threeparttable} 
\usepackage{titlesec}
\usepackage{tikz}
\usetikzlibrary{shapes.geometric, arrows}
\usepackage{enumitem}
\tikzstyle{startstop} = [rectangle, rounded corners, minimum width=3cm, minimum height=1cm, text centered, draw=black]
\tikzstyle{process} = [rectangle, minimum width=3cm, minimum height=1cm, text centered, draw=black]
\tikzstyle{chart} = [rectangle, minimum width=3cm, minimum height=1cm, text centered, draw=black, dashed]
\tikzstyle{decision} = [diamond, minimum width=3cm, minimum height=1cm, text centered, draw=black]
\tikzstyle{arrow} = [thick,->,>=stealth]
\usepackage{bm}

\DeclareRobustCommand{\uvec}[1]{{
  \ifcsname uvec#1\endcsname
     \csname uvec#1\endcsname
   \else
    \bm{\hat{\mathbf{#1}}}
   \fi
}}
\usepackage{cool}
\usepackage{subcaption}
\usepackage{amsmath}
\usepackage{lscape}
\usepackage{txfonts}
\usepackage{hyperref}
\hypersetup{colorlinks=true,citecolor=blue}

\usepackage{csquotes}

\usepackage{multirow}
\usepackage{tabularx}
\usepackage{siunitx}

\usepackage{orcidlink}
\usepackage{comment}
\usepackage{verbatim}
\usepackage{url}
\usepackage{graphicx}
\usepackage{subfig}

\usepackage{xcolor}    
\usepackage{soul}      
\usepackage{array}
\usepackage{xspace}

\usepackage{fontawesome}

\usepackage{titlesec}
\makeatletter
\providecommand{\sf@counterlist}{}
\makeatother
\begin{document}

\title{GPU-Accelerated Gravitational Lensing $\&$ Dynamical (\texttt{GLaD}) Modeling for Cosmology and Galaxies }

\author{Han~Wang\inst{1,2}\orcidlink{0000-0002-1293-5503} \and Sherry~H.~Suyu\inst{2,1}\orcidlink{0000-0001-5568-6052} \and Aymeric~Galan\inst{2,1} \orcidlink{0000-0003-2547-9815} \and Aleksi~Halkola\inst{3}\and Michele~Cappellari\inst{4}\orcidlink{0000-0002-1283-8420}\and
Anowar~J.~Shajib \inst{5,6,7}\and Miha~Cernetic\inst{5} \orcidlink{0000-0002-5088-1745} 
}

\institute{
Max-Planck-Institut f\"ur Astrophysik, Karl-Schwarzschild-Str. 1, 85748 Garching, Germany \\
{\tt e-mail: wanghan@mpa-garching.mpg.de}
\and
Technical University of Munich, TUM School of Natural Sciences, Physics Department, James-Franck Str. 1, 85748 Garching, Germany
\and
Pyörrekuja 5 A, 04300 Tuusula, Finland
\and
Sub-Department of Astrophysics, Department of Physics, University of Oxford, Denys Wilkinson Building, Keble Road,
Oxford, OX1 3RH, UK
\and
Department of Astronomy \& Astrophysics, University of Chicago,
Chicago, IL 60637, USA
\and
Kavli Institute for Cosmological Physics, University of Chicago,
Chicago, IL 60637, USA
\and
Center for Astronomy, Space Science and Astrophysics, Independent University, Bangladesh, Dhaka 1229, Bangladesh
}

\titlerunning{GPU-Accelerated Gravitational Lensing $\&$ Dynamical (\texttt{GLaD}) Modeling for Cosmology and Galaxies}

\authorrunning{H.~Wang et al.} \date{Received / Accepted}

\abstract{Time-delay distance measurements from strongly lensed quasars provide a robust and independent method for determining the Hubble constant ($H_0$). This approach offers a crucial cross-check against $H_0$ measurements obtained from the standard distance ladder in the late universe and the cosmic microwave background in the early universe. However, the mass-sheet degeneracy in strong lensing models may introduce significant systematic uncertainty, limiting the precision of $H_0$ estimates. Dynamical modeling highly complements strong lensing to break the mass-sheet degeneracy, as both methods model the mass distribution of galaxies but rely on different sets of observational constraints. In this study, we develop a methodology and software framework for efficient joint modeling of stellar kinematic and lensing data. Using simulated lensing and kinematic data of the lensed quasar system RXJ1131$-$1131 as a test case, we demonstrate that approximately 4\% precision on $H_0$ is achievable with high-quality and signal-to-noise data. Through extensive modeling, we examine the impact of the presence of a supermassive BH in the lens galaxy and potential systematic biases in kinematic data on $H_0$ measurements. Our results demonstrate that either using a prior range for BH mass and orbital anisotropy, as motivated by studies of nearby galaxies, or excluding the central bins in the kinematic data, can both effectively mitigate potential biases on $H_0$ induced by the BH. By testing on mock kinematic data with values that are systematically biased, we emphasize the importance of using kinematic data with systematic errors under sub-percent control, which is currently achievable. Additionally, we leverage GPU parallelization to accelerate Bayesian inference, reducing a previously month-long process by an order of magnitude. This pipeline offers significant potential for advancing cosmological and galaxy evolution studies with large datasets. }


\maketitle
\section{Introduction}
\label{sec:intro}
The Hubble constant, $H_0$, sets the local expansion rate of the universe and plays a crucial role in understanding its age and size. Previous studies have reported a significant $5\sigma$ tension between $H_0$ measurements from the cosmic microwave background (CMB), which gives $H_0 = 67.4 \pm 0.5 \, \rm km \, s^{-1} \, Mpc^{-1}$ \citep[e.g.,][]{Aghanim2020}, and local distance indicators, such as supernovae (SNe) and Cepheid variables, which yield $H_0 = 73.0 \pm 1.0 \, \rm km \, s^{-1} \, Mpc^{-1}$ \citep[e.g.,][]{SHOES}. However, recent measurements from the Chicago-Carnegie Hubble Program \citep[e.g.,][]{chicago_H0}, which are also based on the SN distance ladder, show no significant tension with the CMB, leaving the true discrepancy uncertain. \citet{riess2024jwstobservationsrejectunrecognized} highlighted that the $H_0$ measurement in \citet{chicago_H0} was based on a subsample selection. Whether the tension is real, or merely a result of systematic uncertainties that were not known and not incorporated in the measurements, remains a topic of debate \citep{Efstathiou2020, 2022Abdalla,Yeung2022,Freedman2023}, but if confirmed, it would indicate the need for new physics beyond the standard cosmological model.

Time-delay cosmography offers a distinct approach, separate from the previously mentioned methods, to measure $H_0$ by analyzing the brightness variations of sources like quasars or supernovae. It constrains cosmological parameters by measuring the time delay between multiple lensed images of the source \citep{Refsdal, lensingReview, lensing2016, TreuSuyuMarshall2020, TreuShajib2023, lensing2022, Oguri2019, Liao+2022, Suyu+2024}. By determining the time-delay distance to the lens system, it is possible to infer the value of $H_0$. However, this approach is affected by the mass-sheet degeneracy (MSD) in strong lensing (SL) \citep[e.g.,][]{1985Falco,1988Gorenstein, 2016Birrer,
 TDCOSMO6}. We categorize MSD into two types: external and internal. Both can potentially bias estimates of $H_0$. The external MSD, which arises from line-of-sight (LoS) effects, can be controlled by studying the environments of the lens galaxies \citep[e.g.,][]{patrickLoS}. The internal MSD arises from the unknown radial profile of the lens galaxies' mass distribution \citep[e.g.,][]{SchneiderSluse2013}. This degeneracy allows for equally well-fitting models of the observed lensing data while introducing a linear bias in the inferred value of $H_0$.

A common strategy to address the internal MSD is to incorporate independent datasets, such as kinematic or weak lensing data \citep[e.g.,][]{Treu_2002, Shajib_2020,TDCOSMO4, Birrer_2021,
TDCOSMO13, ShajibRXJ1131, breakMSDWeak}. These additional observations help detect changes in the mass density slope induced by the internal MSD in SL, both in the inner regions covered by the kinematic measurements---depending on the field of view (FoV) and the signal-to-noise ratio---and in the outer regions covered by weak lensing, up to \(\sim 8 R_{\rm Ein}\) from the lens galaxy’s centroid. This enables a more robust constraint on the mass distribution and, consequently, on \( H_0 \).

With high-resolution kinematic maps provided by the James Webb Space Telescope (JWST) Near-Infrared Spectrograph integrated field unit (NIRSpec IFU) \citep[]{NIRSPEC}, we can obtain more precise stellar velocity dispersion measurements over 2D space compared to previous facilities. \citet{2020Akin} developed a pipeline that enables self-consistent joint modeling by simultaneously fitting lensing and dynamical data to infer $H_0$ value. This code combines lensing mass modeling through pixelated source reconstruction \citep{2010SuyuHalkola_Dpie2} with dynamical mass models based on the Jeans equations in an axisymmetric geometry \citep[]{JAM_cly}. \citet{TDCOSMO13} applied this joint modeling approach to simulated JWST-like kinematic datasets for the lensed quasar system $\rm RXJ1131-1231$ (hereafter referred to as RXJ1131 for simplicity). They explicitly modeled the internal MSD using an isothermal profile with an extended core. Their results demonstrated the power of combining SL with kinematics, showing that the internal MSD can be effectively broken. They successfully recovered the mock input value of $H_0$ with a precision of $4\%$ for a single-lensed quasar system. 

 The H0LiCOW collaboration reported an $H_0$ measurement with $2.4\%$ precision by combining six lensed quasar systems \citep{Howlicow_wong}. These analyses tested two specific mass models, the composite model (baryonic component + dark matter) and the elliptical power-law model, while performing lens modeling without explicitly accounting for the internal MSD. Intuitively, explicitly modeling the internal MSD makes the adopted mass model more flexible, allowing for a broader range of mass distributions. High-resolution spatial kinematics can help distinguish between these more flexible models. However, H0LiCOW used slit kinematics, which primarily served to validate the best-fit mass models rather than to differentiate between them, as slit kinematics alone is insufficient to break the MSD and measure distances with a few-percent uncertainty on an individual lens basis. If mass model assumptions are relaxed and an internal mass sheet—maximally degenerate with $H_0$—is incorporated, the precision of the $H_0$ constraint from the six lensed quasar systems degrades to $5\%$ or $8\%$, depending on whether external priors from non-time-delay lenses are used for orbital anisotropy \citep[]{TDCOSMO4}.
 The TDCOSMO collaboration continues to investigate potential degeneracies and biases in the measurement of $H_0$ caused by the internal MSD in lens modeling \citep[e.g.,][]{TDCOSMO1, TDCOSMO5,TDCOSMO6, TDCOSMO7, TDCOSMO8}, previously studied by the H0LiCOW collaboration. As part of TDCOSMO, \citet{ShajibRXJ1131} conducted a joint modeling analysis to explicitly break the internal MSD using spatially resolved kinematics from KCWI (an integral field spectrograph at Keck Observatory \citep[]{KCWI}). Their study yielded a value of $H_0 = 77.1_{-7.1}^{+7.3}~\rm km~s^{-1},Mpc^{-1}$, achieving a precision of approximately $9.5\%$, from a single time-delay lens system. This analysis was constrained by the kinematic resolution of KCWI.\footnote{The kinematic data exhibit a signal-to-noise ratio of $23~\AA^{-1}$ in the rest-frame wavelength range $3985-4085~\AA$ across 41 bins, with a seeing effect of $0.96\arcsec$ in full width at half maximum (FWHM).} The diffraction-limited resolution of JWST will offer significantly greater precision, further enhancing kinematic constraints.

The TDCOSMO collaboration aims to constrain $H_0$ to within 2\% by combining spatially resolved kinematics data obtained from the JWST NIRSpec IFU for seven gravitational lenses. In order to achieve this level of precision and accuracy, extensive tests have been conducted, including examining the impact of the FoV on kinematics, comparing different mass models, such as the composite and power-law models and evaluating various dark matter profiles, including the standard NFW profile and its generalized form \citep[e.g.,][]{2020Akin, TDCOSMO13}. Additionally, the influence of the deprojected 3D shape of lens galaxies has been investigated \citep[]{ShajibRXJ1131, 2025Huang}. Exploring these effects requires substantial computational resources, making joint modeling highly demanding. For a single lensed-quasar system such as RXJ1131, Bayesian inference using the Markov chain Monte Carlo (MCMC) method takes a month to complete using traditional CPU-based methods.  

In this paper, we present \texttt{GLaD} (Gravitational Lensing and Dynamics), a GPU-accelerated joint modeling code for time-delay cosmography and galaxy studies, built upon \citet{2020Akin}, and the \texttt{GLEE} software \citep{2010SuyuHalkola_Dpie2, Suyu2012}, for lensing modeling, along with \texttt{JamPy}\footnote{\url{https://pypi.org/project/jampy/}} \citep{JAM_cly,JAM_sph} for dynamical modeling.\footnote{\texttt{GLaD} can be performed on the lens galaxy or the lensed background source galaxy.  The \texttt{GLaD} modeling presented here focuses on the lens galaxy, in contrast to the \texttt{GLaD} modeling of the lensed source in \citet{Chirivi2020}.} \texttt{GLaD} significantly reduces the Bayesian inference runtime from several months to just a few days. Furthermore, we probe two additional effects, the mass of the black hole (BH) in the lens galaxy and the possible systematic error in the kinematics measurement from the IFU data, which may bias $H_0$ inference. On the one hand, since lens galaxies are typically massive elliptical galaxies with high velocity dispersions $\sigma_{\rm disp} > 200~\rm km~s^{-1}$ \citep[]{IFUKnabel}, with corresponding BH mass $M_{\rm BH} > 10^{8}~{\rm M_{\odot}}$ \citep[]{Kormendy2013review,2013CPMA}, the presence of a massive BH may be detectable in high-resolution kinematic data. On the other hand, kinematic measurements are susceptible to systematic errors, especially when different methods are used to derive velocities from IFU data. For example, using stellar population synthesis models can introduce errors based on assumptions about star formation history and metallicity. Additionally, inferred velocities can vary depending on the chosen stellar libraries.
These factors must be mitigated to attain the precision and accuracy required for cosmography. \citet{Knabel2025} recently conducted a detailed study on the accuracy of kinematic measurements, demonstrating that percent-level precision is achievable using cleaned stellar libraries—stellar libraries refined to exclude spectra affected by artefacts or poor data quality. Previously, kinematic accuracy was limited to the few-percent level. In this work, we assess the impact of systematic errors by analyzing a worst-case hypothetical scenario, assuming a 5\% uncertainty in kinematic measurements of $H_0$, even though the actual effect is expected to be much smaller, around 1\%. We highlight the importance of the current developments for kinematic measurements.

We perform all the tests described above using \texttt{GLaD} on simulated lensing and kinematic data for the RXJ1131 system. This system has the most precise time-delay measurements, with an accuracy of approximately $1.6\%$, among the six systems in the H0LiCOW sample. Additionally, the lens galaxy in RXJ1131, with a redshift of $z = 0.295$, is the closest among these systems and will provide the most accurate kinematic measurements. Furthermore, the galaxy’s central velocity dispersion of $\sigma_{\rm disp} \geq 300~\rm km~s^{-1}$ \citep[]{2014Suyu, ShajibRXJ1131} strongly suggests the presence of a supermassive BH.

This paper is organized as follows. In Sect.~\ref{section:Overview of the lens and dynamical modeling}, we provide an overview of lensing theory and introduce the MSD in lens modeling. We also present the dynamical modeling approach. In Sect.~\ref{Method}, we describe the GPU-accelerated components of the joint modeling and provide a detailed overview of the modeling workflow. In Sect.~\ref{sect:Simulated Mock Datasets}, we describe the simulated lensing and kinematic datasets for the lensed quasar system RXJ1131. In Sect.~\ref{sect:Analysis and Discussion of the Joint Modeling Results}, we present the results of the joint modeling and discuss the effects of BH mass and potential systematic errors in the kinematic map. In Sect.~\ref{sect:Summary and Outlook}, we summarize our work and present concluding remarks and an outlook. Throughout this paper, we adopt a standard cosmological model with $H_0 = 82.5~\text{km}~\text{s}^{-1}~\text{Mpc}^{-1}$, a matter density parameter of $\Omega_{\rm m} = 0.27$, and a dark energy density of $\Omega_\Lambda = 0.73$. Our choice of cosmology is motivated by the time-delay distance measurements of RXJ1131 from \citet{2014Suyu}. Note that our conclusions are independent of the choice of cosmological model. Additionally, all runtime comparisons across different modeling approaches are conducted using 64-bit floating-point precision. All tests are performed on a 2.10 GHz, 16-core Intel(R) Xeon(R) Silver 4110 CPU and an NVIDIA A100 GPU.

\section{Overview of the lens and dynamical modeling}
\label{section:Overview of the lens and dynamical modeling}
In Sect.~\ref{subsection:SL}, we provide a brief overview of the SL formalism in the context of time-delay cosmography. In Sect.~\ref{subsect:Internal MSD}, we introduce the MSD, a major source of systematic uncertainty in SL modeling that limits the precision of $H_0$ measurements. The internal MSD arises from the unknown size and brightness of source galaxies, as well as the uncertain mass distribution of lens galaxies. These uncertainties impact the measurement of the time-delay distance $D_{\Delta \rm t}$, which is directly proportional to $H_0^{-1}$. Similarly, the external MSD, caused by unknown mass distributions along the LoS, introduces an additional uncertainty in $D_{\Delta \rm t}$ measurement, as discussed in Sect.~\ref{subsection:External Mass Sheet Degeneracy}. In this section, we also show that when using joint modeling with a non-fixed cosmological model, the external MSD does not impact dynamical modeling. In Sect.~\ref{subsection:Stellar dynamics under the internal MSD}, we provide a brief overview of stellar dynamical modeling, assuming an axisymmetric mass distribution and employing the Jeans Anisotropic Modeling (JAM) approach \citep{JAM_cly, JAM_sph}.

\subsection{Strong lensing}
\label{subsection:SL}
In the SL scenario, massive foreground galaxies act as gravitational
lenses, warping spacetime and bending light from background sources. This causes each light beam to follow a slightly different path, resulting in multiple images of the background sources. Image $i$ arrives at the observer with a time delay compared to the unlensed case:
\begin{equation}
    t_i(\boldsymbol{\theta}^{\,},\boldsymbol{\beta}^{\,} ) = \frac{(1 + z_{\rm d})}{c}\frac{D_{\rm d} D_{\rm s}}{D_{\rm ds}} \phi_{\rm L}(\boldsymbol{\theta}^{\,},\boldsymbol{\beta}^{\,} )
    \label{eq:time delay}
\end{equation}
where $\boldsymbol{\theta}$ is the angular image position, $\boldsymbol{\beta}$ the background source position, $z_{\rm d}$ the lens redshift, $D_{\rm d}$, $D_{\rm s}$ and $D_{\rm ds}$ the angular diameter distance to the lens, the source and the distance between the lens and source. The Fermat potential $\phi_{\rm L}$ is written in terms of 
\begin{equation}
\phi_{\rm L}(\boldsymbol{\theta}^{\,},\boldsymbol{\beta}^{\,} ) = \frac{(\boldsymbol{\theta}^{\,} - \boldsymbol{\beta}^{\,} )^2}{2} - \psi_{\rm L} (\boldsymbol{\theta}^{\,})
\label{eq:fermat potential}
\end{equation}
The difference in light travel time at an image position $\boldsymbol{\theta_i}$, relative to another observed image position $\boldsymbol{\theta_j}$, arises from two components of $\phi_{\rm L}$. The first component in Eq.~\ref{eq:fermat potential} represents the geometric excess path length, while the second accounts for the gravitational time delay caused by the 2D lens potential $\psi_{\rm L}$. Thus, the time delay between the observed multiple images $i$ and $j$ can be derived as:
\begin{equation}
    \Delta t_{ij} =  \frac{(1 + z_{\rm d})}{c}\frac{D_{\rm d} D_{\rm s}}{D_{\rm ds}} \left[\phi_{\rm L}(\boldsymbol{\theta}^{\,}_{i},\boldsymbol{\beta}^{\,}) - \phi_{\rm L}(\boldsymbol{\theta}^{\,}_{j},\boldsymbol{\beta}^{\,} )\right].
    \label{eq: time delay between multiple images}
\end{equation}
We define the normalization factor in Eq.~\ref{eq: time delay between multiple images} as the time-delay distance $D_{\Delta \rm t}$ \citep{Suyu2010}, which is proportional to $H_0^{-1}$:
\begin{equation}
    D_{\Delta \rm t} \equiv (1 + z_{\rm d}) \frac{D_{\rm d} D_{\rm s}}{D_{\rm ds}} \propto  \frac{1}{H_0}
    \label{eq:time delay distance}
\end{equation}
By measuring the time delays $\Delta t_{ij}$ and the positions of the lensed images $\boldsymbol{\theta}_{ij}$, we can reconstruct $\phi_{\rm L}$ and infer $H_0$ using Eq.~\ref{eq: time delay between multiple images}.

\subsection{Internal mass sheet degeneracy}
\label{subsect:Internal MSD}
The source position $\boldsymbol{\beta}$ is not directly observable, and it can undergo an arbitrary affine transformation:
\begin{equation}
\boldsymbol{\beta}_{\rm int} = \lambda_{\rm int}\boldsymbol{\beta} - \boldsymbol{a_{0}},
\label{eq:src scaling}
\end{equation}
where $\lambda_{\rm int}$ and $\boldsymbol{a_{0}}$ affect the scaling and position of the source. These undetectable changes in $\boldsymbol{\beta}$ can be induced by an affine transformation of the projected dimensionless surface mass density $\kappa_{\rm gal}$ of the lens galaxy:
\begin{equation}
\kappa_{\rm int} = (1-\lambda_{\rm int}) + \lambda_{\rm int} \kappa_{\rm gal},
\label{eq:kappa int}
\end{equation}
leaving observables such as image positions and the morphology of lensed images invariant under this transformation, which is known as the internal MSD \citep{Falco1985}. 
In other words, suppose we model the surface mass density of the lens galaxy as $\kappa_{\rm gal}$ (e.g., using a power-law profile), then $\kappa_{\rm int}$ that accounts for the internal mass sheet would fit lensed image positions and morphology equally well.
This transformation propagates to the lens potential via Poisson’s equation:
\begin{equation}
\nabla^2 {\psi_{\rm L, int}} = 2 \kappa_{\rm int},
\label{eq:poisson equation}
\end{equation}
where the transformed lens potential is given by
\begin{equation}
\psi_{\rm L,int} (\boldsymbol{\theta}) = \frac{ 1-\lambda_{\rm int}}{2} |\boldsymbol{\theta}|^2 + \boldsymbol{a_0} \cdot \boldsymbol{\theta} + \lambda_{\rm int}\psi_{\rm L, gal}(\boldsymbol{\theta}) + c_0,
\label{eq: transformed MST}
\end{equation}
where $c_0$ is an arbitrary constant.
Substituting $\psi_{\rm L,int}$ into Eqs.~\ref{eq:time delay} and \ref{eq: time delay between multiple images} cancels out the arbitrary additive constant $c_0$ and yields the rescaled time-delay distance:
\begin{equation}
\begin{aligned}
D_{\Delta \rm t, int} &= \frac{D_{\Delta \rm t,gal}}{\lambda_{\rm int}} = (1 + z_{\rm d}) \frac{D_{\rm d,int} D_{\rm s, int}}{D_{\rm ds,int}} \\
&\propto \frac{1}{\lambda_{\rm int} H_0}
\end{aligned}
\label{eq:D_delta_int}
\end{equation}
where $D_{\Delta \rm t,gal}$ is associated with $\kappa_{\rm gal}$ and $D_{\Delta \rm t, int}$ with $\kappa_{\rm int}$. The distances $D_{\rm s, int}$, $D_{\rm ds, int}$, and $D_{\rm d, int}$ are considered in the context of the internal MSD associated with $D_{\rm \Delta t, int}$. Note that the distance to lens galaxy is influenced by internal MSD, such that,
\begin{equation}
    D_{\rm d, gal} \neq D_{\rm d, int}.
\end{equation}
However, the change of $D_{\rm d}$ induced by internal MSD is minor under $1\%$ shift, given the single aperture kinematics \citep[see Sect.~4.2]{TDCOSMO6}. With 2D resolved kinematics, the impact on $D_{\rm d, int}$ could be more significant.

The internal MSD alters the mass density slope of lens galaxies. This occurs because, aside from the renormalization factor $\lambda_{\rm int}$ in the second term of Eq.~\ref{eq:kappa int}, the first term results in the addition of a constant sheet to the initial $\kappa_{\rm gal}$. Therefore, if the intrinsic radial profile of the mass distribution in lens galaxies were known, the internal MSD would cease to be a degeneracy. However, in practice, the underlying mass distribution may not be known to sufficient precision, making the class of mass models $\kappa_{\rm int}$ indistinguishable from $\kappa_{\rm gal}$ when relying solely on lensing data. In time-delay cosmography, this means that $D_{\Delta \rm t,int}$  yields
 a linearly scaled $\lambda_{\rm int} H_0$.

Dynamical modeling provides an independent measurement of the mass distribution in lens galaxies, as its constraints come from kinematic data, which are entirely different observables from those in lensing analyses. Moreover, galaxy dynamics measures the intrinsic density distribution in 3D rather than the projected mass surface density. Combining dynamical modeling with lensing modeling allows us to constrain the scaling factor $\lambda_{\rm int}$, meaning we can determine which $\kappa_{\rm int}$ models within the internal MSD framework are favored. This approach helps break the internal MSD degeneracy (see Sect.~\ref{subsection:Stellar dynamics under the internal MSD}).

\subsection{Internal and External mass sheet degeneracy}
\label{subsection:External Mass Sheet Degeneracy}
Unlike internal MSD, which has relatively strong effects on small scales, such as altering mass density slopes of lens galaxies, external MSD merely performs the renormalization of the underlying mass convergence distribution. We use a class of $\kappa_{
\rm int}$ to represent the mass distributions of lens galaxies, as they are all viable choices until distinguished by kinematic data. In the external MSD regime, $
\kappa_{
\rm int}$ scales as:

\begin{equation}
    \kappa = (1 - \kappa_{\rm ext})  \kappa_{\rm int} + \kappa_{\rm ext}
\end{equation}
where $\kappa_{\rm ext}$ indicates the mass perturbations along the LoS that do not dynamically affect the mass distribution of lens galaxies at the primary lens plane.

Taking into account the influence of the external MSD, $D_{\Delta  \rm t,int}$ is rescaled by

\begin{equation}
    D_{\Delta \rm t} = \frac{D_{\Delta  \rm t,int}}{(1-\kappa_{\rm ext})}.
    \label{eq:Dt external internal}
\end{equation}
The inferred \( D_{\Delta \rm t} \) is a distance that can be compared with the distance calculated from the cosmological models and test the cosmology, whereas \( D_{\Delta \rm t,int} \) cannot be used to test assumed cosmological models directly because it has not been corrected for the external convergence. The external convergence $\kappa_{\rm ext}$ can be estimated by examining the lens environment using photometric and spectroscopic surveys, as well as through ray-tracing methods in cosmological simulations \citep[e.g.,][]{Suyu2010, Greene2013,2014Suyu, Rusu2017}. 
We investigate whether the renormalization factor \( (1 - \kappa_{\rm ext}) \) from the external MSD affects the dynamical modeling. We derive the 2D surface mass density \( \Sigma\) as  
\begin{equation}
\Sigma = \Sigma_{\rm crit} \kappa = \Sigma_{\rm crit} \left[ (1 - \kappa_{\rm ext}) \kappa_{\rm int} + \kappa_{\rm ext} \right],
\label{Eq:sigma_external_internal}
\end{equation}  
where \( \Sigma_{\rm crit} \) is the critical density. In the framework of internal and external MSD, we express \( \Sigma_{\rm crit} \) in terms of \( D_{\rm \Delta t} \) \footnote{ $D_{\Delta \rm t,int}$ represents the actual distance, i,e, the distance that can be directly compared to the predictions from cosmological models.} as  
\begin{equation}
   \Sigma_{\rm crit} = \frac{c^2}{4\pi G} \frac{D_{\rm  \Delta t}}{(1 + z_{\rm d})D_{\rm d}^2 },
   \label{Eq:critical_sfd}
\end{equation}  
where \( D_{\rm d} \) remains fully invariant under external MSD when modeling is performed without fixing the cosmology, using lens dynamics and time-delay cosmography jointly \citep[]{2015Jee, TDCOSMO6}:  
\begin{equation}
      D_{\rm d} = D_{\rm d,int}.
    \label{eq:Dd_ext}
\end{equation}  
By substituting Eqs.~\ref{Eq:critical_sfd} and \ref{eq:Dd_ext} into Eq.~\ref{Eq:sigma_external_internal}, we find that the factor \( (1 - \kappa_{\rm ext}) \) cancels out in the first term of Eq.~\ref{Eq:sfd final}. As a result, the 2D surface mass density $\Sigma$ is written as  
\begin{align}
\Sigma
&= \frac{c^2}{4\pi G} \frac{D_{\Delta \rm t}}{(1 + z_{\rm d})D_{\rm d}^2 } 
   \left[ (1 - \kappa_{\rm ext})  \kappa_{\rm int} + \kappa_{\rm ext} \right] \nonumber \\
&= \frac{c^2}{4\pi G} \frac{D_{\rm \Delta t, \rm int}}{(1 + z_{\rm d})(1 - \kappa_{\rm ext})D_{\rm d,int}^2 } 
   \left[ (1 - \kappa_{\rm ext})  \kappa_{\rm int} + \kappa_{\rm ext} \right] \nonumber \\
&= \frac{c^2}{4\pi G} \frac{1}{(1 + z_{\rm d})D_{\rm d,int}^2 } 
   \left[ D_{\Delta \rm t, \rm int} \kappa_{\rm int} + D_{\Delta \rm t} \kappa_{\rm ext} \right] \nonumber \\
&= \Sigma_{\rm int} + \Sigma_{\rm ext}
\label{Eq:sfd final}
\end{align}
In the lensing and dynamical modeling, we focus on modeling $\Sigma_{\rm int}$ in Eq.~\ref{Eq:sfd final}. We sample the distance \( D_{\Delta \rm t, \rm int} \) in the joint modeling, rather than \( D_{\Delta \rm t} \), to ensure that the dynamical modeling remains unaffected by the external convergence \( \kappa_{\rm ext} \). Sampling \( D_{\Delta \rm t} \) directly will introduce a scaling factor of \( (1 - \kappa_{\rm ext}) \) into the dynamical analysis. The second term $\Sigma_{\rm ext}$ in Eq.~\ref{Eq:sfd final} is essentially a constant accounting for all the perturbations along LoS that do not affect the dynamics of the lens galaxy.

\subsection{Stellar dynamics }
\label{subsection:Stellar dynamics under the internal MSD}
Here we briefly revisit the theoretical framework for the dynamical modeling. Stars within a galaxy can be characterized by the collisionless Boltzmann equation \citep[e.g.,][eq.~4-13b]{Binney1987} which is a differential equation of the phase-space density $f(\boldsymbol{x}, \boldsymbol{v})$ at the position $\boldsymbol{x}$ with velocity $\boldsymbol{v}$,
\begin{equation}
  \frac{\partial f}{\partial t} +  \sum_{i=1}^{3} v_i \frac{\partial f}{\partial x_i} - \frac{\partial \psi_{\rm D,int}}{\partial x_i} \frac{\partial f}{\partial v_i} = 0
  \label{eq:cbe}.
\end{equation}
This equation describes stars embedded in a 3D gravitational field of the lens galaxy, with $
\psi_{
\rm D,int}$ being the deprojection of the 2D lensing potential $
\psi_{
\rm L,int}$ (up to a constant factor), ensuring phase-space density conservation. The phase-space density is not accessible for galaxies, and we can only measure the velocities $v$ along the LoS, and velocity dispersions $\sigma$ using the spectroscopy for distant galaxies $z > 0.1$. To solve the Eq.~\ref{eq:cbe}, we reduce the number of the degree of freedom by assuming an axisymmetric mass distribution ($\partial \psi_{\rm D,int}/\partial \phi = \partial f/\partial \phi = 0$, with $\phi$ being the azimuthal angle in the spherical coordinate system) and the spherically-aligned velocity ellipsoids. The choice of spherically-aligned velocity ellipsoids is due to the fact that lens galaxies are generally massive slow rotators. These galaxies exhibit a near-spherical mass distribution in their central regions, as opposed to a flat mass distribution characterized by cylindrically-aligned velocity ellipsoids. We multiply velocities along radial $v_r$, polar $v_{\theta}$ and azimuthal direction $v_{\phi}$ with Eq.~\ref{eq:cbe} and integrate over all velocity space, obtaining two Jeans equations \citep[e.g.,][eqs.~1, 2]{Bacon1983}
\begin{equation}
  \frac{ \partial \left( \rho_{*} \overline{v_r^2}\right)}{\partial r} + \frac{ ( 1 +\beta_{\rm ani}) \rho_{*} \overline{v_r^2} - \rho_{*} \overline{v_{\phi}^2}} { r} = -\rho_{*} \frac{\partial \psi_{\rm D,int}}{\partial r},
  \label{eq:J1}
\end{equation}
\begin{equation}
  \frac{ (1 - \beta_{\rm ani})\partial \left(\rho_{*} \overline{v_{r}^2}\right)}{\partial \theta} 
  + \frac{ ( 1 -\beta_{\rm ani}) \rho_{*} \overline{v_r^2} - \rho_{*} \overline{v_{\phi}^2}} { \tan \theta} 
  = -\rho_{*} \frac{\partial  \psi_{\rm D,int}}{\partial \theta},
  \label{eq:J2}
\end{equation}
with the following notations
\begin{equation}
     \rho_{*}\overline{v_k v_j} =  \int v_k v_j f \text{d}^3 v,
\end{equation}
\begin{equation}
    \beta_{\rm ani} = 1-\overline{v_{\theta}^2}/ \overline{v_r^2}
\end{equation}
where $\rho_{*} = \int f \text{d}^{3} \boldsymbol{v}$ represents an estimate of the luminosity density of the stellar tracer from which the observed kinematics are derived, $\rho_{*}\overline{v_k v_j}$ represents the second intrinsic velocity moment in the spherical coordinate, and $\beta_{\rm ani}$ denotes the orbital anisotropy. The anisotropy presents stellar motion preference regarding the direction. The anisotropy $\beta_{\rm ani} >0$ indicates most stars inside the galaxies move along the radial direction. In contrast, $\beta_{\rm ani} <0$ indicates the tangential motions dominate the galaxies.

To derive the line-of-sight velocity moments $\overline{v_{\rm LoS}^2}$ from the Jeans equations (Eqs.~\ref{eq:J1} and \ref{eq:J2}), it is essential to reconstruct the intrinsic 3D luminosity and mass density distributions of the lens galaxy. It is a common strategy to first apply multiple Gaussian expansion (MGE; \citealt{Emsellem1994, Cappellari2002mge}) to the observed 2D surface brightness (SB) and mass convergence, then deproject them later using the inclination angle (see Appendix.~\ref{app:MGE to surface brightness and surface mass convergence}). This yields the MGEs of the 3D luminosity density $\rho_{*}(r, \theta)$ and the 3D total mass density $\rho_{\rm int}(r, \theta)$. They are then substituted into the Jeans equations (\ref{eq:J1}) and (\ref{eq:J2}) to compute the intrinsic second velocity moments $\overline{v_r^2}$, $\overline{v_\theta^2}$, and $\overline{v_\phi^2}$. These moments correspond to the diagonal elements of the velocity dispersion tensor, assuming a spherically aligned velocity ellipsoid where all off-diagonal elements vanish.

The next step is to convert the intrinsic second velocity moments from spherical coordinates to Cartesian coordinates ($x^{\prime}$, $y^{\prime}$, $z^{\prime}$), with the $z^{\prime}$-axis aligned along the LoS direction (see Sect.3.1 in \citet{JAM_sph}). We then derive $\overline{{v_z^{\prime}}^2}$ in terms of $\overline{v_r^2}$, $\overline{v_\theta^2}$, and $\overline{v_\phi^2}$. In real observations, we measure integrated light from stars at various positions along the LoS. Therefore, we compute the luminosity-weighted $\overline{v_{\rm LoS}^2}$ at the spaxel located at $(x^{\prime}, y^{\prime})$ as follows: 
\begin{equation} 
\overline{v_{\rm LoS}^2} = \frac{\int_{-\infty}^{\infty} {\text{d}z}^{\prime}\rho_{*}\overline{v_{z^{\prime}}^2}}{\int_{-\infty}^{\infty} \rho_* dz'}. \label{eq:VLos} 
\end{equation}

In the end, we convolve $\overline{v_{\rm LoS}^2}$ values (see Eq.~\ref{eq:VLos})  with the kinematic point spread function $\rm PSF_{\rm kin}$ to account for the atmosphere and instrument effect, weighted by the SB of lens galaxies $I(x^{\prime}, y^{\prime})$, and integrated over the region associated in each of the Voronoi bins \citep{Voronoi}, yielding the predicted $\left[\overline{v_{\rm LoS}^2}\right]_{l}^{\rm pre}$ to compare with the observed kinematic data ${v_{\rm rms}}_{,l}$ at bin $l$
\begin{equation}
   \left[\overline{v_{\rm LoS}^2}\right]_{l}^{\rm pre} = \frac{\int_{\rm Bin} ~{\rm dx^{\prime}~dy^{\prime}}~I~(x^{\prime}, y^{\prime})~\overline{v_{\rm LoS}^2} \otimes \rm PSF_{\rm kin}} {\int_{\rm Bin} ~{\rm dx^{\prime}~dy^{\prime}}~I~(x^{\prime}, y^{\prime}) \otimes \rm PSF_{\rm kin}}
\end{equation}
and 

\begin{equation}
     {v_{\rm rms}^{\rm pre}}_{,l} = \sqrt{\left[\overline{v_{\rm LoS}^2}\right]_{l}^{\rm pre}}.
     \label{eq:Vrms}
\end{equation}
Note that the value of ${v_{\rm rms}}_{,l}$ is related to the distance to the lens galaxy:  
\begin{equation}
    {v_{\rm rms}}_{,l} \propto \frac{1}{\sqrt{D_{\rm d}}}.
    \label{Eq:scale Dd to Vlos}
\end{equation}  
This relationship arises because, for a given angular size, the physical size of the lens galaxy increases with distance:  
\begin{equation}
    r_{\rm phy} \propto D_{\rm d} \theta.
\end{equation}  
In dynamical equilibrium, a larger system with the same total mass exhibits lower $v_{\rm rms}$, following the relation:  
\begin{equation}
  {v_{\rm rms}}_{,l} \propto \sqrt{\frac{G M}{r_{\rm phy}}}.  
\end{equation}  
Since $r_{\rm phy}$ increases with $D_{\rm d}$, $v_{\rm rms}$ decreases accordingly, leading to the inverse square-root dependence in Eq.~\ref{Eq:scale Dd to Vlos}. The distance $D_{\rm d}$ can thus be constrained from the dynamical modeling, together with the time-delay distance $D_{\Delta \rm t, int}$.

The goodness of the dynamical modeling is evaluated by
\begin{equation}
    \chi^2_{\rm dyn} = (\boldsymbol{v_{\rm rms}} -\boldsymbol{v_{\rm rms}^{\rm pre}})^T \boldsymbol{\Sigma_\text{kin}^{-1}} (\boldsymbol{v_{\rm rms}} - \boldsymbol{v_{\rm rms}^{\rm pre}}),
   \label{eq:chi2_dyn}
\end{equation}
where $\boldsymbol{\Sigma_\text{kin}^{-1}}$ is the covariance matrix of the measured uncertainties of the kinematic data. We refer readers to \citet{JAM_sph} for the detailed construction of the 3D gravitational potential $\psi_{\rm D,int}$ from MGEs and the calculation process of $\overline{v_{\rm LoS}^2}$.

\section{Method}
\label{Method}
 In this section, we highlight the aspects of joint modeling that benefit from GPU parallelization. Given the large-scale matrix computations inherent in the modeling process, GPUs outperform CPUs by efficiently handling repetitive, computationally intensive operations. Our joint modeling code \texttt{GLaD}, is implemented in \texttt{JAX} \citep[e.g.,][]{jax2018github}, a high-performance numerical computing library for Python that enables automatic differentiation and Just-In-Time compilation for accelerated computations on GPUs. In Sect.~\ref{subsect:GPU acceleration in lensng modeling}, we briefly introduce SL modeling and demonstrate the speed improvements achieved with GPU on extended image modeling. Additionally, we present a newly implemented NFW profile following \citet{Oguri2021} that directly incorporates ellipticity into the surface mass density. In Sect.~\ref{subsect:GPU acceleration in dynamics modeling}, we describe a fast 1D MGE implementation optimized for GPUs following \citet{MGE_shajib} and a non-adaptive integral solver on a fixed grid to compute the intrinsic second velocity moments in the spherical-aligned JAM. In Sect.~\ref{subsect:Joint modeling}, we provide a detailed overview of the joint modeling code structure and discuss the use of Bayesian inference to obtain best-fit models. In Sect.~\ref{subsect:Bayesian information criterion}, we introduce the Bayesian Information Criterion (BIC) to adjust the weighting of the posterior distribution in joint modeling, since the number of stellar kinematics data points is significantly smaller than that of the lensing data. Without BIC reweighting, the lensing and dynamical (LD) likelihood would be dominated by the lensing information.

\subsection{GPU acceleration in lensing modeling}
\label{subsect:GPU acceleration in lensng modeling}

\subsubsection{Lensing modeling}
We start our joint formalism with the SL part. The observables in the lensed quasar scenario are: i) images positions of the lensed quasar $\boldsymbol{\theta}$, ii) the time delay between images $\Delta t_{ij}$, and iii) the extended images of the host galaxy, which are adopted as constraints to construct the mass models of the lens galaxies.

For modeling i), we use the observed image position $\boldsymbol{\theta}$ to constrain the lens surface mass density $\kappa_{\rm int}$. We determine the deflection angle $\boldsymbol{\alpha_{\rm int}}$ via the lens equation in SL, \begin{equation}
    \boldsymbol{\theta} = \boldsymbol{\beta} -\boldsymbol{\nabla}  \psi_{\rm L,int} (\boldsymbol{\theta}) = \boldsymbol{\beta} - \boldsymbol{\alpha_{\rm int}}(\boldsymbol{\theta}),
    \label{eq:lens_equation}
\end{equation}
and $\boldsymbol{\alpha_{\rm int}}$ is related to $\kappa_{\rm int}$ by

\begin{equation}
 \boldsymbol{\alpha_{\rm int}} = \frac{1}{\pi} \int d^{2} \theta^{\prime} \kappa_{\rm int} ( \boldsymbol{\theta^{\prime} }) \frac{\boldsymbol{\theta} - \boldsymbol{\theta^{\prime}}}{\left| \boldsymbol{\theta} - \boldsymbol{\theta^{\prime}} \right|^{2}}.
\end{equation}
Adopting Eq.~\ref{eq:lens_equation}, we map the observed multiple image positions $\boldsymbol{\theta}$ to the source plane, compute the magnification-weighted average as the modeled source position, and then map it back to the image plane, obtaining $\boldsymbol{\theta}^{\rm pre}$. Magnification weighting improves the accuracy of source position estimation in SL by giving greater importance to highly magnified images, which provide more precise constraints on the lens model. The goodness of the image position modeling is evaluated by 
\begin{equation}
     \chi^2_{\rm img} =   
    \sum_j^{N_{\rm img}} \frac{(\boldsymbol{\theta}_j -\boldsymbol{\theta}_{j}^{\rm pre})^2}{\sigma_{\boldsymbol{\theta},j}^2},
    \label{eq:chi2_img}
\end{equation}
where $\sigma_{\boldsymbol{\theta},j}$ is the positional uncertainty of image $j$.

In modeling (ii), we derive the lens potential $\psi_{\rm L,int}$ from the mass density $\kappa_{\rm int}$ using Eq.~\ref{eq:time delay}. This allows us to model the time delay (tmd) between the observed images. With lensed image $j$ as the reference image, the fit quality for $\Delta t_{ij}$ is assessed by
\begin{equation}
   \chi^2_{\rm tmd} =  \sum_i^{N_{\rm img}-1} \frac{(\Delta t_{ij} - \Delta t_{ij}^{\rm pre})^2}{\sigma_{\Delta t_{ij}}^2},
   \label{eq:chi2_tmd}
\end{equation}
where $\sigma_{\Delta t_{ij}}$ is the time-delay uncertainty. Galaxy-scale lenses typically form either quadruple or double image systems with $N_{\rm img} = 4~\rm or~2$. In such cases, models (i) and (ii) can be calculated in under 0.1 seconds on a 2.10 GHz CPU, achieving the best-fit model within several minutes. Consequently, GPU acceleration is not necessary for these computations, and we continue to perform image position and time-delay modeling using the CPU with \texttt{GLEE}.

Extended image modeling is the bottleneck in SL analysis, as it involves handling approximately $\mathcal{O}(10^4)$ data points across the magnified arcs. We represent the source intensity distribution on a grid of pixels using the vector $\boldsymbol{s}$, which has a dimension of $N_{\rm s}$, corresponding to the number of source pixels. Based on the assumed $\kappa_{\rm int}$ and the PSF introduced by the telescope, we construct an operator $\boldsymbol{\rm f}$, following \citet{2006esr}. This operator utilizes Eq.~\ref{eq:lens_equation} to map the light intensity of the extended source from the source plane to the image plane, followed by convolution with the PSF, producing the predicted lensed extended source $\boldsymbol{d}_{\rm esr}^{\rm pre}$ with a dimension of $N_{\rm d}$ (i.e., predicted intensity values of the $N_{\rm d}$ pixels on the image plane),
\begin{equation}
  \boldsymbol{d}_{\rm esr}^{\rm pre} ={\boldsymbol{\rm f}}~\boldsymbol{s} + \boldsymbol{n}  
  \label{eq:dpre}
\end{equation}
with
\begin{equation}
    \boldsymbol{\rm f} = \boldsymbol{B}~\boldsymbol{L}
\end{equation}
where $\boldsymbol{B}$ the blurring matrix accounting for the PSF effect and $\boldsymbol{L}$ presenting the mapping process from source plane to image plane, $\boldsymbol{n}$ is the noise of the observed data and characterized by the covariance matrix $\boldsymbol{C_{\rm d}}$. 

The pixelated source $\boldsymbol{s}$ is reconstructed by maximizing the posterior probability of $\boldsymbol{s}$, given the data.
\begin{equation}
    P(\boldsymbol{s}~|~\boldsymbol{d}_{\rm esr}, \lambda,\boldsymbol{\rm f}, \boldsymbol{\rm g}) = \frac{\mathcal{L}( \boldsymbol{d}_{\rm esr}~|~ \boldsymbol{s}, \boldsymbol{\rm f}) P(\boldsymbol{s}~|~\boldsymbol{\rm g}, \lambda)}{P( \boldsymbol{d}_{\rm esr}~|~\lambda, \boldsymbol{\rm f}, \boldsymbol{\rm g})}, 
\label{eq:bayes1}
\end{equation}
where the regularization operator $\boldsymbol{\rm g}$ and constant $\lambda$ define the method used to enforce smoothness in the reconstructed source and the strength of the smoothness. The most frequently applied regularization in the SL is \textit{curvature} which minimizes the second derivatives of the source intensity distribution. The analytical form of the most probable source reconstruction $\boldsymbol{s}_{\rm MP}$ is 
\begin{equation}
    \boldsymbol{s}_{\rm MP} = (\left[\boldsymbol{F} + \lambda \boldsymbol{\rm g}\right])^{-1}~\boldsymbol{D}
    \label{Eq:s_MP}
\end{equation}
with $\boldsymbol{F}$ 
\begin{equation}
    \boldsymbol{F} = \boldsymbol{\rm f}^{\rm T} \boldsymbol{C_{\rm d}^{-1} }\boldsymbol{\rm f}
\end{equation}
and $\boldsymbol{D}$ 
\begin{equation}
  \boldsymbol{D} =  \boldsymbol{\rm f}^{\rm T} \boldsymbol{C_{\rm d}^{-1} } \boldsymbol{d}_{\rm esr}
\end{equation}
\citep{2006esr}.
We substitute the Eq.~\ref{Eq:s_MP} into Eq.~\ref{eq:dpre}, inferring $\boldsymbol{d}_{\rm esr}^{\rm pre}$ and then compare it with the intensity of the observed extended arcs $\boldsymbol{d}_{\rm esr}$ in the image plane. The goodness of the extended image modeling is evaluated by the Bayesian evidence, which marginalizes over all possible values of the regularization constant 
 $\lambda$ and the pixel values on the source grid $\boldsymbol{s}$,
\begin{align}
    P(\boldsymbol{d}_{\rm esr} ~|~ \boldsymbol{\rm f}, \boldsymbol{\rm g}) &= \int \mathrm{d}\lambda \, P(\boldsymbol{d}_{\rm esr}~|~\boldsymbol{\rm f}, \lambda, \boldsymbol{\rm g})  \notag \\
    &\simeq P(\boldsymbol{d}_{\rm esr}~|~\boldsymbol{\rm f}, \hat{\lambda}, \boldsymbol{\rm g}) \notag \\
    &= \int \mathrm{d}\boldsymbol{s} \, P(\boldsymbol{d}_{\rm esr}~| ~\boldsymbol{\rm f}, \boldsymbol{s}, \hat{\lambda}, \boldsymbol{\rm g}) P(\boldsymbol{s} | \hat{\lambda}, \boldsymbol{\rm g})
    \label{eq:evi_ext}.
\end{align}
The distribution of possible \( \lambda \) values is approximated by a delta function centered at the optimal regularization constant \( \hat{\lambda} \), which justifies the validity of the approximation in Eq.~\ref{eq:evi_ext} \citep[]{2006esr}. The explicit expression of \( P(\boldsymbol{d}_{\rm esr} |~\boldsymbol{\rm f}, \hat{\lambda}, \boldsymbol{\rm g}) \) is given in \citet{2006esr}, see Eq.~(19). 

The steps outlined above represent the core processes of extended image modeling, which involve extensive manipulation of large matrices. This is why the use of GPUs can provide considerable advantages. The matrix sizes are displayed in Tab.~\ref{tab:matrix_size}. Since the source plane is unobservable, the different source grid resolutions yield the best-fit model in slightly different regions of the parameter space. To account for this degeneracy, the modeling with a series of different source grid resolutions is performed in the SL cosmography analysis and the impact of the grid resolution is marginalized over.

\begin{figure}
  \includegraphics[width=0.8\linewidth]{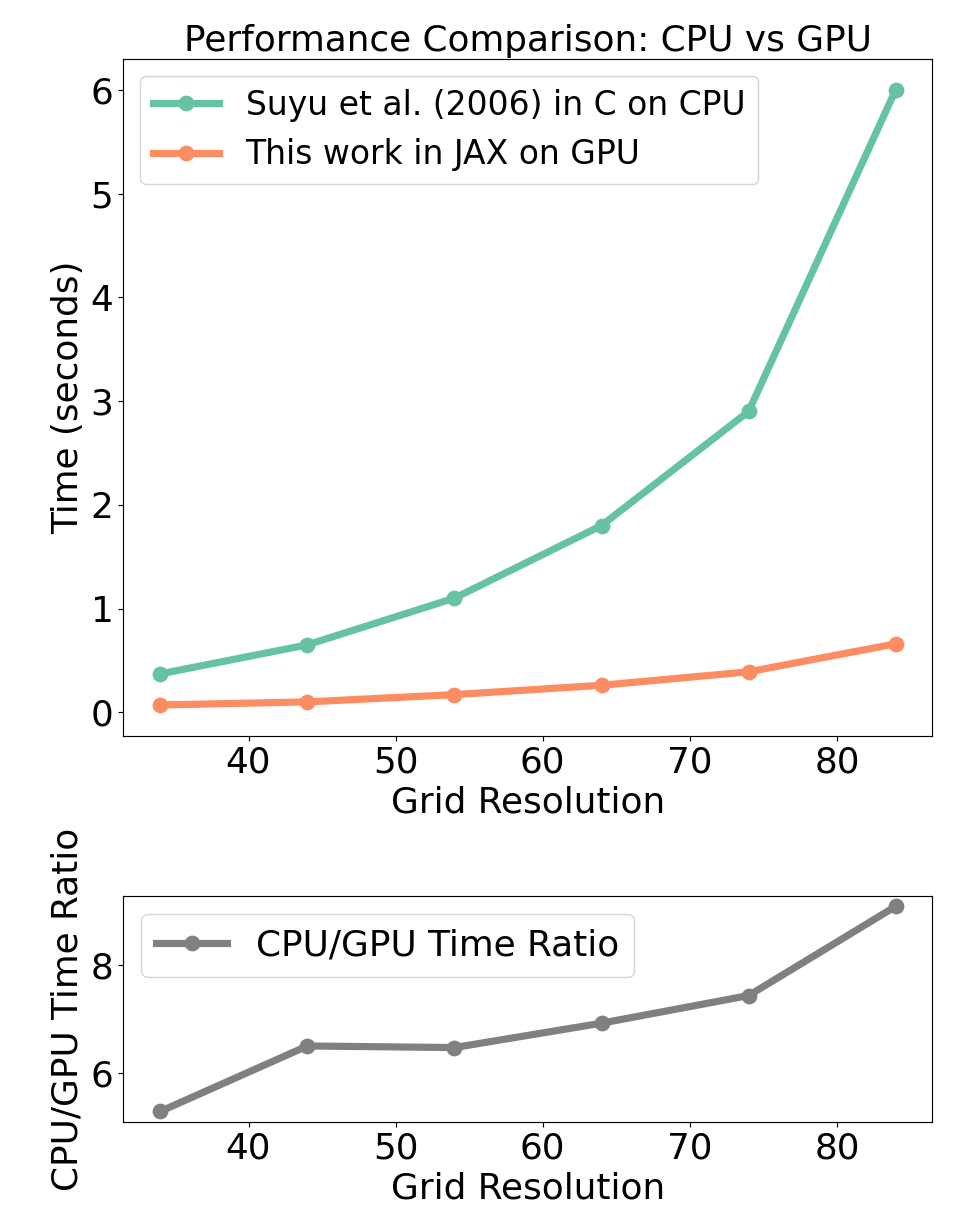}
  \caption{The time comparison between CPU and GPU for extended image modeling is performed using various source resolutions commonly adopted in practice. The computation time is for a single iteration of source and image intensity reconstruction given values for lens mass model parameters. The computations take place on a 2.10 GHz CPU and an A100 GPU, respectively. }
  \label{fig:src_GPU_CPU}
\end{figure}

We present the runtime comparison of extended image modeling in \texttt{GLEE}, implemented in $\texttt{C}$ on a CPU, and our implementation in \texttt{JAX} on a GPU, across various source grid resolutions, as shown in Fig.~\ref{fig:src_GPU_CPU}. We achieve greater acceleration with higher grid resolutions due to larger matrix sizes being more effective at fully saturating the massive parallel computing capability of the GPU.
\begin{table}
  \caption{The matrices size in the extended image modeling}
\centering
\begin{tabular}{lc}
\hline
\hline \\[-0.3em]
Matrix & size \\ 
\hline \\[-0.3em]
$\boldsymbol{B}$ &  $(N_{\rm d},  N_{\rm d})$ \\
$\boldsymbol{L}$ &  $(N_{\rm d},  N_{\rm s})$ \\
$\boldsymbol{C_{\rm d}}$&   $(N_{\rm d},  N_{\rm d})$\\
$\boldsymbol{\rm g}$ &  $(N_{\rm s},  N_{\rm s})$  \\   
\hline
\end{tabular}
\tablefoot{For the galaxy-scale lenses, the number of pixels on the extended arc $N_{\rm d}$ is commonly $\sim \mathcal{O}(10^4)$ and the number of source pixels $N_{\rm s}$ is  $\sim \mathcal{O}(10^3)$. }
\label{tab:matrix_size}
\end{table}

\subsubsection{Dark matter profile $\kappa_{\rm enfw}$}

We implement a dark matter profile following \citet{Oguri2021} on the GPU, directly introducing ellipticity into the density mass profile \( \kappa_{\rm eNFW} \), in contrast to the classical approach, which incorporates ellipticity in the potential. Since all lensing properties of \( \kappa_{\rm eNFW} \) have analytical expressions, computing \( \kappa_{\rm eNFW} \) and \( \boldsymbol{\alpha}_{\rm eNFW} \) on a large grid of approximately $\mathcal{O}(10^3) \times \mathcal{O}(10^3)$ takes a negligible amount of time, approximately $10^{-5}~\text{sec}$ on a GPU. In contrast, performing the same computation on a CPU, following the approach of \citet{2002Golse}, takes approximately 7 seconds. The detailed expressions for \( \boldsymbol{\alpha}_{\rm eNFW} \) and \( \psi_{\rm eNFW} \) are provided in Appendix~\ref{app:Implementation of the enfw profile}.

\subsection{GPU acceleration in dynamical modeling}
\label{subsect:GPU acceleration in dynamics modeling}
As discussed in Sect.~\ref{subsection:Stellar dynamics under the internal MSD}, the MGE is commonly used in dynamical modeling as a prerequisite for JAM. Without accounting for the internal MSD, the SB and mass density of the lens galaxies are sufficient for decomposition up to $3r_{\rm eff}$ in dynamic modeling. However, when considering the internal MSD, which represents a constant mass sheet added to the galaxy mass distribution, this additional mass can extend over a significantly larger region. To accurately account for the internal MSD, the mass profile must be decomposed over a larger area, approximately $\sim 50\arcsec$ for lens system RXJ1131 \citep[]{TDCOSMO13, ShajibRXJ1131}.

In \citet{TDCOSMO13}, the authors applied the 2D MGE fitting method \citep{MGE_cap}\footnote{The adopted approach is the function \texttt{mge\_fit\_sectors} from the \texttt{MgeFit} package (\url{https://pypi.org/project/mgefit/}). } to model the light and mass convergence map of the lens galaxy. In both cases, the maps are characterized by smooth profiles such as Sérsic, power-law, and NFW profiles, without any subtle angular structures. Since the maps primarily describe variations with radius, applying the 2D MGE fitting method is unnecessary in this case. The 2D MGE fitting method requires solving a non-linear least-squares minimization problem, which becomes computationally expensive when performed over a broad region extending $\sim 50\arcsec$ from the lens galaxy center. Moreover, producing the light and mass convergence maps in 2D across a wide area with \( \mathcal{O}(10^3) \times \mathcal{O}(10^3) \) pixels is rather time-consuming. In total, it takes $\mathcal{O}(10)$ s per sampling step. The MGE 2D fit is primarily used to capture more detailed structures in galaxies from optical imaging directly, rather than relying on maps derived from profiles.

In this work, we instead adopt the 1D MGE fitting method. We implement a fast Gaussians decomposition to 1D profile following \citet{MGE_shajib} on GPU. In this approach, an integral transform with a Gaussian kernel is introduced:

\begin{equation}
    f(\sigma) = \frac{1}{\text{i} \sigma^{2}} \sqrt{\frac{2}{\pi}} \int_{C} z~ \text{F}(z)~ \exp{\left( \frac{z^{2}}{2\sigma^{2}} \right)}~dz,
    \label{eq:integral_transfer}
\end{equation}
where $\text{F}(z)$ represents any mass or light profiles that need to be decomposed using Gaussians. The transformed integral $f(\sigma)$ can be approximated using the Euler algorithm:
\begin{equation}
    f(\sigma) = \sum_{n=0}^{2P} \eta_{n} \Re(\text{F}(\sigma \chi_{n})),
\end{equation}
where $\eta_{n}$ and $\chi_{n}$ can be complex-valued and are independent of $f(\sigma)$. These values can be precomputed at the start. The standard deviations $\sigma_{n}$ are chosen to be logarithmically spaced within the fitting region, resulting in:
\begin{equation}
    F(r) = \sum_{n=0}^{N} A_{n} \exp{\left( \frac{R^{2}}{2\sigma_{n}^{2}} \right)},
\end{equation}
where the amplitude $A_{n} = w_{n} f(\sigma_{n}) \Delta (\log \sigma)_{n}/\sqrt{2\pi}$, with $w_{n}$ representing fixed weighting factors and $R = \sqrt{x^2 + y^2/q^2}$. This MGE approach fits each mass or light density profile using 21 Gaussians to recover the profile within $\sim 0.5\%$ accuracy and runs in approximately \SI{2.e-4}{\second} on a single GPU.

We present the runtime of the 1D MGE fitting implemented in \texttt{JAX} in Tab.~\ref{tab:runtime_comparison} and compare it with the \texttt{NumPy} version from \citet{MGE_shajib}. In this case, GPU acceleration does not provide a significant speedup, achieving a runtime comparable to that of a single mass profile. However, performance gains are realized when the models contain multiple 1D profiles of the same type. By leveraging the Just-in-Time (\texttt{@jit}) compiler and the \texttt{vmap} function in \texttt{JAX}, MGE fitting can be applied simultaneously to these profiles, improving efficiency. For readers interested in the speed comparison with the commonly used \texttt{MgeFit} package, we also provide a runtime comparison. In general, switching to the MGE 1D fit results in negligible computation time on both CPU and GPU.

 We reimplement part of the \texttt{jam.axi.proj} function from the \texttt{JamPy} package\footnote{\url{https://pypi.org/project/jampy/}} to compute $\overline{v_{\rm LoS}^2}$, the second velocity moment along the $z^{\prime}$-axis on the plane of the sky. The main computational bottleneck lies in solving the Jeans equations (Eqs.~\ref{eq:J1} and \ref{eq:J2}) to derive $\overline{v_r^2}$, $\overline{v_\theta^2}$, and $\overline{v_\phi^2}$ (see Sect.~5.1 in \citet{JAM_sph}). These computations involve numerical integrals, which are evaluated using adaptive quadrature methods in Shampine (2008). The integration region is initially divided into four subrectangles, and the integral in each subregion is computed using Gauss-Kronrod quadrature. If the estimated error in any subregion exceeds a predefined threshold, that subregion is further subdivided into four smaller subrectangles, and the process is repeated iteratively until the desired accuracy is achieved.

To enhance computational efficiency with the Just-in-Time (JIT) compiler, we modified the algorithm to use a fixed fine mesh. Specifically, the entire integration region is pre-divided into 64 subregions, with each subregion further subdivided into four smaller subrectangles, where Gauss-Kronrod quadrature is applied to compute the integral. The fractional error of $\boldsymbol{v_{\rm rms}^{\rm pre}}$ compared to the results from the \texttt{JamPy} package is, on average, $10^{-5}$, well within the relative error tolerance of 0.01 set by \texttt{JamPy}. This level of accuracy is sufficient given the relatively simple mass and light profiles used in this paper to compute $\boldsymbol{v_{\rm rms}^{\rm pre}}$. However, for more complex mass potentials and luminosity density tracers, a finer integration grid may be required to achieve the same level of precision. 

Switching to the non-adaptive integral solver enables the simultaneous computation of $\overline{v_r^2}$, $\overline{v_\theta^2}$, and $\overline{v_\phi^2}$ at the required positions, significantly reducing the computation time from approximately $\sim$\SI{10}{\second} to $\sim$\SI{0.3}{\second} for over 200 points in polar coordinates on an A100 GPU, assuming a composite mass model. This model consists of baryonic and dark matter components, a BH, and a mass sheet to account for internal MSD (see Tab.~\ref{tab:runtime_comparison}).

\subsection{Joint modeling}
\label{subsect:Joint modeling}
In this section, we provide a detailed description of the joint modeling approach for time-delay cosmography. The input data $\boldsymbol{d}_{\rm LD}$ consists of both lensing and kinematic observations. The lensing data include the lens light, quasar image positions, the extended image of the host galaxy and the time delays between multiple observed images. The kinematic data comprise the spatially resolved kinematics map of the lens galaxy.

 We use two Chameleon profiles to model the lens light in the optical image, which consists of two isothermal profiles with different core radii $\omega_{\rm c}$ and $\omega_{\rm t}$,
\begin{equation}
\begin{split}
I_{\rm cham} (x, y) = \frac{I_0}{1 + q} &\left( \frac{1}{\sqrt{x^2 + \frac{y^2}{q^2} + \frac{4{\omega_{\rm c}}^2}{(1+q)^2}}} - \right. \\
&\left. \quad \frac{1}{\sqrt{x^2 + \frac{y^2}{q^2} + \frac{4{\omega_{\rm t}}^2}{(1+q)^2}}} \right).
\label{eq: Chameleon}
\end{split}
\end{equation}

The goodness of the lens light fitting is evaluated by 
\begin{equation}
    \chi_{\text{light}}^2 = \sum^{N_\text{p}}_{j = 1} \frac{\left(I_{j}-I_j^{\rm pre} \otimes \text{PSF}\right)^2}{\sigma_{\text{light,} j}^2},
  \label{Eq:chi2 of lens_light}
\end{equation}
where $I_j$ is the surface brightness in the pixel of the lens galaxy, and the PSF is the point spread function. The number of pixels $N_{\rm p}$ used for lensing light modeling in Eq.~\ref{Eq:chi2 of lens_light} excludes those used for modeling extended arcs (which already account for the lens light).

We adopt parameterised mass profiles $\kappa_{\rm int}$ in the joint modeling. There are two mass classes. 
\begin{itemize}[label=\textbullet]
 \item {$\kappa_{\rm int,comp} = (1-\lambda_{\rm int}) + \lambda_{\rm int}(\Upsilon_{\ast} \cdot I_{\rm light}  + \kappa_{\rm enfw} +\kappa_{\rm BH})$}
    \item {$\kappa_{\rm int,epl} = (1-\lambda_{\rm int}) + \lambda_{\rm int}\kappa_{\rm epl} $}.
   
\end{itemize}
In the first scenario, we model the baryonic component and dark matter of the lens galaxies separately. The baryonic component is represented by scaling the lens light profile  \( I_{\rm light} \), with a constant factor $\Upsilon_{\ast}$, while the dark matter is modeled using \( \kappa_{\rm enfw} \) (see Eq.~\ref{eq:kappa_enfw}). \( I_{\rm light} \) consists of two Chameleon profiles. The BH mass is included as a point mass \( \kappa_{\rm BH} \).  In the second scenario, we use an elliptical power-law (EPL) profile \( \kappa_{\rm epl} \) to represent the total mass (see Appendix~\ref{app: Implementation of the EPL profile}).  Because the EPL profile has a softening scale $r_{\rm scale} = 0.01\arcsec$ that is set to a small value, the mass distribution diverges in the center, eliminating the need to add a separate point mass to represent the BH. In addition, we adopt an external shear to account for the tidal stretch from neighboring galaxies with external potential, expressed in polar coordinates $(R,\phi)$ as
\begin{equation}
    \psi_{\rm ext} = \frac{1}{2} \gamma_{\rm ext}R^2\cos{(2\phi -2\phi_{\rm ext})},
\end{equation}
where $\gamma_{\rm ext}$ represents the strength of the external shear, and the shear angle $\theta_{\rm ext}$ represents the stretching orientation of the images. We do not list the external shear in the above $\kappa_{\rm int}$ set-up because it 
adds zero contribution to the mass density with $\kappa_{\rm shear} = \frac{1}{2} \nabla^{2} \psi_{\rm ext} = 0 $. 

In order to explicitly characterize the internal MSD, we adopt a dual pseudo-isothermal elliptical density (dPIE) profile \citep[]{2007dpie1,2010SuyuHalkola_Dpie2}, with a substantial core radius $r_{\rm core} = 45\arcsec$ and truncated at $r_{\rm tr} = 45.09\arcsec$. This profile mimics a flat mass sheet up to a radius of $\sim 20\arcsec$ before tapering down to zero. The extended arc observed at $1.65 \arcsec$ from the galaxy center implies that the lensing-only modeling remains unaffected by this additional mass sheet, rendering the distance $D_{\Delta \rm t, int}$ completely degenerate with $\lambda_{\rm int}$ \citep[]{TDCOSMO13}. The expression for $\lambda_{\rm int}$ is:
\begin{equation}
\begin{aligned}
  \lambda_{\rm int}  &= 1 - \kappa_{\rm dPIE}  \\
  & = 1-  \frac{a_{0}}{2} \frac{r_{\rm tr}^2}{r_{\rm tr}^2 - r_{\rm core}^2} \left( \frac{1}{\sqrt{R^2 + r_{\rm core}^2}}-\frac{1}{\sqrt{R^2 + r_{\rm tr}^2}} \right),
  \label{eq:dpie_kappa}
\end{aligned}
\end{equation}

where $a_{0}$ is a normalisation parameter and $R^2 = x^2 + y^2$. In the region where $R \ll r_{\rm core}$, we obtain an approximately constant mass sheet 
\begin{equation}
  \lambda_{\rm int} \simeq  1 - \frac{a_0}{2} \frac{r_{\rm tr}^2}{r_{\rm tr}^2 - r_{\rm core}^2} \left( \frac{1}{r_{\rm core}}-\frac{1}{ r_{\rm tr}} \right).
\end{equation}
In the region where $R \gg r_{\rm tr}$, we have $\lambda_{\rm int} \simeq 1$, indicating that the added mass sheet effectively vanishes at large scales. We emphasize that the values of $r_{\rm core}$ and $r_{\rm tr}$ are carefully selected based on extensive testing to represent the worst-case scenario. While the internal MSD remains unaffected from a lensing perspective, its impact on the kinematic data is significant enough to impose constraints on $\lambda_{\rm int}$. In addition, the dPIE profile, which features a well-defined truncation radius, declines more rapidly than the mass-sheet profile used in \citet{Blum2020}. This makes it a more suitable choice, as it helps reduce the likelihood of having negative values for $\kappa_{\rm int}$ in the outermost regions. Negative mass convergence is unphysical and is therefore rejected by JAM. 

Ideally, a mass profile for $\lambda_{\rm int}$ that characterizes the internal MSD for lens system RXJ1131 would remain constant out to $\sim 20\arcsec$, and then drop sharply to zero beyond that radius. This behavior would entirely prevent the emergence of negative $\kappa_{\rm int}$. However, such an ``ideal'' mass sheet lacks well-defined lensing properties and is difficult to fit using the MGE. In this work, we adopt a dPIE profile with a fixed $r_{\rm core}$ to represent the mass sheet. The relatively gradual decline of the dPIE profile in the outermost region and the fixed $r_{\rm core}$ can lead to negative $\kappa_{\rm int}$, resulting in an asymmetric probability distribution for $D_{\rm \Delta t,int}$ (see Sect.~\ref{subsec:Break the MSD using Joint Modeling}).

\begin{figure*}
   \centering
  \includegraphics[width=0.85\linewidth]{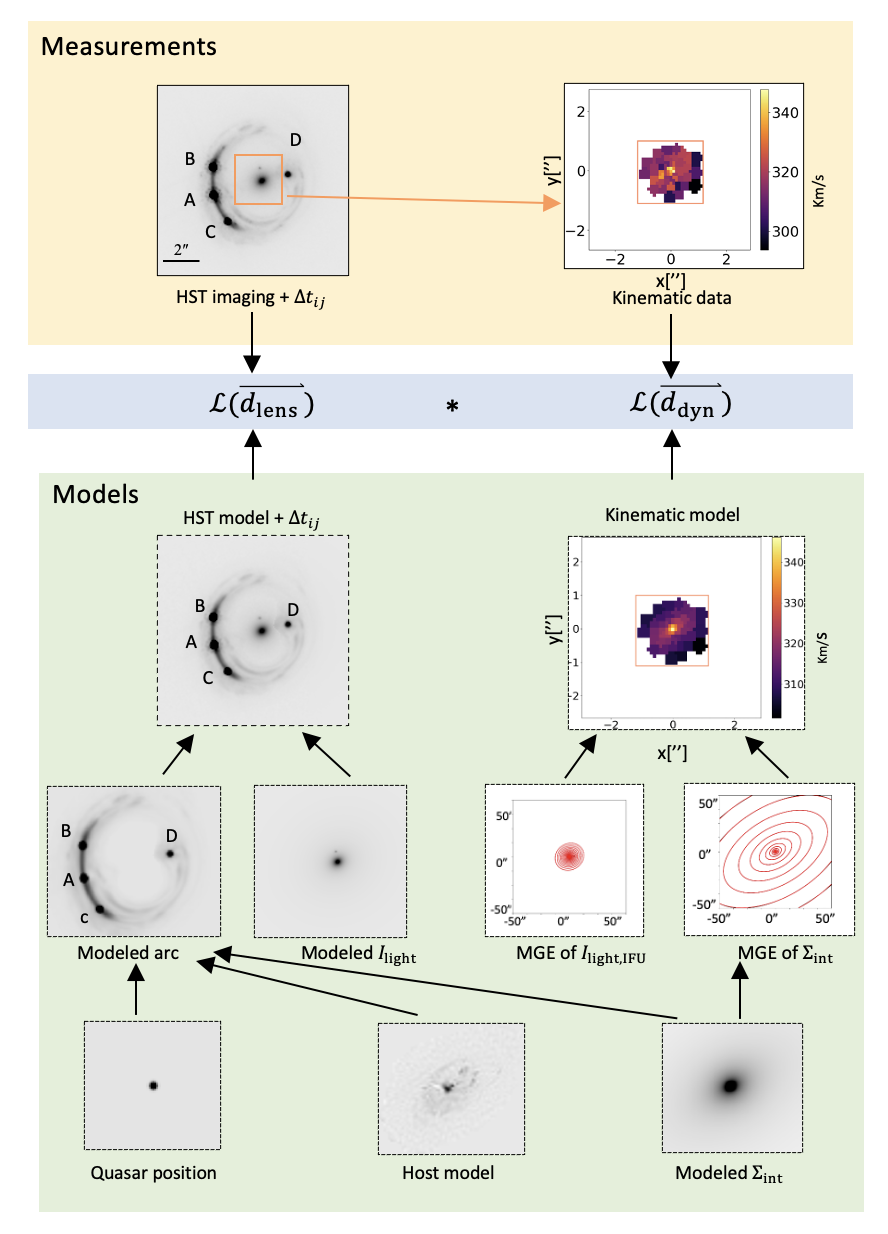}
  \caption{Workflow for joint modeling using RXJ 1131 as an example. The input datasets consist of photometric images and the spatial kinematics of the lens galaxy. The red contours in the middle right of the green panel represent the iso-contours of both the light and mass density distributions of the lens galaxy, derived from the MGE method (see Sect.~\ref{subsect:GPU acceleration in dynamics modeling}). The modeled $I_{\rm light}$ represents the light fitted from optical imaging, whereas $I_{\rm light,IFU}$ corresponds to the light near the spectral absorption lines in the IFU data. In the paper, $I_{\rm light}$ is equivalent to $I_{\rm light,IFU}$. We employ an MCMC sampler to simultaneously sample the parameter space $\boldsymbol{\eta}_{\rm LD}$, for both lensing and dynamical modeling. }
  \label{fig:flowchart_modeling}
\end{figure*}

Using the chosen mass density model, either a composite or power-law model, along with \( I_{\rm light} \), we perform lensing and dynamical modeling simultaneously (see Fig.~\ref{fig:flowchart_modeling}). Both the light and mass density profile of the lens galaxy must have the same position angle $\varphi_{
\rm PA}$, to maintain the axisymmetric assumption. In our joint modeling, we fix this position angle to the mock input value. On the lensing side, we model the extended arc, lensing light, image positions, and time delays. For dynamical modeling, we decompose \( I_{\rm light} \) and \( \Sigma_{\rm int} \) into multiple Gaussian components. The MGE is carried out up to \( 50\arcsec \) from the lensing centroid, corresponding to approximately 200 kpc, ensuring that the mass density \( \kappa_{\rm int} \), transformed by the internal mass sheet, remains physically meaningful at large distances. We restrict our analysis to models with strictly positive total mass density, as negative values would yield unphysical predictions for \( \boldsymbol{v_{\rm rms}^{\rm pre}} \). The predicted kinematic map is then computed by feeding the MGEs of \( I_{\rm light} \) and \( \Sigma_{\rm int} \) into the JAM framework (see Sect.~\ref{subsection:Stellar dynamics under the internal MSD}). In practice, the light $I_{\rm light,IFU}$ near the spectral absorption lines in the IFU data should be provided to JAM to trace the stellar population responsible for these lines. In this paper, we work on the simulated kinematic data. We instead use the best-fit lens light model from the F814W filter in the infrared band. Since the lens galaxy in RXJ1131 is an early-type elliptical galaxy, the infrared band effectively characterizes the dominant stellar populations.
 
The best-fit model is determined through joint modeling within a Bayesian framework. We sample the posterior distribution of parameters \( P(\boldsymbol{\eta}_{\rm LD}|\boldsymbol{d}_{\rm LD}) \) (see Eq.~\ref{eq: Posterior_LD}) using the Metropolis-Hastings Markov Chain Monte Carlo (MCMC) method,

\begin{equation}
\begin{split}
    P(\boldsymbol{\eta}_{\rm LD}|\boldsymbol{d}_{\rm LD}) \propto 
    \mathcal{L}(\boldsymbol{d}_{\rm LD}~|~ \boldsymbol{\eta}_{\rm LD}) P(\boldsymbol{\eta}_{\rm LD})\\
    =  \mathcal{L} (\boldsymbol{d}_{\rm L}~|~ \boldsymbol{\eta}_{\rm LD})\mathcal{L} (\boldsymbol{d}_{\rm D}~|~ \boldsymbol{\eta}_{\rm LD})P(\boldsymbol{\eta}_{\rm LD})\end{split}
\label{eq: Posterior_LD}
\end{equation}
where $\boldsymbol{d}_{\rm L}$ presents the lensing data, $\boldsymbol{d}_{\rm D}$ the kinematic data, and $P(\boldsymbol{\eta}_{\rm LD})$ the prior for the lensing and dynamical parameters. The goodness-of-fit for a model is defined as
\begin{equation}
    \chi^2_{\rm LD} = \chi^{2}_{\rm light} -2 \log P(\boldsymbol{d_{\rm esr}} |~\boldsymbol{\rm f}, \hat{\lambda}, \boldsymbol{\rm g}) + \chi^{2}_{\rm img} + \chi^{2}_{\rm tmd}  + \chi^{2}_{\rm dyn}.
    \label{eq:chi_general}
\end{equation}
The MCMC sampling is conducted on the CPU, where $\boldsymbol{\eta}_{\rm LD}$, involving approximately 10 parameters, is drawn and then transferred to the GPU for extended image, lens light and dynamical modeling. Since the image-position and time-delay modeling involves processing a relatively small dataset, it is kept on the CPU. Although data transfer between the CPU and GPU incurs some latency, the number of transferred data points in our case is on the order of $\sim \mathcal{O}(10)$, resulting in a negligible transfer time.

We achieve a 20× speedup per sampling step using \texttt{JAX} on a single A100 GPU. Tab.~\ref{tab:runtime_comparison} presents the runtime for each step using a composite mass model. Additionally, we include the runtime of the \texttt{JAX} code on a CPU for readers interested in evaluating the parallelization performance gains of \texttt{JAX} on different hardware. We note that JAX is primarily optimized for GPU. On CPUs, its compilation overhead, lack of CPU-specific optimizations, and execution graph transformations can make it slower than \texttt{NumPy}.

\begin{table*}[ht]
    \caption{Time comparison for one-step sampling in joint modeling}
    \centering
    \renewcommand{\arraystretch}{1.1}
    \begin{threeparttable} 
    \begin{tabularx}{\textwidth}{Xllll}
    \toprule
     & Process & Type of Implementation & Runtime (CPU) &  Runtime (GPU)\\
    \midrule
    \multirow{2}{*}{Previous}& Extended Image & \citet{2006esr} in \texttt{C}& \SI{2}{\second} & --\\  
     work  & MGE 1D fit (1 profile; total)  &\citet{MGE_shajib} in \texttt{NumPy} & $\sim$\SI{2.e-4}{\second};  \SI{1.e-3}{\second} &--  \\
        & MGE 1D fit (1 profile; total) & \texttt{mge.fit\_1D(linear = True)} in \texttt{NumPy} \footnotemark[1] & $\sim$ \SI{3.e-3}{\second};  \SI{0.02}{\second}& --\\

                          & $\overline{v_r^2}$, $\overline{v_\theta^2}$, $\overline{v_\phi^2}$ calculations         & Integral solver (adaptive) in \texttt{NumPy} \footnotemark[2]& \SI{13}{\second}&--\\
          &  $\boldsymbol{v_{\rm rms}^{\rm pre}}$ calculation& \texttt{jam.axi.proj} in \texttt{NumPy} \footnotemark[2] &\SI{14}{\second}&--  \\
    \midrule
    \multirow{2}{*}{This paper}  & Extended Image & Follows \citet{2006esr} in \texttt{JAX} & \SI{10}{\second} & \SI{0.21}{\second}\\  
    
    & MGE 1D fit (1 profile; total) & Follows \citet{MGE_shajib} in \texttt{JAX} & $\sim$ \SI{0.13}{\second}; \SI{0.52}{\second} & $\sim$\SI{2.e-4}{\second}; \SI{6.e-4}{\second}\\
&  $\overline{v_r^2}$, $\overline{v_\theta^2}$, $\overline{v_\phi^2}$ calculations         & Integral solver (non-adaptive) in \texttt{JAX} & \SI{118}{\second} &  \SI{0.32}{\second} \\
 &  $\boldsymbol{v_{\rm rms}^{\rm pre}}$ calculation& \texttt{jam.axi.proj} in \texttt{JAX} & \SI{119}{\second}&\SI{0.33}{\second}\\
    \bottomrule
    \end{tabularx}
    \begin{tablenotes}
            \item[1] \url{https://pypi.org/project/mgefit/}
            \item[2] \url{https://pypi.org/project/jampy/}
    \end{tablenotes}
     \end{threeparttable}
    \tablefoot{ Joint modeling is based on a composite mass model with a \(64 \times 64\) source grid (see Sect.~\ref{subsect:Joint modeling} for the adopted profiles). The computations were performed on a 2.10 GHz CPU and an NVIDIA A100 GPU. This table presents the runtime for the MGE 1D fit, both for a single mass or light profile (denoted as ``1 profile'') and for the decomposition of all mass and light profiles in the modeling (denoted as ``total''). The comparison of all MGE 1D fits is conducted using the same number of 21 Gaussians and the same number of log radii. The running time of the integral solver shows the calculation time for the second velocity moment on the diagonal of the tensor, which is the most time-consuming part for deriving $\boldsymbol{v_{\rm rms}^{\rm pre}}$. We adopt the same number of Gaussians for testing, i.e. 42 Gaussians for lens light and 95 Gaussians for the composite mass model. Note that \texttt{JAX} is primarily designed to maximize parallelization performance on GPUs. We present the running time of \texttt{JAX} code on a CPU to isolate the impact of GPU acceleration. In practice, the code is intended to run on GPUs.}
    \label{tab:runtime_comparison}
\end{table*}

\subsection{Bayesian information criterion (BIC)}
\label{subsect:Bayesian information criterion}
In this section, we introduce a BIC method to distinguish the goodness of mass models of lens galaxies with different $\boldsymbol{\eta_{\rm LD}}$. The BIC is an approximation to the Bayesian evidence. 
\begin{equation}
\begin{split}
    P_{\rm LD}(\boldsymbol{d}_{\rm LD}|\mathcal{M}) = \int P_{\rm LD}(\boldsymbol{d}_{\rm LD}|\mathcal{M}, \boldsymbol{\eta}_{\rm LD}) P_{\rm LD}(\boldsymbol{\eta}_{\rm LD}|\mathcal{M}) \, d\boldsymbol{\eta}_{\rm LD}\\
    \approx \exp(-\mathrm{BIC}/2),
\end{split}
\end{equation}
where $\mathcal{M}$ is the constructed mass model with parameters  $\boldsymbol{\eta_{\rm LD }}$. The BIC is defined as
\begin{equation}
    {\rm BIC} = k \ln (n) - 2\rm ln (\mathcal{L}),
\end{equation}
where $k$ is the number of parameters in the model, $n$ is the number of data points, and $\mathcal{L}$ is the maximum likelihood of the model. The BIC penalizes models with a higher number of parameters, effectively balancing goodness of fit with model simplicity. The likelihood in our case is the product of the lensing modeling $ \mathcal{L} (\boldsymbol{d}_{\rm L}~|~ \boldsymbol{\eta}_{\rm LD})$ and dynamical modeling $\mathcal{L} (\boldsymbol{d}_{\rm D}~|~ \boldsymbol{\eta}_{\rm LD})$. The likelihood is easily overwhelmed by the lensing data due to the large number of pixels on the extended arcs. In this work, we focus on using spatially resolved kinematics data to break the internal MSD and constrain $\lambda_{\rm int}$. The lensing-only modeling cannot constrain $\lambda_{\rm int}$ or distinguish between a composite and an elliptical power-law model. Therefore, we exclude $\mathcal{L} (\boldsymbol{d}_{\rm L}~|~ \boldsymbol{\eta}_{\rm LD})$ and the lensing data points, relying on $\mathcal{L} (\boldsymbol{d}_{\rm D}~|~ \boldsymbol{\eta}_{\rm LD})$, the number of kinematic data points, and the number of parameters $\boldsymbol{\eta}_{\rm LD}$ adopted in the joint modeling to weight the posterior distribution.

We identify $\mathcal{M}_{\rm min}$ as the model with the lowest $\rm BIC_{\rm min}$, which corresponds to the minimal $\chi^2_{\rm dyn}$ from the dynamical modeling (since $k$ and $n$ remain the same). The probability ratio of a model $\mathcal{M}_i$ to the model $\mathcal{M}_{\rm min}$ given the data $\boldsymbol{d}_{\rm LD}$ is
\begin{equation}
    \frac{P_{\rm LD}(\boldsymbol{d}_{\rm LD}|\mathcal{M}_i)}{P_{\rm LD}(\boldsymbol{d}_{\rm LD}|\mathcal{M}_{\rm min})} = \exp{ \{ -  ({\rm BIC}_{i} - {\rm BIC}_{\rm min})/2 \} }.
    \label{eq:BIC}
\end{equation}
After normalizing for $N_{\rm m}$ models, we obtain the weighting factor for each model $\mathcal{M}_i$,
\begin{equation}
    f_{\text{BIC},i} = \frac{\exp{ \{ -  ({\rm BIC}_{i} - {\rm BIC}_{\rm min})/2 \} }}{\sum_{i=1}^{N_{\rm m}} \exp{\{-({\rm BIC}_{i} - {\rm BIC}_{\rm min})/2\}}},
    \label{eq:BIC2}
\end{equation}
with ${\rm BIC}_{i} - {\rm BIC}_{\rm min} >= 0$. As discussed in Sect.~\ref{subsect:GPU acceleration in lensng modeling}, the preferred lensing mass parameters vary across different parameter spaces depending on the source resolution. The choice of source pixelization introduces uncertainties in the BIC for a given lens mass parametrization (see Appendix.~\ref{app: Joint modeling with ideal kinematic data across varying source grid resolutions}). To quantify this uncertainty, we compare the BIC values across different source grids and measure the root-mean-square scatter $\sigma_{\rm BIC}$. Following \citet{BirrerH0LiCOW11} and \citet{2020Akin}, we incorporate this uncertainty into the model weighting by convolving $f_{\rm BIC}$ in Eq.~\ref{eq:BIC} with a Gaussian distribution of variance $\sigma_{\rm BIC}^2$, thereby obtaining the updated model weights:
\begin{equation}
    f^*_{\mathrm{BIC}, i}(\mathrm{BIC}_{i}) = h(\mathrm{BIC}_{i}, \sigma_{\mathrm{BIC}}) \otimes f_{\mathrm{BIC}, i}(\mathrm{BIC}_{i}),
\end{equation}
where
\begin{equation}
   h(\mathrm{BIC}_{i}, \sigma_{\mathrm{BIC}}) = \frac{1}{\sqrt{2\pi \sigma_{\mathrm{BIC}}^2}} \exp\left( -\frac{\mathrm{BIC}_i^2}{2\sigma_{\mathrm{BIC}}^2} \right)
\end{equation}

\section{Simulated mock datasets}
\label{sect:Simulated Mock Datasets}

\begin{figure*}
  \centering
  \includegraphics[width=.31\linewidth]{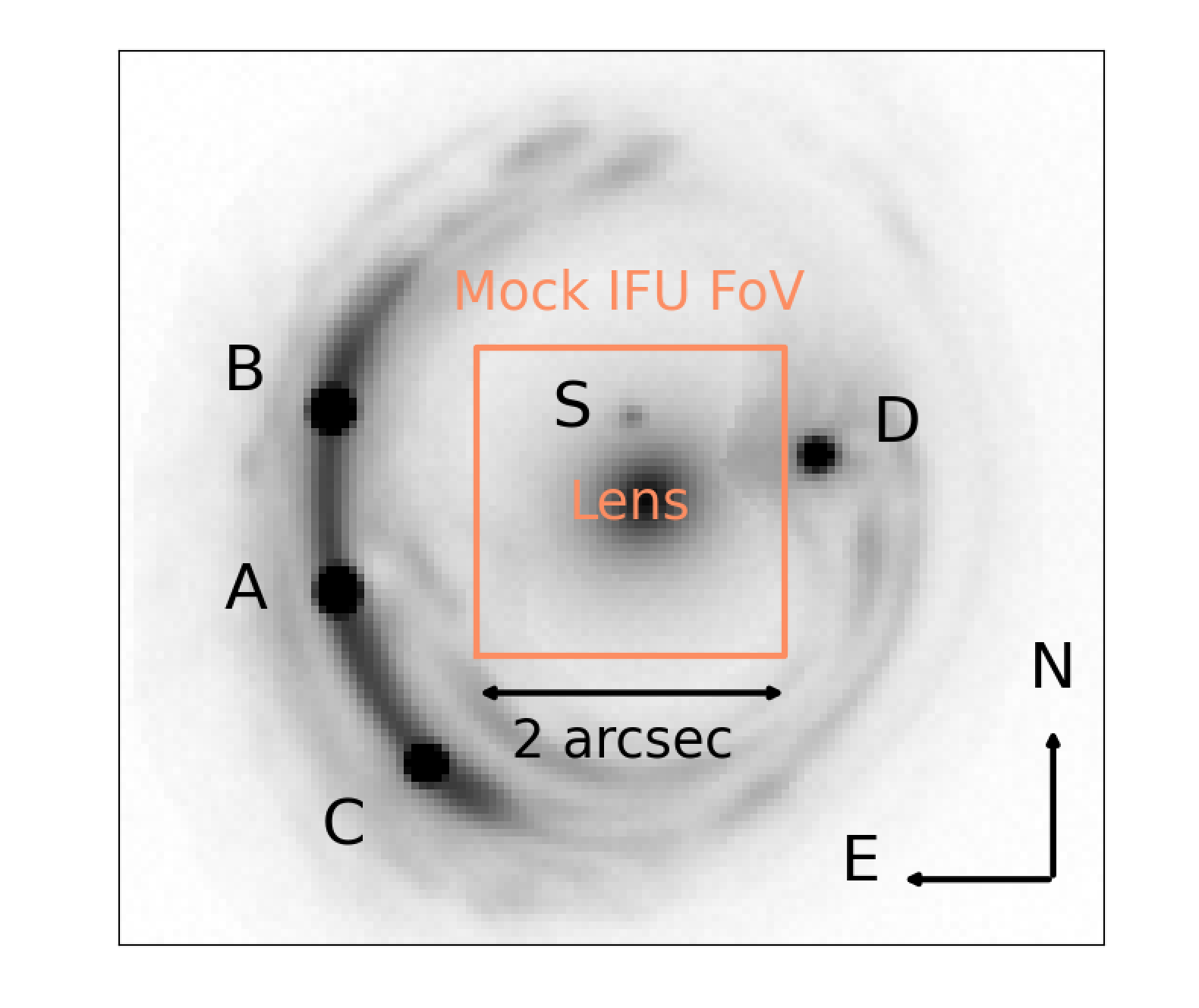}
  \includegraphics[width=.33\linewidth]{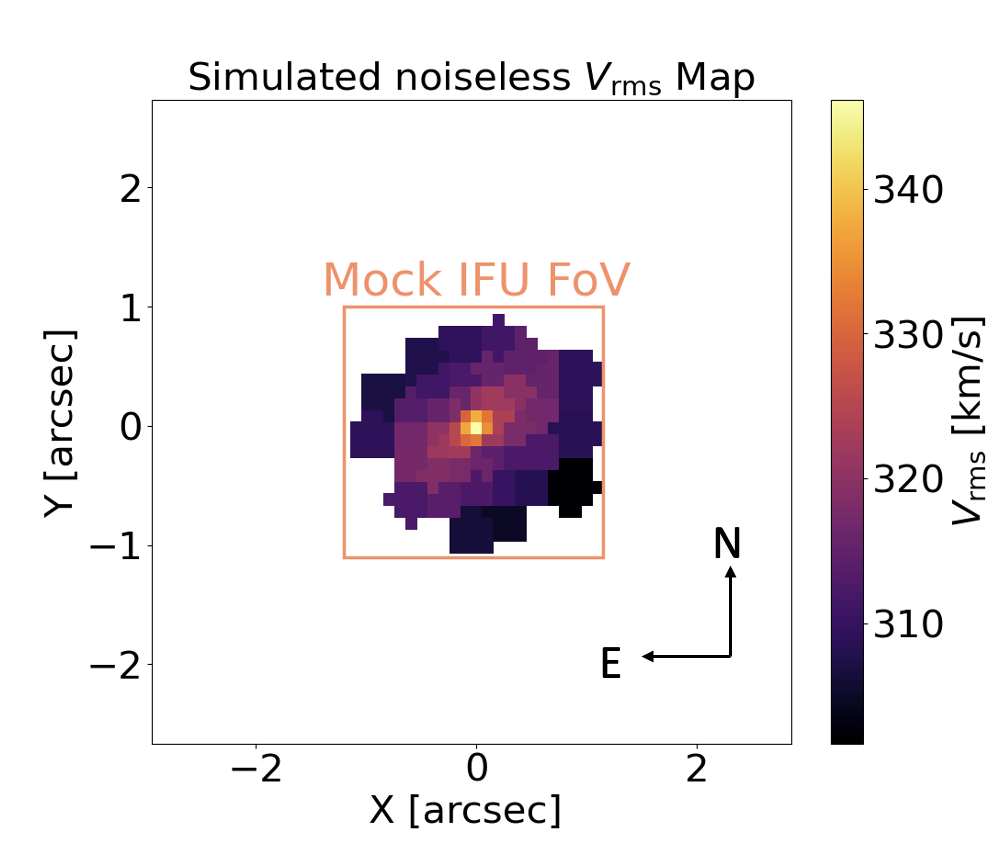}
  \includegraphics[width=.33\linewidth]{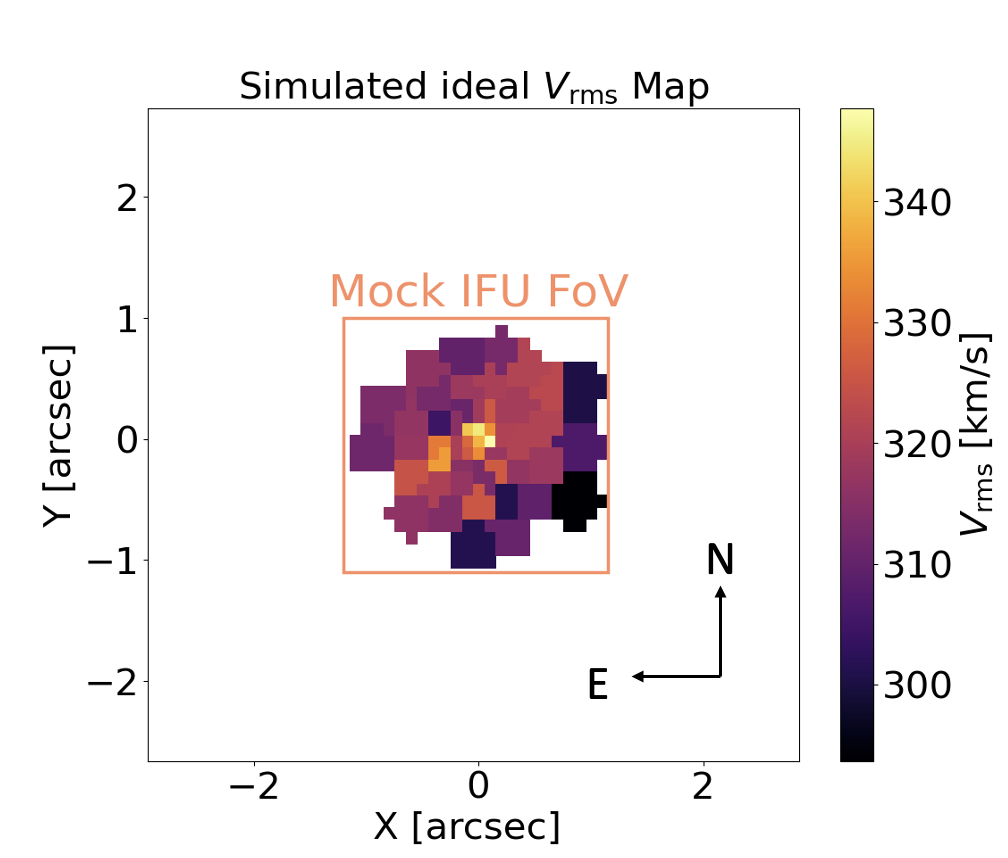}
  \caption{Mock data sets of lensing imaging and kinematic data. Note that we observe a faint satellite galaxy above the lens galaxy at the same redshift as the lens galaxy. We neglect the satellite when mocking up the IFU data. The satellite is too small to extract useful kinematic information from the IFU datacube other than the redshift (see \citet{ShajibJWST}). More importantly, based on the previous study, the satellite has a negligible effect on mass modeling and cosmological distance inference \citep[]{2014Suyu}. The mock IFU data with 52 bins is presented in the same reference frame as the simulated HST imaging, with north oriented upwards and east to the left.}
    \label{fig:mock_lensing_kinematics}
\end{figure*}

RXJ1131 was discovered by \citet{2003Sluse}. The lens galaxy is located at a redshift of $z_{\text{lens}} = 0.295$, while the lensed source galaxy is at a redshift of $z_{\text{s}} = 0.654$, both confirmed through spectroscopy \citep[e.g.,][]{Sluse2007}. The lens is accompanied by a faint satellite galaxy S (see Fig.~\ref{fig:mock_lensing_kinematics}), which JWST NIRSpec has confirmed to be at the same redshift as the lens (see \citet{ShajibJWST}). Imaging data was collected from the Hubble Space Telescope (HST) Advanced Camera for Surveys (ACS) with an exposure time of 1980 seconds. Time-delay measurements for RXJ1131 were made through a dedicated optical monitoring campaign under the COSMOGRAIL program \citep[e.g.,][]{2013Tewes}. These measurements, based on frequent observations (every 3 days) over more than 9 years and involving over 700 epochs using meter-class telescopes and new curve-shifting techniques, reported an approximately $3\%$ precision time delay by \citet{2013Tewes, 2015Liao, 2017Bonvin}. Microlensing-induced time-delay shifts, as analyzed by \citet{2018TieAndKochanek}, have been found to be negligible within the context of the extended delay, as discussed by \citet{chan2018}.

To generate the mock HST imaging of RXJ1131, we use the best-fit mass model obtained from lensing-only modeling of the HST F814W-band imaging, with a source grid resolution of $64 \times 64$. The mass model consists of a composite profile, where the baryonic component is represented by two Chameleon profiles (see Eq.~\ref{eq: Chameleon}) scaled by a constant and the dark matter halo is characterized by $\kappa_{\rm enfw}$. Additionally, the model includes an external shear and a fixed BH mass. The lens galaxy in RXJ1131 exhibits a high central velocity dispersion $\sigma_{\rm disp}$ with $\sigma_{\rm disp}=320 \pm 20~\rm km~s^{-1}$ \citep[]{2014Suyu, ShajibRXJ1131}. By applying the scaling relation between $\sigma_{\rm disp}$ and $M_{\rm BH}$
\citep[e.g.,][]{Kormendy2013review,2013CPMA}, we estimate the BH mass to be between $10^{9}~{\rm M_{\odot}}$ and $10^{10}~{\rm M_{\odot}}$.
Kormendy (2013) predicts \( M_{\rm BH} \approx 2.4 \times 10^9 M_\odot \), while the version by \citet{2013CPMA} gives \( M_{\rm BH} \approx 3.0 \times 10^9 M_\odot \). We set a higher BH mass of \( M_{\rm BH} = 5 \times 10^9~M_{\odot} \) in the simulated kinematic data to explore its effects in cosmography inference. This value remains a reasonable estimate, as suggested by Fig.~16 of \citet{Kormendy2013review} and Fig.~1 of \citet{2013CPMA}. We do not add any mass sheet to the best-fit model, ensuring that $\lambda_{\rm int}^{\rm mock} = 1$, indicating no MSD in the simulated data. We randomly select an external convergence value of $\kappa_{\rm ext}^{\rm mock} = 0.079$ as the ground truth based on the probability distribution function obtained from ray tracing through the Millennium Simulation for the composite mass model \citep[e.g.,][]{2014Suyu}.

To simulate the kinematics map, we follow the approach presented in \citet{2020Akin}.
We use the best-fit lensing light map for the kinematic mock data and assume a Poisson noise-dominated region. The relative pixel intensities are then converted into a relative 2D signal-to-noise map. We adopt \texttt{VorBin}\footnote{\url{https://pypi.org/project/vorbin/}} package \citep{Voronoi} to apply the adaptive spatial binning to the signal-to-noise ratio map, with a target signal-to-noise ratio of 50 per bin. We simulate the data with a high signal-to-noise ratio to ensure that by combining high-quality kinematic data, the internal MSD can be effectively broken. Considering the light contamination from nearby quasar images and the extended host galaxy at the Einstein radius of $\theta_{\rm E} \simeq 1.65~\arcsec$, the simulated binned map covers a small field of view (FoV) ranging from $-1 \arcsec$ to $1 \arcsec$ relative to the lens centroid (see Fig.~\ref{fig:mock_lensing_kinematics}). For simplicity, we neglect the satellite when mocking up the IFU map as well as during the modeling of the SL and stellar kinematic data. We assume a single Gaussian kinematic $\rm PSF_{\rm kin}$ with a FWHM of $0.14\arcsec$, which corresponds approximately to the PSF measured from JWST NIRSpec data of RXJ1131 (see \citet{ShajibJWST}). We generate the noiseless kinematic map with \texttt{JamPy}\footnote{\url{https://pypi.org/project/jampy/}} package based on the mass and light distribution from the best-fit lens model (refer to the best-fit parameters in Tab.~\ref{tab:modeled true values}) and the simulated binned map.

We simulate two kinematic data sets. The first is an ideal kinematic dataset where only statistical errors are added to the noiseless kinematic map:  
\begin{equation}
    {v_{\rm rms, ideal}}_{,l} = {v_{\rm rms}}_{,l} + {\delta v_{\rm stat}}_{,l}
    \label{eq:idea data}
\end{equation}
where ${\delta v_{\rm stat}}_{,l} = \text{Gaussian}[0,  0.02{v_{\rm rms}}_{,l}]$. We assume a statistical error of approximately $2\%$ of the bin values for each Voronoi bin $l$. In the second kinematic dataset, we introduce a 5\% systematic bias to test the impact of potential misfits in the kinematic data:
\begin{equation}
    {v_{\rm rms, biased}}_{,l} = {v_{\rm rms}}_{,l} + {\delta v_{\rm stat}}_{,l} + 0.05 {v_{\rm rms}}_{,l}.
    \label{eq:shift data}
\end{equation}
Systematic errors can arise during the kinematics extraction process, as the measured kinematics may be biased by different methods, such as stellar population synthesis and the use of various stellar libraries such as, X-Shooter \citep[]{2022Verroa, 2022Verrob}, MILES \citep[]{E-MILES}, and Indo-US \citep[]{indo_US}. However, by carefully cleaning the stellar libraries before measuring the kinematics, these systematic errors can be controlled within a sub-percent level \citep[see][]{Knabel2025}.  We test here an overly high level of systematics of 5\% in order to illustrate the impact of a systematic shift in kinematics on the distance inference, even though we anticipate sub-percent level kinematic shifts in reality.

\begin{table*}
    \centering
    \caption{Model parameters and prior for joint modeling}
    \begin{tabular}{ccccc}
    \toprule
    Description & Parameters & Mock input  & prior & prior range\\
    \hline
    \textbf{Flat $\Lambda$CDM} \\
    Hubble constant [$\rm km~s^{-1}~Mpc^{-1}$]&$H_0$ &82.5&Flat& [50, 120]  \\             Matter density parameter&$\Omega_{\rm m}$&0.27&Flat& [0.05, 0.5]    \\

    \textbf{Distances} \\
    Model time-delay distance [Mpc] &$D_{\rm \Delta t, int}$ &1823& Flat & [1000, 4000] \\
    Model lens distance [Mpc]    &$D_{\rm d}$ &775& Flat & [600, 1000] \\
    \textbf{Composite} \\
    Position Angle [$^\circ$] &  $\varphi_{\rm PA}$    & 30      &- &-  \\
    Stellar M/L    &  $\Upsilon_{\ast}$ &1.95& Flat&[0.5, 3.5]\\
    Axis ratio     &   $q_{\rm enfw}$ &    0.56 & Flat &[0.2, 1.0]\\
    \textbf{Characteristic density [-]} & $\rho_{\rm s}$ &0.24 & Flat& [0., 1.]\\
   Scale radius [$\arcsec$] & $r_{\rm s}$ & 23.0& Gaussian& [23.0, 2.6]\\
    External shear strength & $\gamma_{\rm ext}$ &0.09 & Flat& [0.0, 0.2]\\
    External shear position angle & $\phi_{\rm ext}$ & 1.42    & Flat &[0.0, $2\pi$]     \\
    BH mass [${\rm M_{\odot}}$]& $M_{\rm BH}$   &$5 \times 10^{9}$& Discrete & $[10^{9}~{\rm M_{\odot}}, 10^{10}~{\rm M_{\odot}}]$\\
    Mass Sheet & $\lambda_{\rm int}$& 1 &Flat& [0.5, 1.5]\\
    External convergence & $\kappa_{\rm ext}$&0.079&-&-
    \\
    \textbf{Dynamics}\\
    Anisotropy & $\beta_{\rm ani}$& 0.15 & Flat& $[-0.3, 0.3]$\\
    Inclination [$^\circ$]& $i$& 84.3& Flat & [80, 90]\\

    \bottomrule

    \end{tabular}
    \def\sym#1{\ifmmode^{#1}\else\(^{#1}\)\fi}
    \centering
    \tablefoot{The value of $D_{\rm d}$ is determined from $z_{\rm d} = 0.295$, assuming $H_0 = 82.5~\rm km~s^{-1}~Mpc^{-1}$ , $\Omega_{\rm m} = 0.27$ and $\Omega_{\Lambda} = 0.73$. On the contrary, the value of $D_{\rm \Delta t, int}$ must be corrected for external convergence (see Eq.~\ref{eq:Dt external internal}) to obtain the true distance. Applying this correction, we find $D_{\Delta \rm t, int} = 1980.14$~Mpc, given the assumed cosmology. The position angle $\theta_{\rm PA}$ defines the orientation of both the light and mass profiles of the lens galaxy, measured counterclockwise from the +$x$-axis and is fixed during modeling process. The scale radius $r_{\rm s}$ in the dark matter profile indicates the slope transition of the density profile from $-1$ (inner region) to $-3$ (outer region), which cannot be well constrained by either SL or kinematic data due to its large distance from the galaxy centroid. For this reason, a Gaussian prior is used in the joint modeling \citep{Gavazzi}. In each joint modeling, the BH mass is fixed, but we explore different models by probing $M_{\rm BH}$
  within the range of [$10^{9}, 10^{10}]$ ${\rm M_{\odot}}$. Note that $H_0$ and $\Omega_{\rm m}$ are not directly modeled in the joint analysis. The prior in the table indicates the range from which the samples are drawn and then evaluated by the posterior of the distances.}
    \label{tab:modeled true values}
\end{table*}

\section{Analysis and discussion of the joint modeling results}
\label{sect:Analysis and Discussion of the Joint Modeling Results}
In this section, we present the results of the joint modeling using mock lensing and kinematic data. In Sect.~\ref{subsec:Set up of the joint modeling}, we outline the joint modeling setup and describe the modeling procedure. In Sect.~\ref{subsec:Break the MSD using Joint Modeling}, we discuss the fitting results of the joint modeling and demonstrate how it breaks the internal MSD, given ideal kinematic data. In Sect.~\ref{subsect:systematic effect}, we analyze the effect of systematic errors in the kinematic map on $H_0$, given the $5\%$ biased kinematic datasets. In Sect.~\ref{subsec:BH effect}, we examine the impact of $M_{\rm BH}$ on $H_0$ inference, given ideal kinematic data. In Sect.~\ref{subsec: Mitigating the impact of the BH}, we present joint modeling based on ideal kinematic data, excluding the central region, to test whether the impact of an unknown black hole mass can be mitigated, thereby reducing potential bias in distance measurements. We also discuss how the $M_{\rm BH}$–$\beta_{\rm ani}$ degeneracy has been addressed in the literature and explore the potential for applying these solutions to lensing and dynamical modeling. 

\subsection{Joint modeling setup and procedure}
\label{subsec:Set up of the joint modeling}
The mass models adopted in our joint modeling are $\kappa_{\rm int,comp}$ and $\kappa_{\rm int,epl}$ (see Sect.~\ref{subsect:Joint modeling}). Both explicitly represent the mass sheet using a dPIE profile. In the composite model, each component of the galaxy is modeled separately with distinct mass profiles. To account for the central BH, we consider the lens galaxy in RXJ1131, whose velocity dispersion has been measured as $\sigma_{\rm disp} = 320 \pm 20$ km\,s$^{-1}$ by \citet{2014Suyu}. Using the $M_{\rm BH} - \sigma_{\rm disp}$ relation, we explore the full range of possible BH masses of $[10^{9}~{\rm M_{\odot}}, 10^{10}~{\rm M_{\odot}}]$ to be conservative. For each composite mass model setup, we fix the BH mass and increment it in steps of $10^9{\rm M_{\odot}}$ across the specified range. In contrast, the EPL mass model treats the galaxy as a single mass component and does not include an additional mass profile for the BH.

For each mass model setup, whether using $\kappa_{\rm int,comp}$ or $\kappa_{\rm int,epl}$, we perform joint modeling across a range of source grids, from $60 \times 60$ to $68 \times 68$ pixels, increasing in steps of 2. This variation accounts for potential degeneracies introduced by the source-grid resolution. These resolutions are sufficient to mitigate parameter degeneracies while ensuring a good fit to the extended arcs (see Appendix~\ref{app: Joint modeling with ideal kinematic data across varying source grid resolutions} for details). 

The lensing constraints in the joint modeling are consistently provided by the same simulated lensing data. For the kinematic input, we consider two sets of simulated data: one ideal (see Eq.~\ref{eq:idea data}) and one with a $5\%$ bias (see Eq.~\ref{eq:shift data}). For each case, we perform 55 joint modeling runs (1 EPL model plus 10 composite models for the 10 fixed BH masses, and each of the 11 models has 5 source-grid resolutions). The final posterior distributions of the lensing and dynamical parameters are then obtained by combining the results of these 55 models using BIC weighting (see Sect.~\ref{subsect:Bayesian information criterion}), for the ideal and $5\%$ biased kinematic datasets, respectively.

\subsection{Breaking the MSD using joint modeling}
\label{subsec:Break the MSD using Joint Modeling}
\begin{figure*}
\includegraphics[width=1.\linewidth]{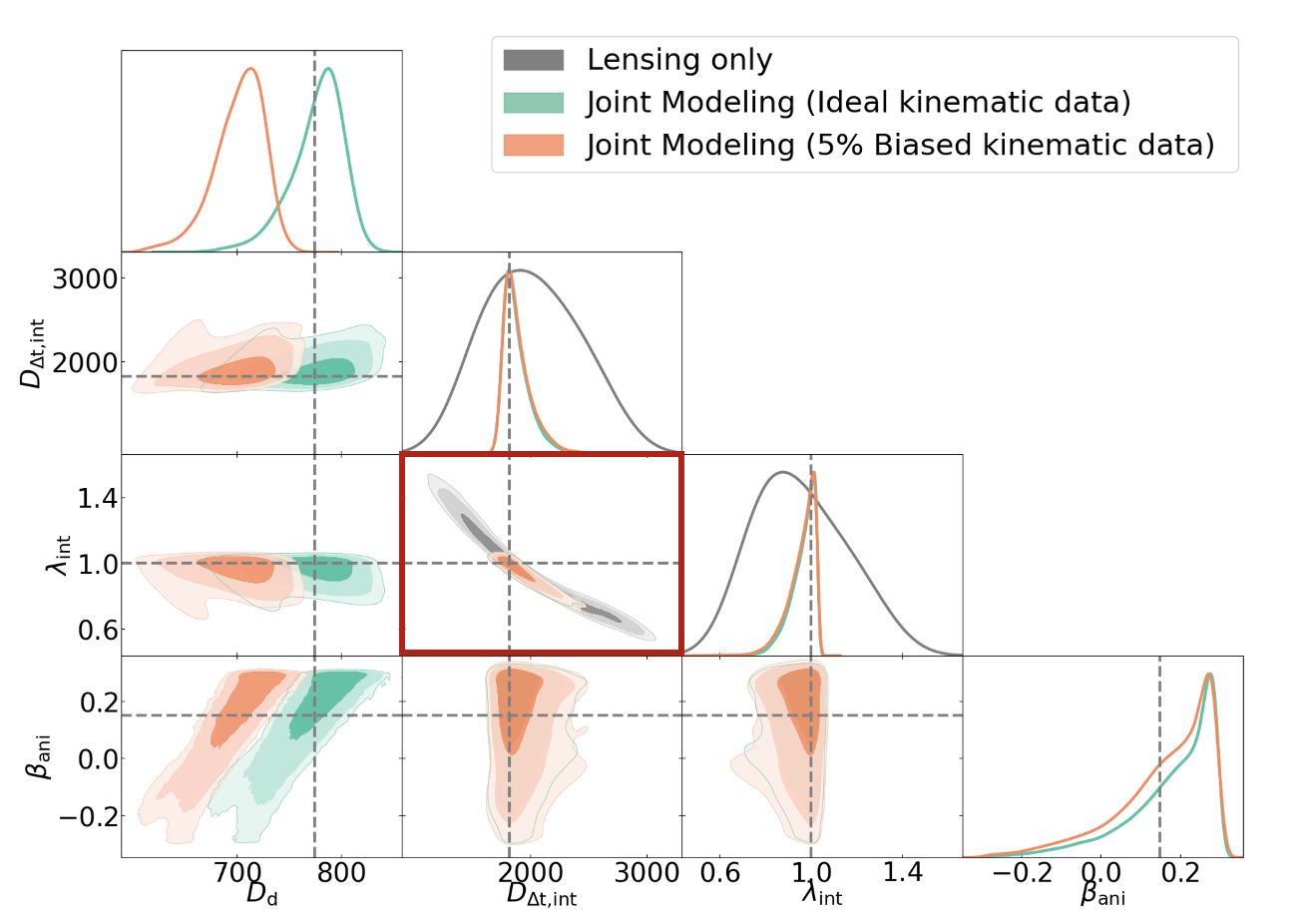}
 \caption{Measurements from joint modeling (combining all mass models) and lensing-only modeling. The shaded contours represent $1\sigma$, $2\sigma$, and $3\sigma$ confidence regions. The green contours correspond to the joint modeling using the ideal kinematic data, while the orange contours represent the joint modeling using the kinematic data with a $5\%$ systematic bias. The grey contours represent the lensing-only modeling. The red box highlights the complete degeneracy between $D_{\Delta \rm t, int}$ and $\lambda_{\rm int}$ in the lensing-only modeling. The green and orange contours illustrate how the internal mass-sheet degeneracy (MSD) is broken when incorporating kinematic data. The inferred $D_{\rm d}$ using the kinematic data with a 5\% bias is biased by $9\%$ (as expected through Eq.~\ref{Eq:scale Dd to Vlos}), while the constraints on $D_{\rm \Delta t,int}$ and $\lambda_{\rm int}$ remain unaffected by this systematic bias. 
 This occurs because $\lambda_{\rm int}$ is constrained by the shape of the $v_{\rm rms}$ profile, which remains unchanged under a systematic 5\% bias in the kinematic map. However, this bias affects the overall amplitude of the profile, leading to a bias in $D_{\rm d}$. In both cases, we observe that $\beta_{\rm ani}$ is degenerate with the distance $D_{\rm d}$ and is not well constrained.}
 \label{fig:Distance_lanbda_beta_ideal_shift}
\end{figure*}

In this section, we illustrate how joint modeling resolves the internal MSD given an ideal kinematic dataset. The time-delay distance $D_{\rm \Delta t, int}$ is entirely degenerate with $\lambda_{\rm int}$ over the prior range of $\lambda_{\rm int}$ when considering lensing-only modeling. The kinematic data aid in constraining $\lambda_{\rm int}$ and in identifying the uniquely preferred $\kappa_{\rm int}$ model within the range $\lambda_{\rm int} \in [0.5, 1.5]$. Consequently, we can break the internal MSD and firmly constrain $D_{\Delta \rm t, int}$ (see the red box in Fig.~\ref{fig:Distance_lanbda_beta_ideal_shift}).

\begin{figure*}
    \centering
    \begin{subfigure}{0.44\textwidth}
        \centering
        \includegraphics[width=\linewidth]{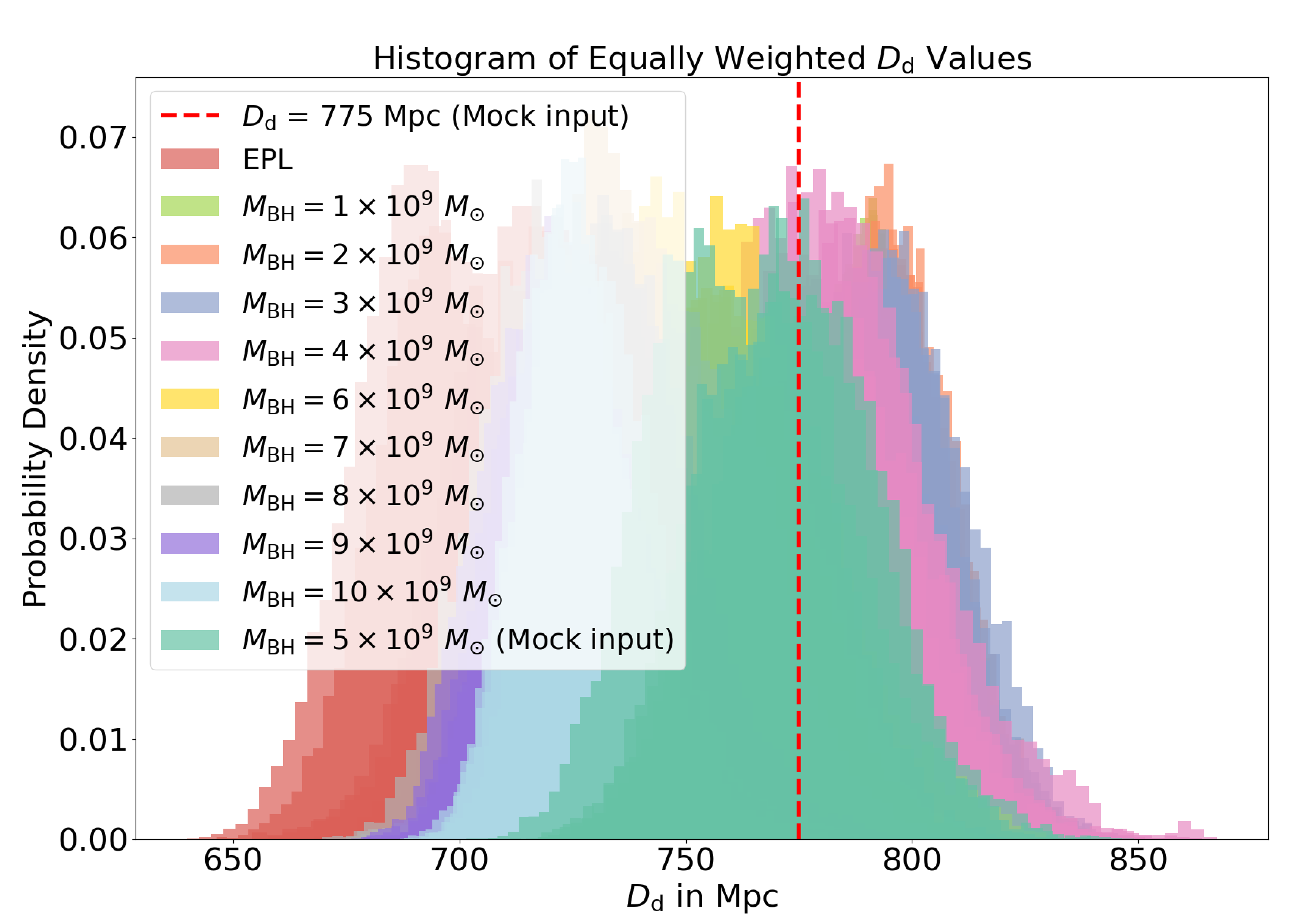}
    \end{subfigure}
    \begin{subfigure}{0.46\textwidth}
        \centering
        \includegraphics[width=\linewidth]{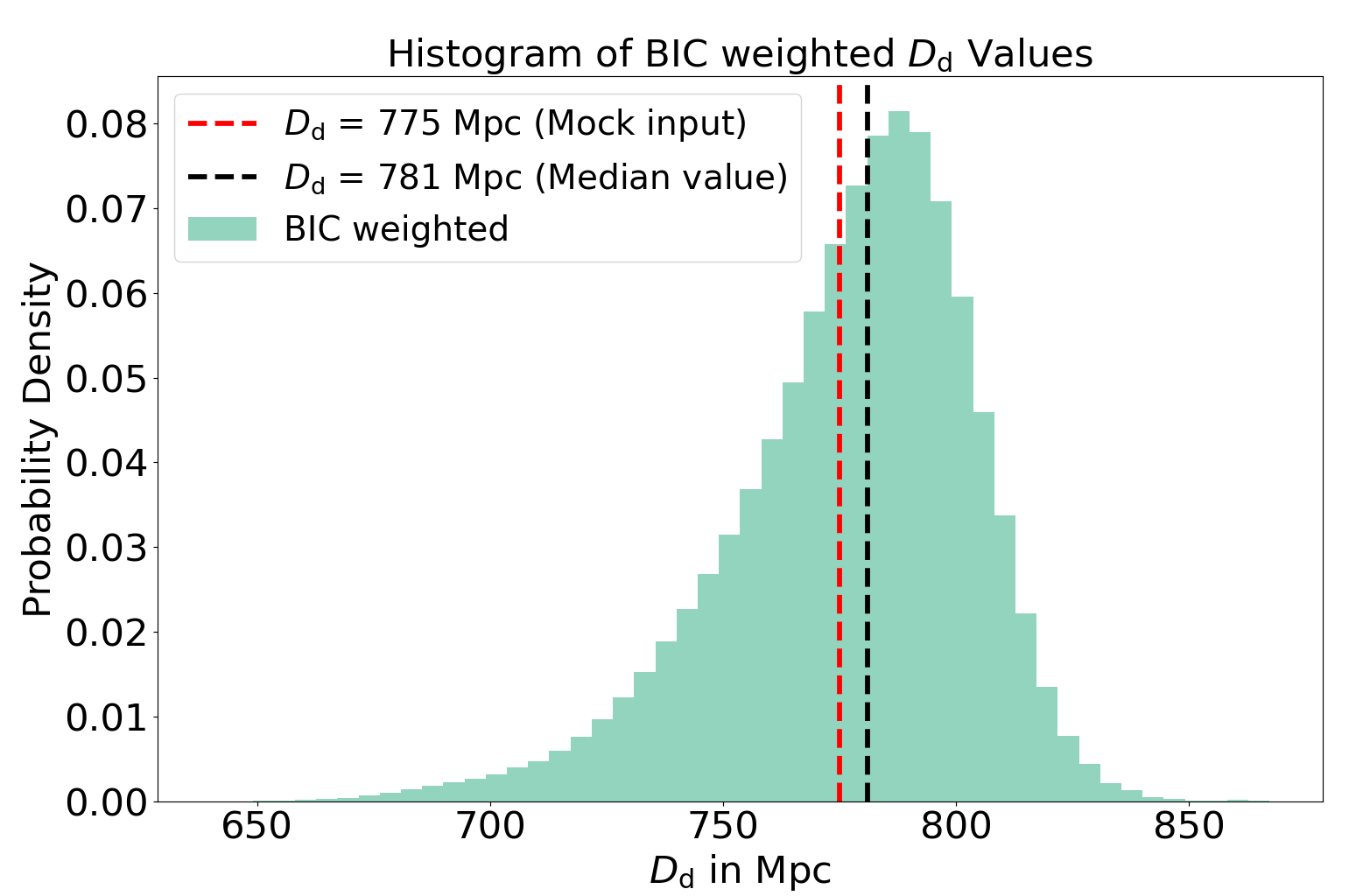}
    \end{subfigure}

    \vspace{0.5cm}

    \begin{subfigure}{0.46\textwidth}
        \centering
        \includegraphics[width=\linewidth]{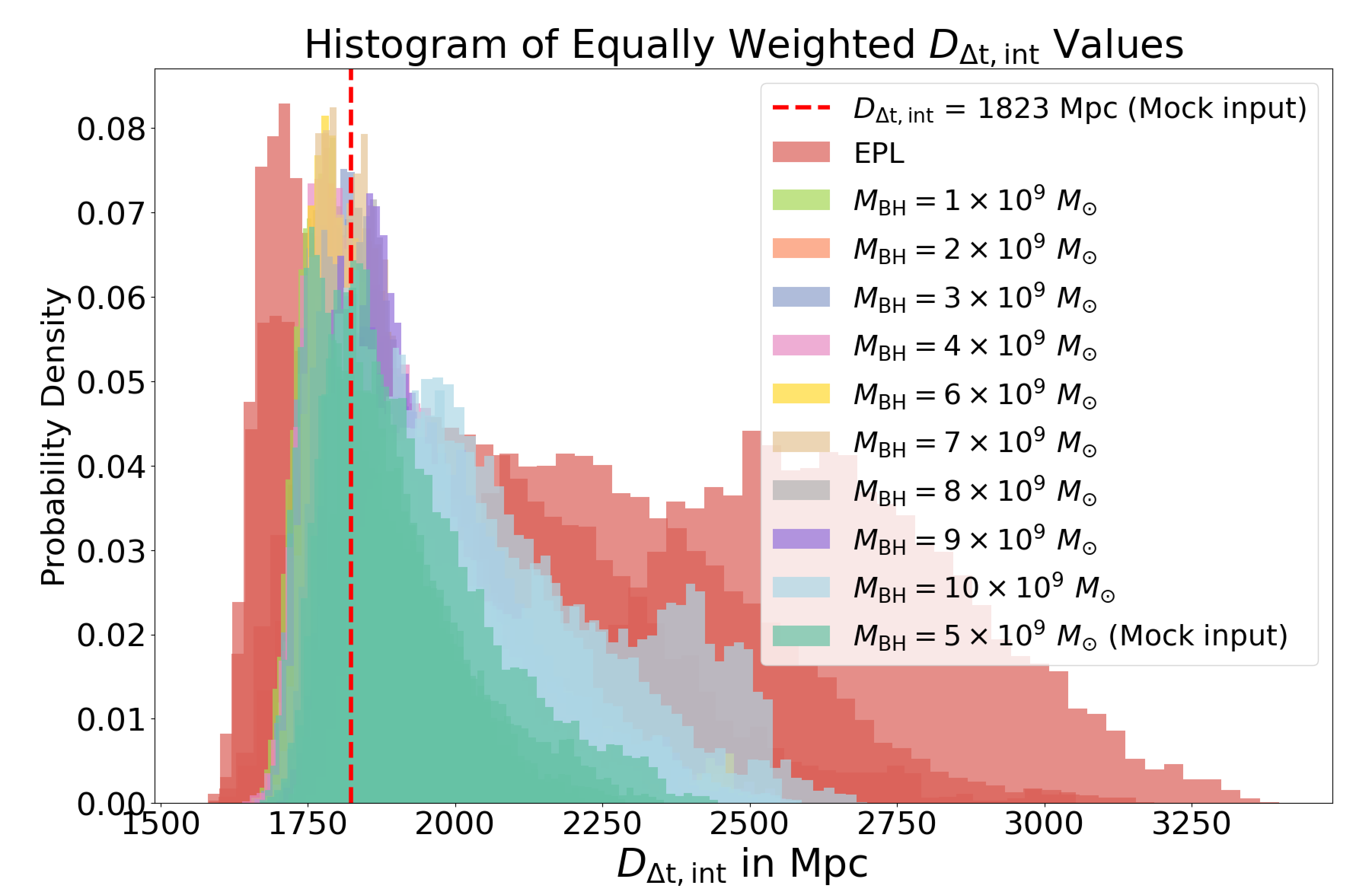}
    \end{subfigure}
    \begin{subfigure}{0.43\textwidth}
        \centering
        \includegraphics[width=\linewidth]{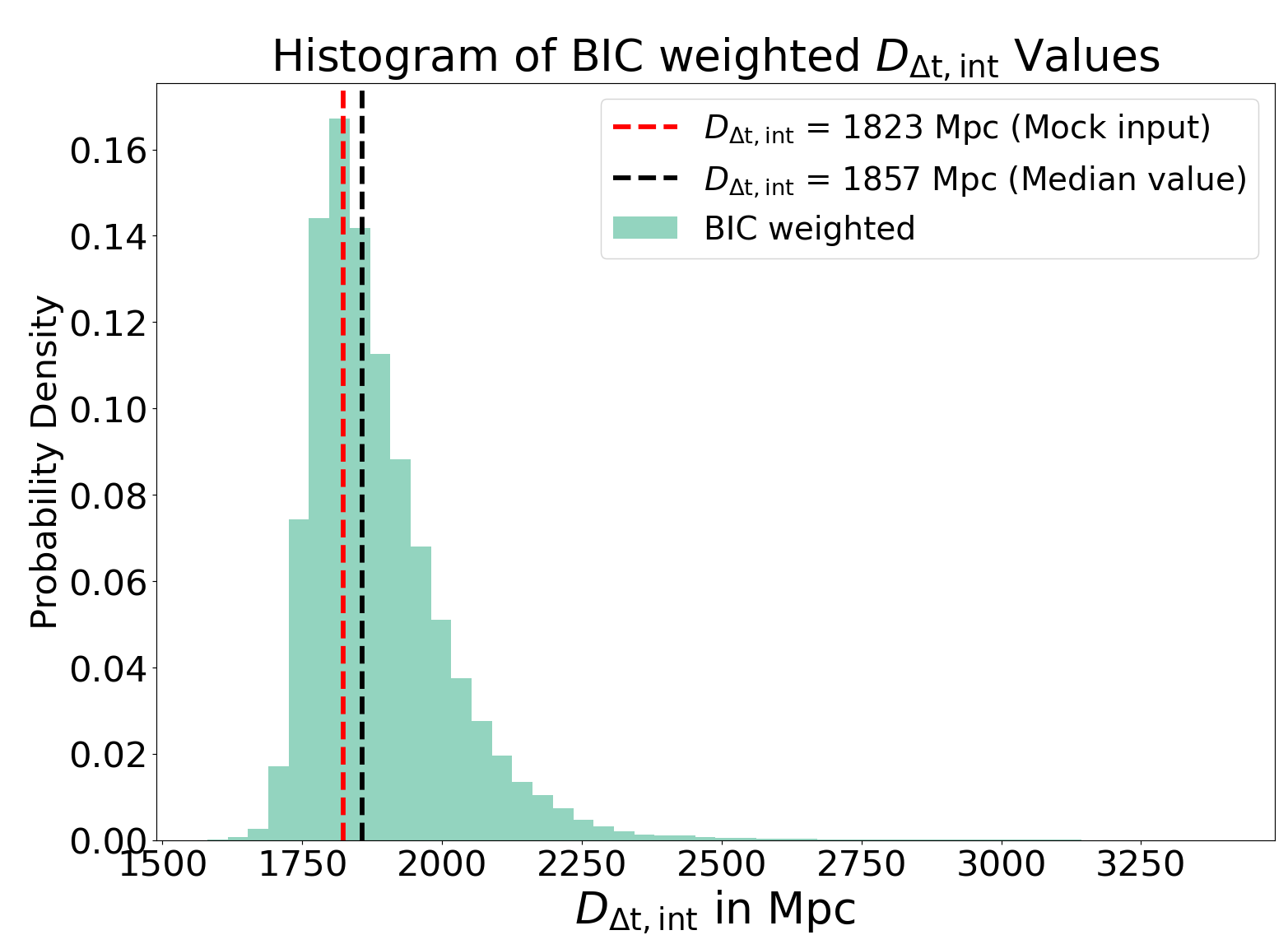}
    \end{subfigure}

    \caption{\textit{Top panels:} Marginalized posterior density distribution of $D_{\rm d}$, based on joint models using ideal kinematic data. Different colors represent posterior densities corresponding to different BH masses, while each color represents five separate posterior distributions that are tightly clustered together, corresponding to models with identical mass parameterizations but different source-grid resolutions. The red color represents EPL mass models, where a small softening scale of $r_{\rm soft} = 0.01\arcsec$ mimics the presence of a massive BH, eliminating the need to explicitly include an additional $M_{\rm BH}$. The left panel shows equally weighted $D_{\rm d}$ posterior densities, whereas the right panel presents the combined $D_{\rm d}$ posterior density weighted by BIC (see Sect.~\ref{subsect:Bayesian information criterion}). \textit{Bottom panels:} Marginalized posterior density distribution of $D_{\rm \Delta t, int}$, based on joint models using ideal kinematic data. The red dashed lines in both panels indicate the mock input values used in the simulated data. The black dashed lines represent the median values in the BIC weighted distribution.
}
    \label{fig:Dd_Dt_equal_BIC}
\end{figure*}

\begin{figure}
    \centering
    \begin{subfigure}{0.5\textwidth}
        \centering
        \includegraphics[width=\textwidth]{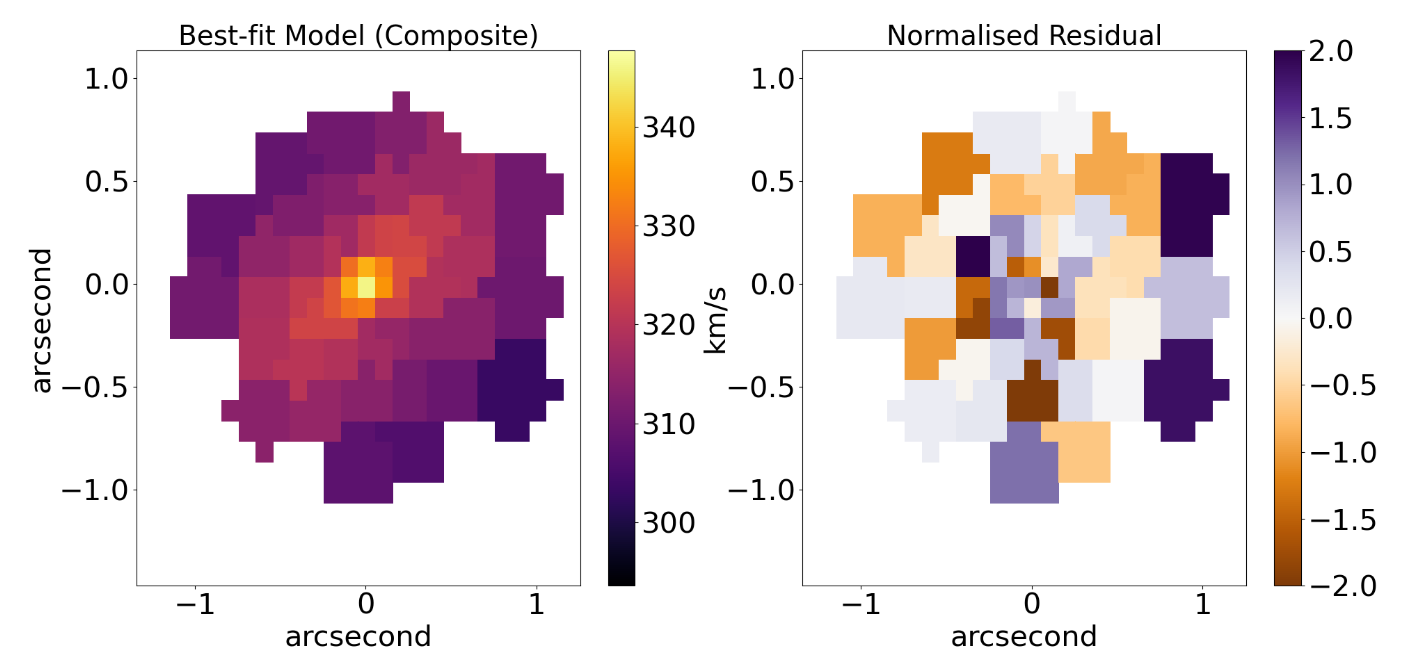}
    \end{subfigure}
    
    \vspace{0.5cm} 

    \begin{subfigure}{0.5\textwidth}
        \centering
        \includegraphics[width=\textwidth]{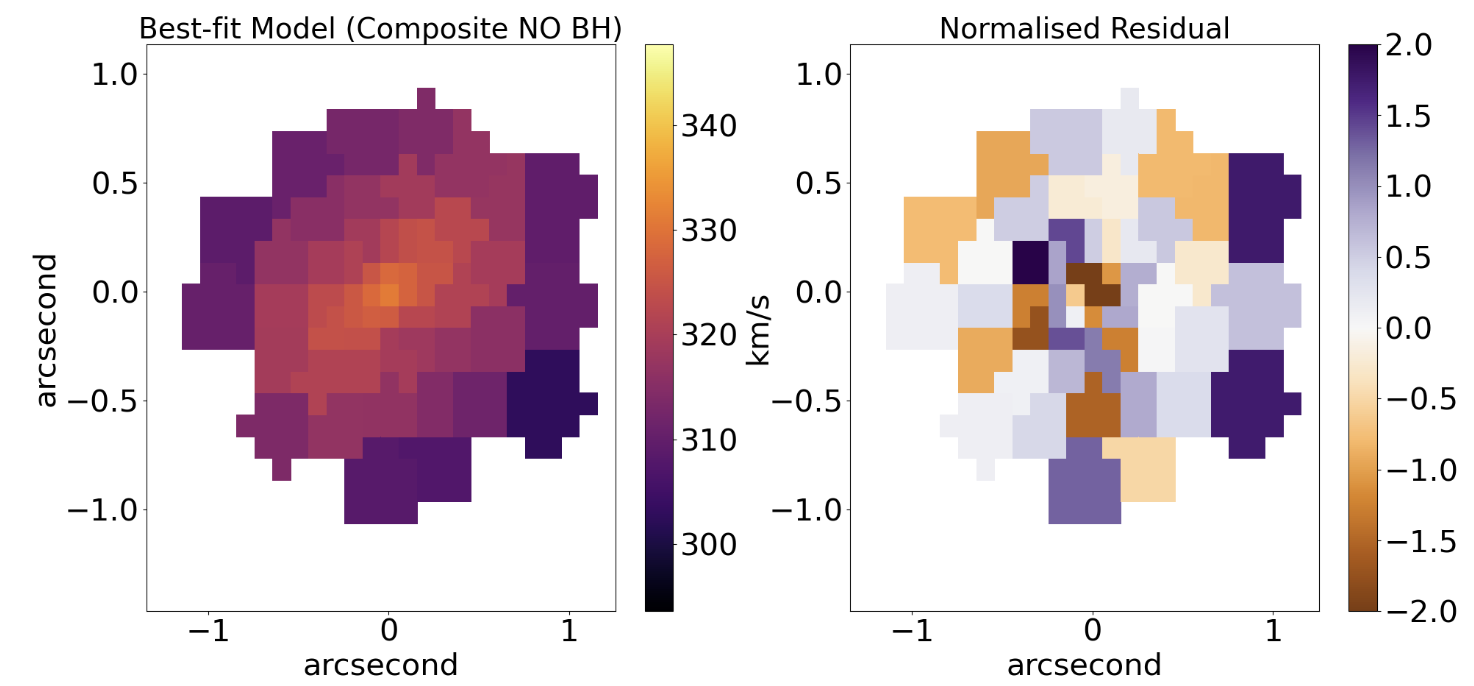}
    \end{subfigure}

    \vspace{0.5cm} 

    \begin{subfigure}{0.5\textwidth}
        \centering
        \includegraphics[width=\textwidth]{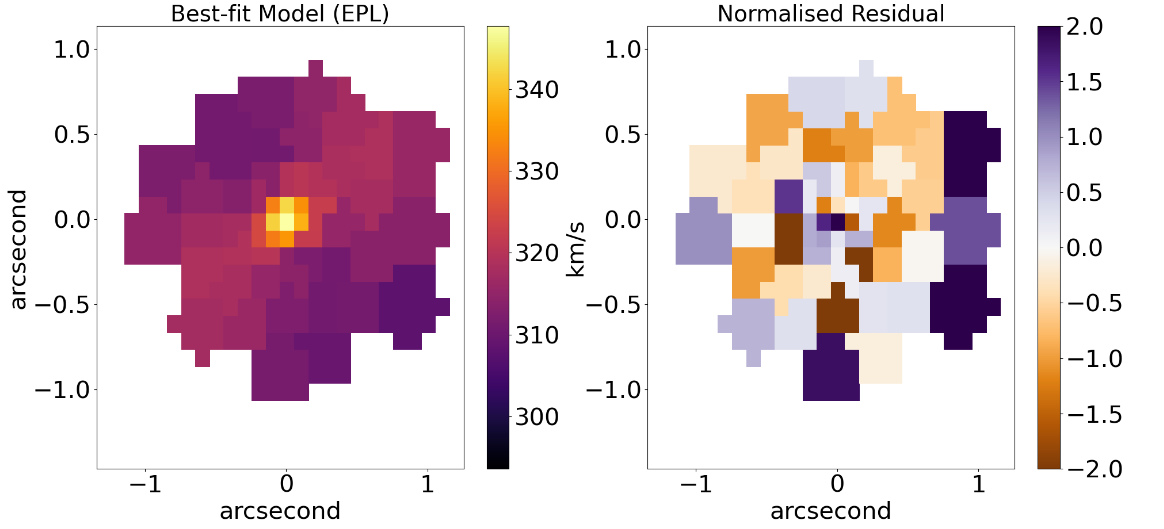} 
    \end{subfigure}
   
    \caption{The best-fit kinematic maps from joint models under different BH mass assumptions. The displayed kinematic bin maps have 52 bins in total. \textit{Upper Panel:} Joint modeling is performed using a grid of $M_{\rm BH}$ values ranging from $10^{9}$ to $10^{10}$ ${\rm M_{\odot}}$. In each model, the BH mass is fixed and incremented in steps of $10^{9}$ ${\rm M_{\odot}}$ within this range. The best-fit kinematic map, corresponding to $M_{\rm BH} = 3 \times 10^{9}~{\rm M_{\odot}}$, achieves $\chi^2_{\rm dyn} = 50$. \textit{Middle Panel:} The best-fit kinematic map from joint modeling that assumes no BH, yielding $\chi^2_{\rm dyn} = 64$. \textit{Lower Panel:} The best-fit kinematic map from joint modeling using the EPL mass profile, resulting in $\chi^2_{\rm dyn} = 58$.}
    \label{fig:Compsite_epl_kinematics_fit}
\end{figure}

We present both equal-weighted and BIC-weighted histograms of $D_{\rm d}$ and $D_{\rm \Delta t,\mathrm{int}}$ in Fig.~\ref{fig:Dd_Dt_equal_BIC}. The BIC does not clearly distinguish between composite mass models with BH masses in the range $M_{\mathrm{BH}} = 2 \times 10^{9}, M_{\odot}$ to $5 \times 10^{9}, M_{\odot}$, as the differences in BIC values $\Delta \mathrm{BIC}$ are comparable to their associated uncertainties $\sigma_{\mathrm{BIC}}$ (see Appendix~\ref{app:BIC weight factor for all mass models}). Outside this mass range, as well as for the EPL model, the BIC is more effective at discriminating between models. The limited effectiveness of BIC weighting is consistent with expectations, as $M_{\rm BH}$ cannot be well constrained by dynamical modeling for galaxies beyond the local universe. The EPL model, in particular, exhibits higher scatter in the probability density distribution of $D_{\rm \Delta t,\mathrm{int}}$ across different source resolutions. This increased scatter is likely due to the fact that the mock kinematic data were generated using a composite mass model. Since the EPL model assumes a single power-law mass profile, it fails to fully capture the complexity of the true lens mass distribution, leading to less reliable constraints on $D_{\rm \Delta t,\mathrm{int}}$. Despite this internal scatter, the EPL model is statistically disfavored under BIC weighting. It consistently underperforms in fitting the mock kinematic data. We obtain a minimal discrepancy of $\Delta \chi_{\rm dyn}^{2} = 8$ of the EPL model across different source grid resolutions relative to the best-fit composite mass model (see Fig.~\ref{fig:Compsite_epl_kinematics_fit}).

We combine all 55 models, obtaining the recovered time-delay distance $D_{\rm \Delta t, int} = 1857_{-78}^{+137}$ Mpc,  
which deviates from the mock input by $1.87\%$, within the 1-sigma uncertainty range.  
Similarly, the recovered lens distance  
$D_{\rm d} = 781_{-29}^{+30}$ Mpc  
shows a deviation of $0.77\%$. 

 The uncertainty in $D_{\rm \Delta t, int}$ is asymmetric, exhibiting a longer tail on the positive side and a shorter tail on the negative side (see Figs.~\ref{fig:Distance_lanbda_beta_ideal_shift} and \ref{fig:Dd_Dt_equal_BIC}). This occurs because, as $\lambda_{\rm int}$ approaches the upper bound of 1.5 in the prior, it implies that $\kappa_{\rm gal}$ is being modified by the addition of a negative constant sheet $(1-\lambda_{\rm int})$ on top of $\lambda_{\rm int} \kappa_{\rm gal}$ (see Eq.~\ref{eq:kappa int}). In regions far from the lensing centroid, $\kappa_{\rm int}$ becomes negative, which is not allowed under the dynamical modeling framework of JAM. As a result, the probability distribution of $D_{\rm \Delta t, int}$ becomes asymmetric (see Sect.~\ref{subsect:Joint modeling}). The lower $1\sigma$ bound of 78 Mpc may be underestimated relative to the true $1\sigma$ interval, assuming that $\lambda_{\rm int}$ follows an ``ideal'' mass sheet profile with a sharp cutoff beyond $\sim 20\arcsec$, which mimics the MST and also allows higher values of $\lambda_{\rm int}$ without leading to negative $\kappa_{\rm int}$. The probability distribution of $D_{
\rm d}$ is also slightly asymmetric, but it is less pronounced than that of $D_{\Delta \rm t,int}$. The asymmetry in $D_{\rm d}$ arises from the influence of $M_{\rm BH}$, which is discussed in more detail in Sect.~\ref{subsec:BH effect}.

\begin{figure*}
    \centering
    \begin{subfigure}{0.44\textwidth}
        \centering
        \includegraphics[width=\textwidth]{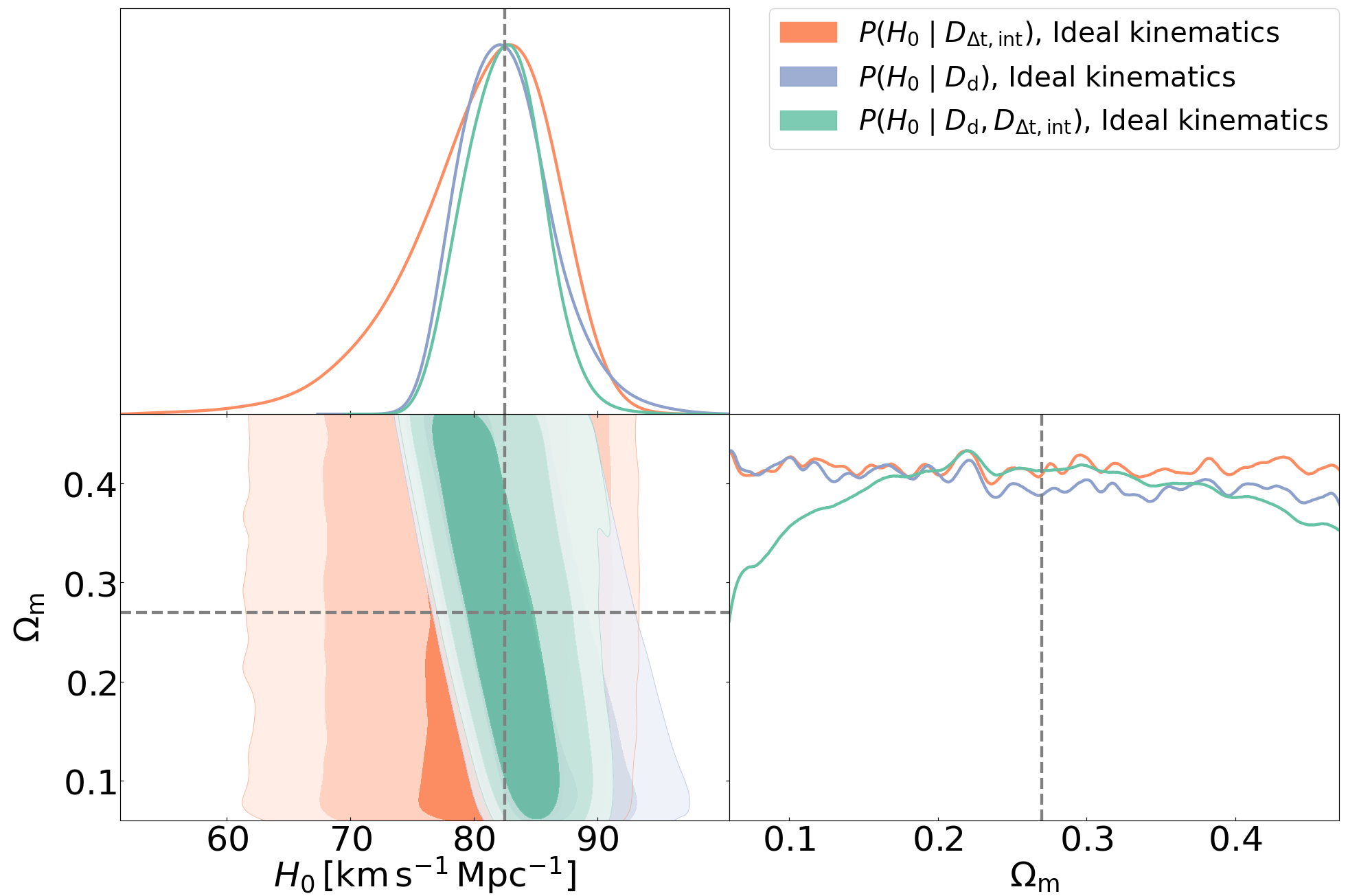}
    \end{subfigure}
    \hfill
    \begin{subfigure}{0.45\textwidth}
        \centering
        \includegraphics[width=\textwidth]{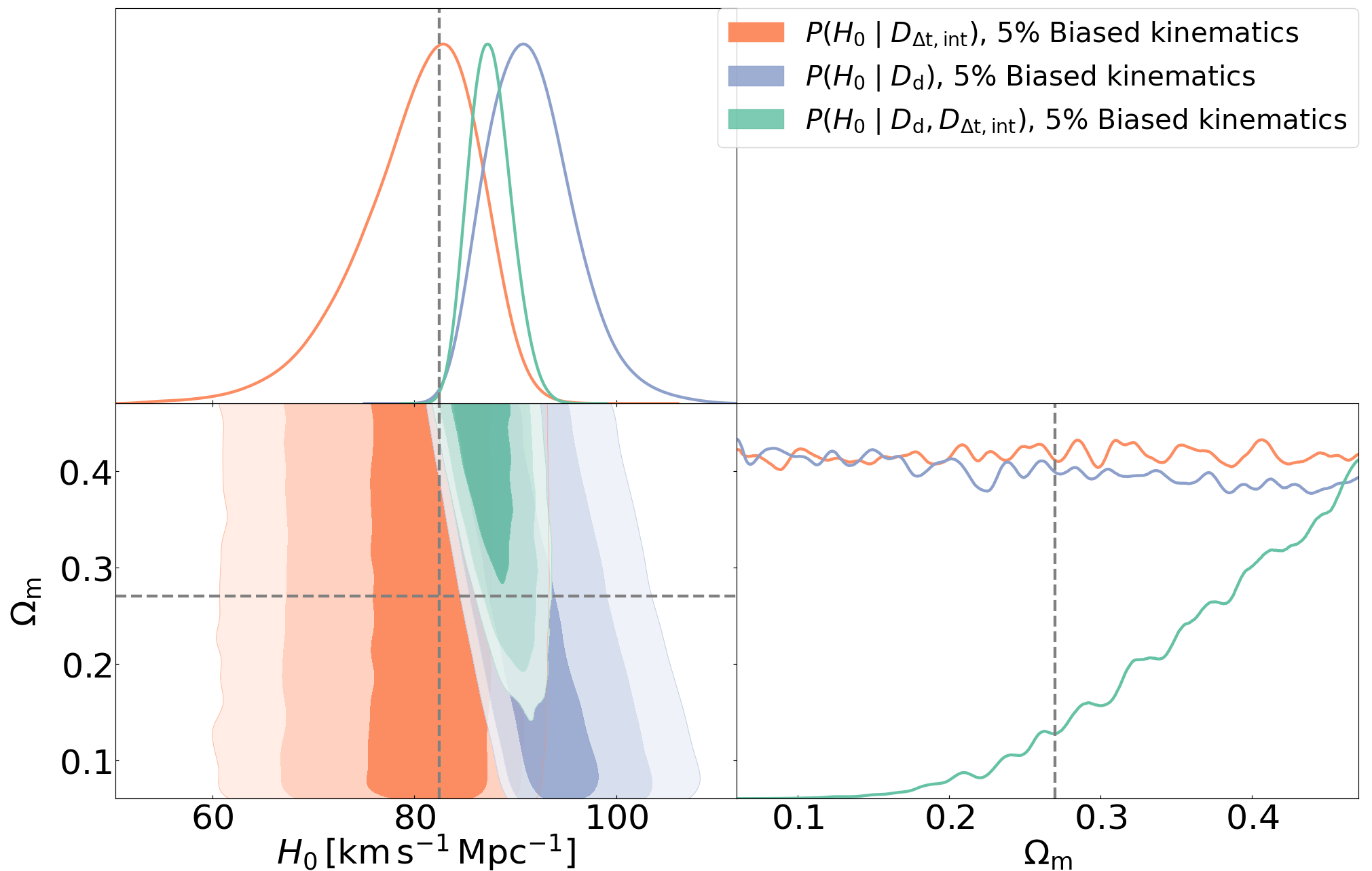}
    \end{subfigure}
    \caption{$H_0$ and $\Omega_{\rm m}$ constraints from our models in flat $\Lambda \rm CDM$ cosmology, for the ideal (left) and kinematic data with 5\% bias (right). The shaded contours represent $1\sigma$, $2\sigma$, and $3\sigma$ confidence regions. The blue (orange) contours represent the constraints based on $P(D_{\rm d})$ ($P(D_{\rm \Delta t, int})$), marginalized over all other parameters in the joint modeling. The green contours represent the constraints based on $P(D_{\rm d}, D_{\rm \Delta t,int})$. The gray dashed lines represent the mock input values in the data sets. The kinematic data, with a $5\%$ systematic bias, affects only $D_{\rm d}$, such that the inferred $H_0$ median value based on $P(D_{\rm d})$ is biased by $13\%$ relative to the mock input value of $H_0$ (see right panel).
}
\label{fig:H0Dddtidealshift}
\end{figure*}

With the posterior probability distribution $
P(D_{\rm \Delta t, int}, D_{\rm d} \mid \boldsymbol{d}_{\rm LD} )$, we infer $H_0$ and $\Omega_{\rm m}$ in a flat $\Lambda \rm CDM$ universe. We adopt uniform priors on $H_0$ between [50, 120]~$\rm~km~s^{-1}~Mpc^{-1}$ and on the matter density parameter\footnote{The inferred cosmological parameter in joint modeling can also be expressed in terms of the dark energy density, $\Omega_{\rm \Lambda}$, instead of $\Omega_{\rm m}$ since $\Omega_{\rm \Lambda}=1-\Omega_{\rm m}$ in flat $\Lambda$CDM cosmology. However, a single quasar system with quad images is not sensitive to cosmological parameters other than $H_0$.} $\Omega_{\rm m}$  between [0.05, 0.5]. We generate $5 \times 10^{5}$ samples for the parameters $\{H_0, \Omega_{\rm m}\}$ and calculate the corresponding $D_{\Delta \rm t,int}$ \footnote{$D_{\Delta \rm t,int}$ is the angular diameter distance calculated from the assumed cosmology.} and $D_{\rm d}$ values using the lens and source redshifts under a flat $\Lambda$CDM cosmology. For each sample, we randomly draw a $\kappa_{\rm ext}$ value from the external convergence distribution inferred by \citet{2014Suyu} and scale the distance using Eq.~\ref{eq:Dt external internal} to obtain $D_{\rm \Delta t, int}$. Subsequently, we weight the samples using $P(D_{\rm \Delta t, int}, D_{\rm d} \mid  \boldsymbol{d}_{\rm LD} )$. From the weighted sample distribution, we obtain constraints on $H_0 = 82.5_{-3.1}^{+3.2} \rm~km~s^{-1}~Mpc^{-1} $ (see Tab.~\ref{tab:H_0_values}). We also present $H_0$ values derived from the posterior probability distribution, marginalized over all parameters, including $D_{\Delta \rm t,  int}$ and $D_{\rm d}$ separately. The value $H_0 = 81.0_{-6.4}^{+4.4}~\rm~km~s^{-1}~Mpc^{-1}$, obtained from $P(D_{\rm \Delta t, int})$, reflects asymmetrical uncertainties inherited from $D_{\Delta  \rm t,  int}$ distribution. However, these skewed uncertainties of the inferred $H_0$ are mitigated by incorporating both distances $P(D_{\Delta  \rm t,  int}, D_{\rm d})$ (see Tab.~\ref{tab:H_0_values} and Fig.~\ref{fig:H0Dddtidealshift}). This demonstrates the advantage of joint modeling, where using two distances improves the constraint on $H_0$. The value of $\Omega_{\rm m}$ inferred by joint modeling is poorly constrained from a single lens system; therefore, it is not included in the table.

\begin{figure}
\includegraphics[width=1.\linewidth]{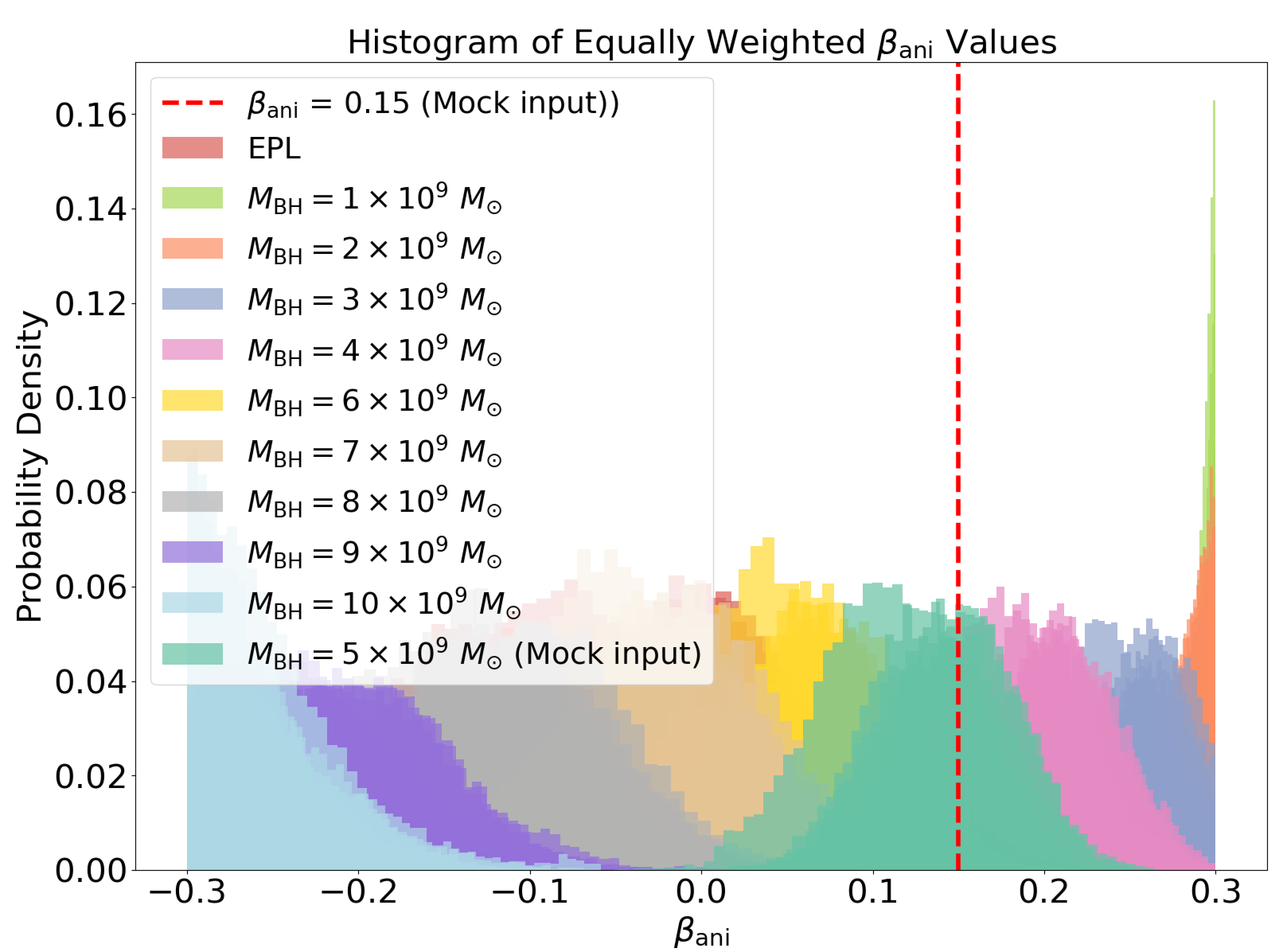}
 \caption{Marginalized posterior density distribution of $\beta_{\rm ani}$, based on joint models using ideal kinematic data. Different colors represent posterior densities corresponding to different BH masses, while the same color indicates models with identical mass parameterization but different source grid resolutions. We observe that as $M_{\rm BH}$ increases, the inferred $\beta_{\rm ani}$ decreases and vice versa. The inferred $\beta_{\rm ani}$ distributions given different $M_{\rm BH}$ spread over the prior range [$-$0.3,0.3], but different $M_{\rm BH}$ yield different goodness of fit to the kinematic data (as shown in Fig.~\ref{fig:Dd_Dt_equal_BIC}). }
 \label{fig:Beta_equal}
\end{figure}

\begin{figure}
  \includegraphics[width=1\linewidth]{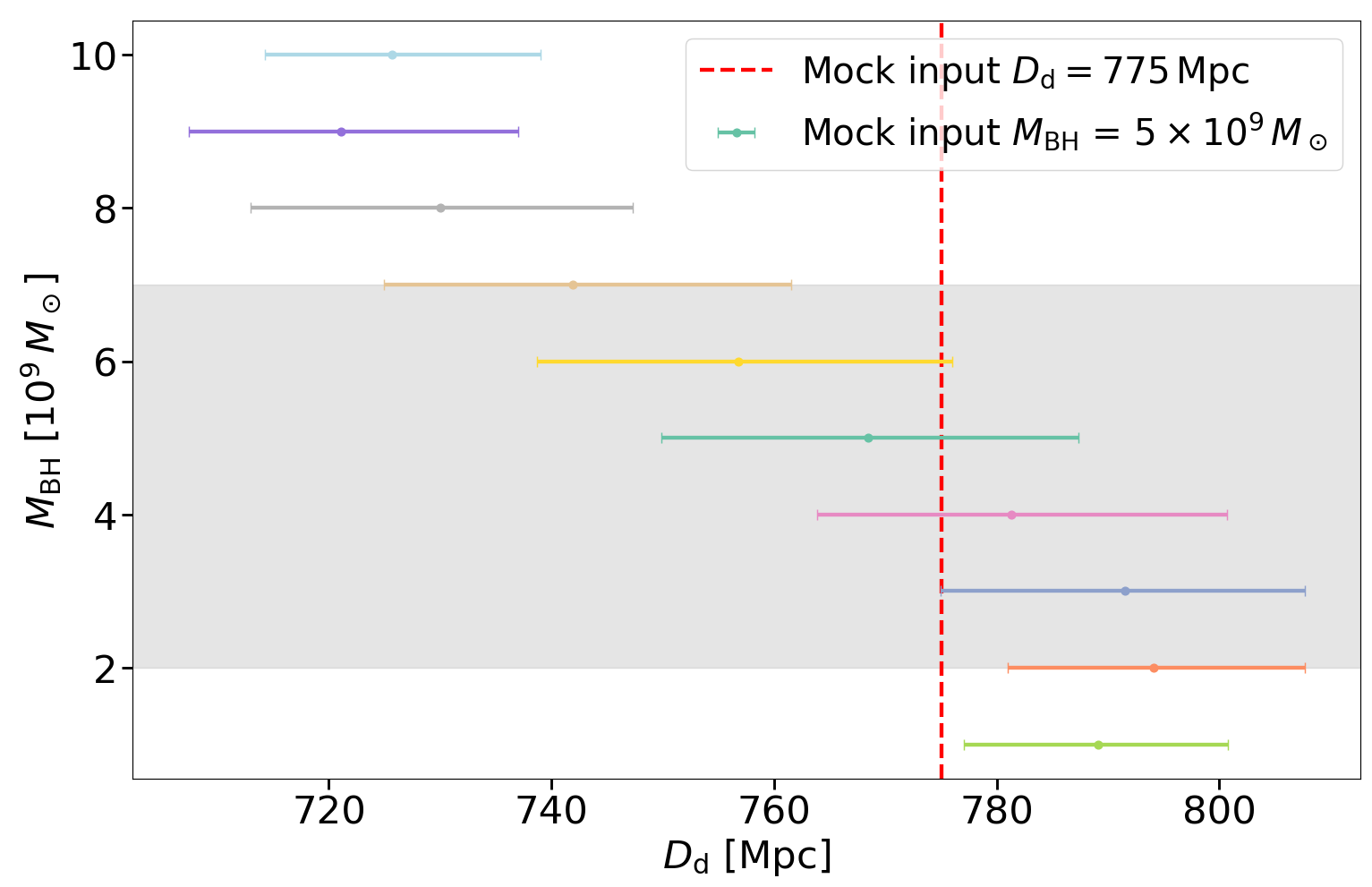}
  \caption{$M_{\rm BH}$ versus $D_{\rm d}$ with $1\sigma$ uncertainties, based on the ideal kinematic data. This provides a clearer visualization of the upper-left panel in Fig.~\ref{fig:Dd_Dt_equal_BIC}. The plot also illustrates how the BH mass affects the inference of $D_{\rm d}$, given that $\beta_{\rm ani}$ is within the prior range of $-0.3$ to $0.3$. The grey shaded region represents models with the BH mass satisfying $f_{\rm BIC}^{*} \geq 0.2$, indicating a significant contribution to $D_{\rm d}$ when combining all models together.}
  \label{fig:Bh_Dd}
\end{figure}

\subsection{The impact of high systematic bias in kinematics data on \texorpdfstring{$H_0$}{H0} measurement}
\label{subsect:systematic effect}
In this section, we perform joint modeling of the kinematic data, incorporating a 5\% systematic bias (see Eq.~\ref{eq:shift data}) to account for measurement-related systematic errors in the kinematic map. We emphasize that this adopted error represents a worst-case scenario, in which the kinematic measurements are not optimally performed. Furthermore, we highlight the importance of achieving sub-percent systematic errors in the kinematic map to ensure the robustness of cosmographic modeling, using the method presented in \citet{Knabel2025}.

The 5\% biased kinematic data also helps break the internal MSD, yielding consistent results for $\lambda = 0.97_{-0.07}^{+0.04}$ and $D_{\rm \Delta t,int} = 1863_{-80}^{+144}$~Mpc, which agree with values inferred from the joint modeling using ideal kinematic data. This demonstrates that the overall systematic bias does not affect the constraints on $D_{\rm \Delta t,int}$ and $\lambda_{\rm int}$ (see Fig.~\ref{fig:Distance_lanbda_beta_ideal_shift}). This is because $\lambda_{\rm int}$ is constrained by the 2D kinematic map, where the shape of the $v_{\rm rms}$ profile breaks the internal MSD and constrains $D_{\rm \Delta t,int}$. The 5\% bias does not alter the shape of the $v_{\rm rms}$ profile, which is why neither $D_{\rm \Delta t,int}$ nor $\lambda_{\rm int}$ is biased. The orbital anisotropy parameter, $\beta_{\rm ani}$, is primarily constrained by the shape of the kinematic map; therefore, its inference remains robust. We obtain $\beta_{\rm ani} = 0.20^{+0.08}_{-0.13}$, which is consistent with the value inferred from the ideal kinematic data.

The systematic bias primarily impacts $D_{\rm d}$ because it changes the amplitude of the $v_{\rm rms}$ overall. Given the relation $v_{\rm rms}^{\rm pre} \sim \frac{1}{\sqrt{D_{\rm d}}}$, a $5\%$ bias in $v_{\rm rms}^{\rm pre}$ results in an expected $\sim9\%$ bias in $D_{\rm d}$. We obtain $D_{\rm d} = 706_{-25}^{+20}$~Mpc, which is $9\%$ lower than the mock input value of $D_{\rm d} = 775$~Mpc, as expected. If the combined kinematics are obtained from a single aperture rather than an IFU, the impact on distances will not be cleanly isolated to $D_{\rm d}$ alone, as the single aperture lacks information on the shape of $v_{\rm rms}$. We anticipate a more severe effect on both $D_{\rm d}$ and $D_{\rm \Delta t,int}$.

The inferred value of $H_0 = 93.6_{-2.1}^{+3.3}\rm~km~s^{-1}~Mpc^{-1}$ from $P(D_{\rm d})$ is biased by 13\% compared to the mock input. However, since the inferred $D_{\rm \Delta t,int}$ remains unbiased, we obtain $H_0 = 87.4_{-2.0}^{+2.2}~\rm km~s^{-1}~Mpc^{-1}$ using $P(D_{\rm \Delta t, int}, D_{\rm d} \mid \boldsymbol{d}_{\rm LD})$, which carries a 6\% bias relative to the mock input (see Fig.~\ref{fig:Distance_lanbda_beta_ideal_shift}). 

Any systematic error affecting the overall kinematic map will be amplified in $D_{\rm d}$ inference \citep{Chen2021}. Although the bias does not impact $D_{\rm \Delta t, int}$ inference, the joint modeling of $H_0$ remains highly susceptible to bias. This crucially highlights the importance of accurately measuring kinematics and controlling systematic uncertainties to the sub-percent level, which is achieved by \citet{Knabel2025}, in order to measure $D_{\rm d}$ and $H_0$ to the percent level.

\subsection{The impact of \texorpdfstring{$M_{\rm BH}-\beta_{\rm ani}$}{MBH-beta ani} degeneracy on $H_0$ measurement in time-delay cosmography}
\label{subsec:BH effect}
The BH in lensing-only modeling is often neglected since SL only provides constraints at Einstein radius, which is far from the galaxy's center. Only in some rare cases, the lensed source image appears close to the galaxy center within $\lesssim 1~\rm kpc$ \citep[e.g.,][]{2023BHmass, 2025BHmass}. Kinematic data can provide some constraints, but its effectiveness is highly limited by the instrument's resolution, particularly for galaxies at galaxies at $z>0.1$. The lens galaxy RXJ1131 at $z = 0.295$ might hold a supermassive BH with $M_{\rm BH}$ in the range of [$10^{9}, 10^{10}]$ ${\rm M_{\odot}}$ which corresponding to a sphere of influence $r_{\rm soi}$ for the BH in the range of $[0.011\arcsec, 0.11\arcsec]$. The simulated kinematic data has a spaxel size with $0.1\arcsec$ convolved with $\rm FWHM = 0.14\arcsec$ of $\rm PSF_{\rm kin}$. For $M_{\rm BH}$ near $10^{10}\,{\rm M_{\odot}}$, the influence of the BH dynamics can be imprinted on the central Voronoi bins.

 We investigate the impact of the BH on distance inference using joint modeling with ideal kinematic data. This is due to the well-known mass–anisotropy degeneracy \citep[e.g.,][]{Binney1982, Treu_2002, 2014Courteau} in spherical systems, where a wide range of density profiles can be matched by varying $\beta_{\rm ani}$ while keeping the observed $v_{\rm rms}$ unchanged. The lens galaxy in RXJ1131, as a slow rotator with a nearly spherical central structure, is affected by this degeneracy \citep{2016Cappellari}.

The joint modeling results show that the inferred values of $\beta_{\rm ani}$ span the full prior range of [$-$0.3, 0.3] (see Fig.~\ref{fig:Beta_equal}), given a black hole mass in the range $M_{\rm BH} \in [10^{9}, 10^{10}]{\rm M_{\odot}}$. This prior range is motivated by studies of nearby massive elliptical galaxies \citep[see review in][Figs.~8, 10]{Cappellari2025}, and is quite conservative in its broad range compared to the typical scatter of anisotropies of galaxies shown in \citet{Cappellari2025}. The anisotropy $\beta_{\rm ani}$ is constrained by the spatial pattern in the kinematic data. However, $M_{\rm BH}$ and $\beta_{\rm ani}$ similarly affect stellar motions in the galaxy centroid, resulting in a trade-off between them. In Fig.~\ref{fig:Beta_equal}, we observe that a heavier $M_{\rm BH}$ leads to a smaller $\beta_{\rm ani}$, and vice versa. A higher $M_{\rm BH}$ deepens the central gravitational potential, allowing more tangential orbits in the dynamical model when reproducing the same observed line-of-sight velocity dispersion, corresponding to $\beta_{\rm ani} < 0$. Conversely, a lower $M_{\rm BH}$ can produce similar velocity dispersions if the stellar orbits are more radial, with $\beta_{\rm ani} > 0$, as radial orbits allow stars to reach higher line-of-sight velocity dispersion near the galaxy center. Both BH mass and $\beta_{\rm ani}$ contribute to accelerating stellar motion, but in different directions.

A trade-off between $M_{\rm BH}$ and $\beta_{\rm ani}$ can lead to a misestimated $\beta_{\rm ani}$, which in turn affects the accuracy of the distance inference, particularly in $D_{\rm d}$. When the BH mass used in the modeling is lower than the mock input value, a higher value of $\beta_{\mathrm{ani}}$ is required to fit the kinematic data. Since we assume a constant $\beta_{\mathrm{ani}}$ across all radii in the joint modeling, this means that even in the outer regions where $v_{\rm rms}^{\rm pre}$ adjustments are unnecessary, $v_{\rm rms}^{\rm pre}$ are still affected by $\beta_{\mathrm{ani}}$. To match the observed kinematics beyond the central region, $D_{\rm d}$ increases to counterbalance the effect introduced by changes in $\beta_{\mathrm{ani}}$. This is because $D_{\rm d}$ acts as a normalization factor for scaling $v_{\rm rms}^{\rm pre}$, following the relation $v_{\rm rms}^{\rm pre} \sim \frac{1}{\sqrt{D_{\rm d}}}$ (see Sect.~\ref{subsection:Stellar dynamics under the internal MSD}). If the BH mass in the joint modeling is heavier than the mock input, the entire trend reverses. This explains the observed correlations in the values of $D_{\rm d}$, $\beta_{\rm ani}$ and $M_{\rm BH}$ (see Figs.~\ref{fig:Dd_Dt_equal_BIC}, \ref{fig:Beta_equal} and \ref{fig:Bh_Dd}). In Fig.~\ref{fig:Bh_Dd}, we observe a negative correlation between the adopted $M_{\rm BH}$ in the joint modeling and $D_{\rm d}$. BH masses in the range of $M_{\rm BH} = 2 \times 10^9~{\rm M_{\odot}}$ to $M_{\rm BH} = 7 \times 10^9~{\rm M_{\odot}}$ are difficult to distinguish based on kinematic data and all contribute to the inference of distances in the BIC framework, given the $M_{\rm BH} = 5 \times 10^9~{\rm M_{\odot}}$ in the mock input. This results in a slight asymmetry in the probability distribution of $D_{\rm d}$. Since we use $P(\boldsymbol{d}_{\rm D} \mid \mathcal{M}_i)$ for BIC weighting in the joint models, some models with lower or larger BH masses can still provide comparably good fits to the kinematic data as the model with the true $M_{\rm BH}$, by appropriately rescaling both $D_{\rm d}$ and $\beta_{\rm ani}$. These models contribute to the extended tails of the inferred $D_{\rm d}$ distribution (see the upper panels in Fig.~\ref{fig:Dd_Dt_equal_BIC}). Nonetheless, the inferred $D_{\rm d}$ recovers the mock input value within the $1\sigma$ uncertainty. For the same underlying reason—but with a more pronounced impact—the constraint on $\beta_{\rm ani}$ is significantly weaker. As a result, $\beta_{\rm ani}$ is not tightly constrained, with an inferred value of $\beta_{\rm ani} = 0.21_{-0.14}^{+0.07}$ (see Tab.~\ref{tab:H_0_values}).

\begin{figure}
\includegraphics[width=1.\linewidth]{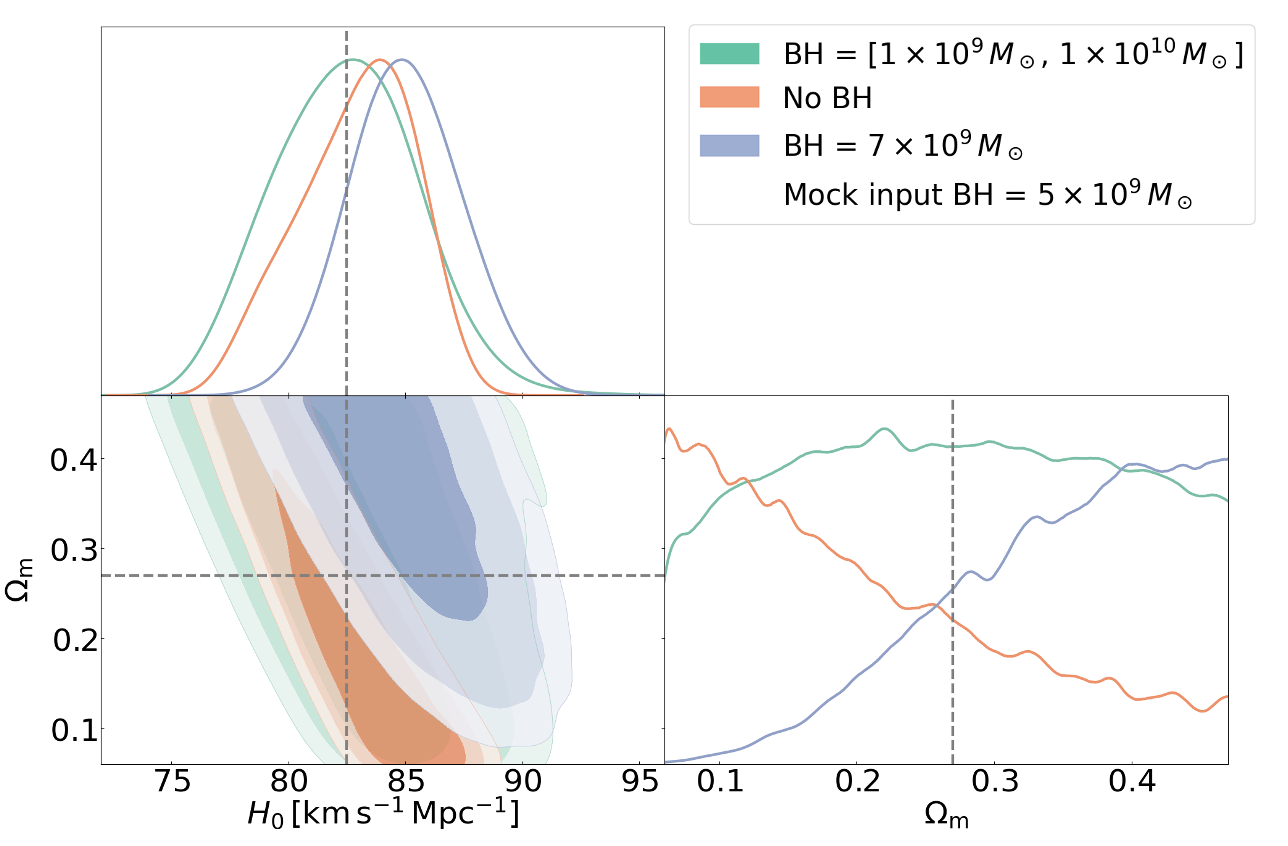}
 \caption{Constraints on $H_0$ and $\Omega_{\rm m}$ from our models in flat $\Lambda \rm CDM$ cosmology, using the ideal kinematic data and different assumptions for $M_{\rm BH}$. The green contour represents the constraints from the joint modeling, considering the possible values of $M_{\rm BH}$, and combining all 55 models weighted by BIC. The orange contour shows the constraints from the joint modeling, excluding the BH mass. The blue contour shows the constraints from the joint modeling with $M_{\rm BH} = 7 \times 10^{9}~{\rm M_{\odot}}$. The BH mass in the simulated kinematics data is set to $M_{\rm BH} = 5 \times 10^{9}~{\rm M_{\odot}}$. }
 \label{fig:BH_effect}
\end{figure}

While the above analysis considers a range of plausible BH masses with BIC weighting, we now turn to the case of a single, incorrect BH mass assumption to examine its impact on the inferred distances. First, we test the scenario where we assume no BH in the composite mass model. The inferred value of $H_0 = 83.2_{-3.0}^{+2.3} \rm \, km~s^{-1}~Mpc^{-1}$ successfully recovers the input $H_0$ value (see Fig.~\ref{fig:BH_effect} and Table~\ref{tab:H_0_values}). However, the best-fit kinematic map exhibits a significantly different pattern compared to the one obtained when the BH is included in the modeling (see Fig.~\ref{fig:Compsite_epl_kinematics_fit}). The difference of $\Delta \chi_{\rm dyn}^2 = 14$  for the dynamical fit indicates a poorer fit compared to the best-fit model that accounts for the BH. The fitted value of $\beta_{\rm ani} = 0.29_{-0.006}^{+0.003}$ reaches the upper bound of the prior range, yet it remains insufficient to fully compensate for the absence of the BH, leading to a suboptimal fit to the kinematic data. The high $\beta_{\rm ani}$ value leads to an excessive increase in velocity dispersions in the outer regions. As a result, this imbalance starts to affect the probability density distribution of $D_{\rm \Delta t, int}$. To compensate for the effect induced by $\beta_{\rm ani}$, $D_{\rm \Delta t, int}$ decreases slightly. 

A possible explanation is that the high $\beta_{\rm ani}$ requires a significantly different mass model than initially assumed to fit the kinematic data, which in turn affects $\lambda_{\rm int}$ and $D_{\rm \Delta t, int}$. We obtain a lower value of $D_{\rm \Delta t, int} = 1770_{-39}^{+54}$ Mpc, with a median value that is 3\% lower than the input value. The reason $H_0$ can still be recovered in this case is that $D_{\rm d}$ is recovered to within 1\% of its input value. However, if the prior range of $\beta_{\rm ani}$ is extended beyond 0.3, its inferred value continues to increase until it adequately fits the kinematic data. As a result, $D_{\rm d}$ increases accordingly to counterbalance the effect of $\beta_{\rm ani}$ in the outer regions. This will ultimately introduce additional bias in $D_{\rm d}$ that exceeds the value reported in Tab.~\ref{tab:H_0_values}, thereby biasing the inferred $H_0$ value.

In the second case, we probe the scenario where an incorrect BH mass of $M_{\rm BH} = 7 \times 10^{9}~{\rm M_{\odot}}$ was assumed. In contrast to the first case, the orbital anisotropy $\beta_{\rm ani} = -0.028_{-0.05}^{+0.06}$ shifts toward the lower bound, and the value of $D_{\rm d} = 742_{-17}^{+20}$~Mpc is lower than the $D_{\rm d}$ obtained using the true $M_{\rm BH} = 5 \times 10^{9}$ (see Fig.~\ref{fig:Bh_Dd}). In this case, the best-fit kinematic map can reach almost the same quality as the model that uses the true BH mass. However, due to the lower value of $D_{\rm d}$, the inferred value of $H_0 = 85.0_{-2.4}^{+2.5}~\rm km~s^{-1}~Mpc^{-1}$ is 3\% higher than
 the mock input value of $H_0 = 82.5~\rm km~s^{-1}~Mpc^{-1}$ (see Fig.~\ref{fig:BH_effect}).

The above tests indicate that a severely misfitted $\beta_{\rm ani}$ can strongly bias $D_{\rm d}$ and mildly influence $D_{\rm \Delta t, int}$ in the extreme case, thereby affecting $H_0$ inference, even when the kinematic data appears to be well-fitted. The value of $\beta_{\rm ani}$ can be accurately recovered when the BH mass is known and vice versa. We find that the fitted value of $\beta_{\rm ani} = 0.13_{-0.04}^{+0.04}$ is well constrained when the true $M_{\rm BH}$ is used in the joint modeling (see Table~\ref{tab:H_0_values}). However, in nearly all cases of lens galaxies, the precise BH mass is unknown. The bias in $D_{\rm d}$ caused by a misfitted $\beta_{\rm ani}$ can be mitigated by performing joint modeling over a range of possible BH masses and using the BIC to downweight models that are disfavored by the kinematic data. It naturally follows that the prior range of $\beta_{\rm ani}$ should be carefully chosen \citet{2024Simon}. Expanding the prior range allows adjustments to $\beta_{\rm ani}$ and $D_{\rm d}$ to always effectively compensate for the presence of the BH. While this results in a well-fitted kinematic model, it significantly biases the inferred $D_{\rm d}$.

\begin{table*}
    \centering
    \caption{Inferred \( H_0 \) and key parameters across models.}
    \small
    \setlength{\tabcolsep}{4pt} 
    \renewcommand{\arraystretch}{1.2} 
    \begin{tabular}{p{4.2cm}cccccccc}
    \toprule
    Model  & $D_{\rm d}$ \newline [Mpc] & $D_{\rm \Delta t, int}$ \newline [Mpc] & $\lambda_{\rm int}$ & $\beta_{\rm ani}$ & $P(H_0~|~D_{\rm d})$ & $P(H_0~|~D_{\rm \Delta t, int})$ & $P(H_0~|~D_{\rm d}, D_{\rm \Delta t, int})$ & $\chi_{\rm dyn}^2$ \\
    \midrule
    \textbf{Full FoV (52 bins)}\\
    Ideal Kinematics with \newline $M_{\rm BH}$ in [$10^{9}, 10^{10}]$ ${\rm M_{\odot}}$  
    & $781_{-29}^{+30}$ & $1857_{-78}^{+137}$ & $0.98_{-0.07}^{+0.04}$ & $0.21_{-0.14}^{+0.07}$ & $83.1_{-2.9}^{+3.7}$ & $81.0_{-6.4}^{+4.4}$ & $82.5_{-3.1}^{+3.2}$ & 50\\
    \\
    Ideal Kinematics with \newline $M_{\rm BH} = 5 \times 10^{9}~{\rm M_{\odot}}$*  
    & $769_{-18}^{+18}$ & $1868_{-80}^{+140}$ &$0.98_{-0.07}^{+0.04}$ & $0.13_{-0.04}^{+0.04}$ & $83.5_{-2.9}^{+3.1}$ & 
    $80.9_{-6.3}^{+4.6}$ & $83.3_{-3.0}^{+3.0}$ & 51\\
    \\
    Ideal Kinematics with \newline no BH  
    & $785_{-11}^{+12}$ & $1770_{-39}^{+54}$ & $1.00_{-0.03}^{+0.02}$ & $0.29_{-0.006}^{+0.003}$ & $81.3_{-3.1}^{+3.1}$ & $85.3_{-3.6}^{+3.0}$ & $83.2_{-3.0}^{+2.3}$ & 64 \\
    \\
    Ideal Kinematics with \newline $M_{\rm BH} = 7 \times 10^{9}~{\rm M_{\odot}}$  
    & $742_{-17}^{+20}$ & $1876_{-78}^{+144}$ &$0.98_{-0.07}^{+0.04}$  & $-0.028_{-0.05}^{+0.06}$ & $86.9_{-1.8}^{+2.0}$ & $80.4_{-6.6}^{+4.6}$ & $85.0_{-2.4}^{+2.5}$ & 53 \\
    \\
     Kinematics with a 5\% bias with \newline $M_{\rm BH}$ in [$10^{9}, 10^{10}]$ ${\rm M_{\odot}}$  
    & $706_{-27}^{+20}$ & $1863_{-82}^{+149}$ & $0.97_{-0.07}^{+0.04}$ & $0.19_{-0.15}^{+0.08}$ & $93.7_{-2.0}^{+3.0}$ & $80.5_{-6.6}^{+4.4}$ & $87.4_{-2.0}^{+2.2}$ & 50 \\
    \\
     \textbf{Full FoV exclude inner region (43 bins)}\\
      Ideal Kinematics with \newline no BH  &
       $777_{-19}^{+18}$    &
        $1804_{-54}^{+96}$    &
         $0.99_{-0.06}^{+0.03}$  &
         $0.23_{-0.06}^{+0.05}$  &
         $82.9_{-2.7}^{+3.0}$  &
         $83.4_{-5.3}^{+3.8}$  &
         $83.3_{-3.2}^{+2.9}$   &   34\\
    \\
     Ideal Kinematics with \newline $M_{\rm BH} = 7 \times 10^{9}~{\rm M_{\odot}}$     &  $752_{-30}^{+34}$   &
       $1903_{-97}^{+176}$    &
       $0.97_{-0.08}^{+0.05}$    &
       $0.019_{-0.15}^{+0.13}$    &
       $85.5_{-2.7}^{+2.7}$ &    
         $79.6_{-7.0}^{+5.1}$  &
        $84.1_{-3.3}^{+3.1}$    &  42  \\
    \bottomrule
    \end{tabular}
    \tablefoot{Important parameters and the inferred $H_0$~[$\rm kms^{-1}~Mpc^{-1}$] from different joint models. In the individual models, we present the marginalized values of $H_0$ constrained by $P(D_{\rm d})$, $P(D_{\rm \Delta t, int})$, and $P(D_{\rm d}, D_{\rm \Delta t, int})$, respectively. We also provide the marginalized distance values for $D_{\rm d}$ and $D_{\rm \Delta t, int}$. The $1\sigma$ uncertainties are calculated from the 16th, 50th, and 84th percentiles of the distribution. The input mock values are $H_0 = 82.5~\rm km~s^{-1}~Mpc^{-1}$, $D_{\rm d} = 775$~Mpc and $D_{\rm \Delta t, int} = 1823$~Mpc. The star symbol denotes joint modeling that includes the BH mass, which is the mock input.}
    \label{tab:H_0_values}
\end{table*}

\subsection{Mitigating the impact of the \texorpdfstring{$M_{\rm BH}$-$\beta_{\rm ani}$}{MBH-beta_ani} degeneracy on $H_0$ measurements}
\label{subsec: Mitigating the impact of the BH}

\begin{figure*}
   \centering
  \includegraphics[width=1.\linewidth]{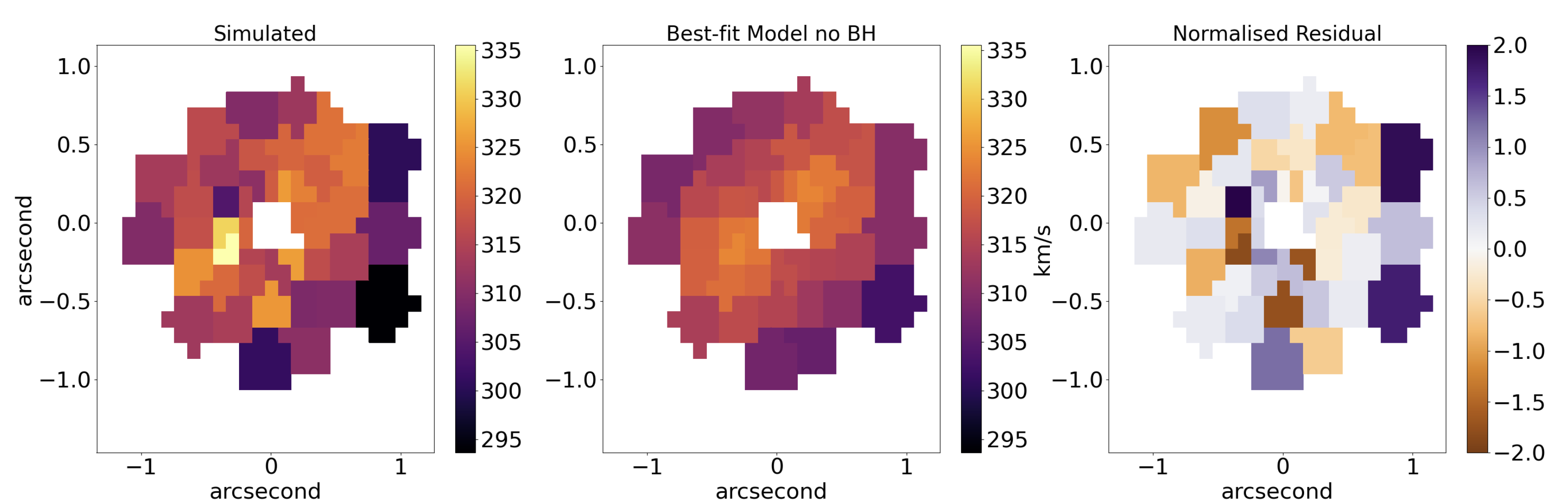}
  \caption{\textit{Left:} The ideal kinematic data excludes the bins within $-0.15\arcsec$ to $0.15\arcsec$ with 43 bins in total. \textit{Middel:} The best-fit kinematic map with $\chi_{\rm kin}^{2} = 34$, given the composite mass model with no BH. We do not display the best-fit kinematic map of the case with \(M_{\rm BH} = 7 \times 10^9 \, M_\odot\)  because they show similar fitting results. \textit{Right:} The normalised residual. }
  \label{fig:kinematics_fit_exclude_outer_region}
\end{figure*}

 We set the BH mass in our simulated datasets to $M_{\rm BH} = 5 \times 10^{9}~{\rm M_{\odot}}$, corresponding to a sphere of influence radius of $r_{\rm soi} = 0.056\arcsec$. As a result, the BH primarily affects the inner region. To account for this, we exclude the nine central bins in the ideal simulated kinematic map within the FoV range of 
\(-0.15\arcsec\) to \(0.15\arcsec\). We then examine whether the joint modeling becomes insensitive to the BH mass, thereby mitigating the bias in the \(D_{\rm d}\) measurement caused by the \(M_{\rm BH}\)-\(\beta_{\rm ani}\) degeneracy.

We perform joint modeling using the ideal kinematic map while excluding the central regions. We reassess the recovery of the \(H_0\) value and evaluate the quality of the kinematic fit for both cases of no BH
and a BH with \(M_{\rm BH} = 7 \times 10^9 \, M_\odot\). In both cases, we observe that $D_{\rm d}$ and $\beta_{\rm ani}$ shifted closer to the mock input values, allowing for an accurate recovery of $H_0$ within $1\sigma$ uncertainties (see Tab.~\ref{tab:H_0_values} and more details in Appendix.~\ref{app: Joint modeling using kinematics data exclude the central bins}). As anticipated, the $1\sigma$ uncertainties are broader than the full ideal kinematic dataset because we adopt 43 bins instead of the complete dataset with 52 bins. Additionally, the kinematic data excluding the central region is effectively recovered through joint modeling with no BH and \(M_{\rm BH} = 7 \times 10^9 \, M_\odot\) (see Fig.~\ref{fig:kinematics_fit_exclude_outer_region}). This suggests that excluding the central kinematic region can help mitigate the impact of a BH with a highly uncertain mass, and reduce potential biases in distance inference. However, in both cases, $\beta_{\rm ani}$ remains poorly constrained. This is due to the fact that the outer regions of galaxies are often significantly influenced by dark matter. The presence of dark matter introduces its own mass–anisotropy degeneracy, which complicates the interpretation of kinematic data in these regions—similar to the effect of the BH in the central regions. As a result, determining $\beta_{\rm ani}$ using only the outer kinematic bins is challenging due to the unknown distribution of dark matter.

To fully resolve the $M_{\rm BH}$–$\beta_{\rm ani}$ degeneracy, one effective way is to use the full line-of-sight velocity distribution (LOSVD). Instead of relying only on the second velocity moments $v_{\rm rms}$, incorporating higher-order moments like $h_3$ and $h_4$ provides stronger constraints on the orbital structure. This approach, however, demands high signal-to-noise data and advanced modeling techniques, such as Schwarzschild’s orbit-superposition method \citep{Cappellari2009, 2014Jens, 2016Jens, 2022Poci}. The lens galaxy of RXJ1131 at $z=0.295$ even with JWST NIRSpec data is unlikely to reach the required resolution and signal-to-noise ratio to perform such modeling.

Another approach, while still relying only on $v_{\rm rms}$ in JAM modeling, is to adopt a physically motivated prior for a more flexible orbital anisotropy. This can be informed by studies of well-observed nearby galaxies, as demonstrated in \citet{2024Simon}. In that work, the author introduced a radially varying anisotropy profile, $\beta_{\rm ani}(r)$, and carefully defined the prior range based on physical expectations. Specifically, the central regions near the black hole are assumed to be more tangentially biased, while the outer regions are expected to be more radially anisotropic. This pattern is supported by theoretical predictions that dry mergers eject stars on radial orbits, which pass close to a central supermassive black hole binary. Such interactions can create a core in the surface brightness profile and lead to a dominance of tangential orbits in the central region, while the stars in the outer regions retain a more radial orbital distribution. This scenario is consistent with both observations and simulations \citep[e.g.][]{Milosavljev2001, 2018DiffuseGalaxy, 2024Saglia}.

In principle, a similar methodology, incorporating well-justified anisotropy priors, could also be useful in lensing-based dynamical analyses. However, the current version of our modeling code supports only a constant anisotropy across radii. Fully resolving the $M_{\rm BH}$–$\beta_{\rm ani}$ degeneracy is beyond the scope of this paper. Instead, the main goal of this work is to show that, once this degeneracy is reasonably mitigated, it does not introduce significant bias on the inferred cosmological distances.

\section{Summary and outlook}
\label{sect:Summary and Outlook}
In this paper, we present a GPU-accelerated code (\texttt{GLaD}) for self-consistent lensing and dynamical modeling, based on \citet{2020Akin} for the lensing part and on \citet{JAM_sph} for the dynamics part. This method combines lensing and dynamical models by solving the Jeans equations in an axisymmetric geometry. The primary purpose of this code is for time-delay cosmography, but it can also be naturally applied to galaxy evolution studies \citep[]{2021Shajib,Dinos1, DinosII, Sahu}.

In time-delay cosmography, accounting for parameter uncertainties is essential. The most time-consuming part of joint modeling is running analyses across a range of source grids to account for parameter uncertainties associated with source grid resolutions. Another computational challenge is solving the Jeans equation to determine the intrinsic second velocity moments. The first issue is naturally optimized using GPU architecture, which excels at accelerating large matrix calculations, while the second is handled with a non-adaptive integral solver. In both cases, we achieve at least an order-of-magnitude speedup.

We simulate the lensing and kinematic data for the lensed quasar system RXJ1131 to test whether \texttt{GLaD} can recover the mock input value. Since the lens galaxy in RXJ1131 exhibits a central velocity dispersion $\geq 300\rm~km~s^{-1}$, we add a $M_{\rm BH} = 5 \times 10^{9}~{\rm M_{\odot}} $ in the mock mass profile. For the kinematic map, we generate one ideal kinematic map with the $2 \%$ statistical error and a biased kinematic map with a $5\%$ systematic bias (as a worst-case scenario) in all the velocities. We use \texttt{GLaD} to perform the joint modeling on the simulated data to test the influence of the systematic error and BH effects in the kinematic map. We found as follows:

\begin{itemize}

\item \texttt{GLaD} achieves a sampling time of $\sim0.5$ seconds per step on a single A100 GPU, reducing the Bayesian inference of the joint modeling in \citet{2020Akin, TDCOSMO13} from month-long to several days.

\item We perform joint modeling using two types of mass models and combine 55 models based on the BIC weighing. As expected, the kinematic data helps break the internal MSD.  Using ideal kinematic data, we achieve $4\%$ uncertainty in the inference of \( H_0 \). 

\item 
Systematic biases in the overall amplitude of spatially resolved kinematic data do not significantly impact the constraints on $\lambda_{\rm int}$ and $D_{\rm \Delta t, int}$, because these parameters are primarily sensitive to the shape of the two-dimensional $v_{\rm rms}$ distribution. Since an overall bias alters only the amplitude and not the shape of $v_{\rm rms}$, its effect on the inference of these parameters is minimal.

\item
The bias in the amplitude of $v_{\rm rms}$ primarily affects the inference of $D_{\rm d}$. A 5\% bias leads to an approximately 10\% bias in $D_{\rm d}$, which in turn results in a 10\% bias in the $H_0$ measurement, given $P(D_{\rm d})$. However, as we emphasized earlier, a 5\% bias in the kinematic data does not bias $D_{\rm \Delta t, int}$. Consequently, when considering $H_0$ given both distances, $P(D_{\rm d},D_{\rm \Delta t, int})$, the bias is reduced to approximately 6\%. We have demonstrated that systematic bias in the kinematic data doubles the error as it propagates to $H_0$ \citep[as also shown by][see Eqs.~20 and 21]{Chen2021}. This highlights the importance of measuring kinematics with sub-percent systematic uncertainty, as recently achieved by \citet{Knabel2025}.

\item The BH mass does not influence the breaking of the internal MSD. Therefore, the measurement of \( \lambda_{\rm int} \) and \( D_{\rm \Delta t, int} \) remains independent of the adopted \( M_{\rm BH} \) in the joint modeling, provided that the kinematic data is well fitted.

\item Given the high BH mass of $5\times10^9\ {\rm M_{\odot}}$ adopted in our mock data, the BH mass plays a crucial role in constraining $\beta_{\rm ani}$ and $D_{\rm d}$. By adjusting $\beta_{\rm ani}$, one can mimic the effect of a massive BH, making it difficult to constrain anisotropy without precise knowledge of the BH mass. Additionally, $\beta_{\rm ani}$ is positively correlated with $D_{\rm d}$, meaning any bias in the inferred $\beta_{\rm ani}$ leads to a corresponding bias in $D_{\rm d}$. As shown in Sect.~\ref{subsec:BH effect}, modeling with an incorrect BH mass results in an inferred $H_0$ that is 3\% higher than the mock input value.

\item In Sects.~\ref{subsec:BH effect} and \ref{subsec: Mitigating the impact of the BH}, we present two approaches to mitigate the impact of the BH on $H_0$ measurements. The first approach involves using insights from nearby galaxies to determine the most probable range for the BH mass. We then perform a series of models with BH mass variations within this range, combining the results using the BIC weights to obtain an unbiased distance and $H_0$ measurement. The advantage of this approach is that it leverages the full kinematic dataset and a well-motivated prior. However, the disadvantage is the need to run multiple models. In the second approach, we bypass the sensitivity of the kinematic data to the BH mass by excluding the central kinematic bins, allowing us to retrieve the $H_0$ value with just one model.
\end{itemize}

\texttt{GLaD} will be applied to the NIRSpec IFU observations of the lens galaxy in RXJ1131. Using the ideal kinematic simulated data, we identified a trade-off between the BH mass and the anisotropy parameter $\beta_{\rm ani}$, as well as the influence of BH mass on $D_{\rm d}$ in this paper. In our simulated kinematic dataset, we used a higher BH mass compared to the value from the $M_{\rm BH}-\sigma_{\rm disp}$ relation. We aim to determine whether these effects are also present in real observations.  If confirmed, this would motivate further exploration of strategies to mitigate potential biases in the inference of $D_{\rm d}$. One promising approach would be to adopt a radially varying anisotropy profile, $\beta_{\rm ani}(r)$, with carefully chosen priors, as proposed in \citet{2024Simon}. This would be especially helpful in tightening the constraints on $\beta_{\rm ani}$ values near the BH.

As for the test on systematic bias in the kinematic data, its impact on future $H_0$ measurements is expected to be minor, given the recent work by \citet{Knabel2025} who demonstrated that systematics errors of kinematic measurements can be controlled at the sub-percent level.

Another test we will explore in the future is the adopted mass sheet and how it interacts with the system. In this paper, we set the mass sheet with a fixed core and truncation radius. We ensure that, with this setup, the lensing data is completely degenerate with respect to different values of $\lambda_{\rm int}$ while the kinematic data are sensitive to $\lambda_{\rm int}$. Future studies could further explore the parameter space for the mass sheet that satisfies the above requirements and marginalize over them to assess the impact on the BH mass and $H_0$.

Our study highlights the speed gains achieved by using a single GPU, and in the future, parallelizing computations across multiple GPUs could further improve efficiency.  Our developments will enable more efficient lensing and dynamical modeling of galaxies with high-quality data for future cosmological and galaxy studies.

\section*{Acknowledgements}
We would like to thank the anonymous referee whose
comments were helpful in improving the paper. We thank Tommaso Treu, Shawn Knabel, Simon Birrer and Xiang-Yu Huang for helpful discussions and feedback on this work. 

HW and SHS thank the Max Planck Society for support through the Max Planck Fellowship for SHS. 
This project has received funding from the European Research Council (ERC) under the European Union's Horizon 2020 research and innovation programme (LENSNOVA: grant agreement No 771776).
This work is supported in part by the Deutsche Forschungsgemeinschaft (DFG, German Research Foundation) under Germany's Excellence Strategy -- EXC-2094 -- 390783311.  

AG acknowledges the Swiss National Science Foundation (SNSF) for supporting this work.

AJS received support from NASA through the STScI grants HST-GO-16773 and JWST-GO-2974.

\bibliography{reference}
\newpage
\begin{appendix}
\onecolumn
\section{MGE to surface brightness and mass convergence}
\label{app:MGE to surface brightness and surface mass convergence}
The observed 2D SB of the lens galaxies, $I(x', y')$ \footnote{Note that we present the general case here. In this paper, we perform 1D MGE fitting (see Sect.~\ref{subsect:GPU acceleration in dynamics modeling}) to model the profile along the radial direction $R$ with a fixed axis ratio $q$, i.e.,  $R = \sqrt{x^2 + y^2/q^2}$.}, is expressed through multiple Gaussians:

\begin{equation}
I(x^{\prime}, y^{\prime}) = \sum_{j=1}^{N} I_{0,j} \exp \left[-\frac{1}{2\sigma^{\prime 2}_{j}} \left( x^{\prime 2} + \frac{y^{\prime 2}}{q^{\prime 2}_{j}} \right) \right],
\label{eq:sb_mge_2D}
\end{equation}
where $I_{0, j}$ is the peak SB, $\sigma^{\prime }_{j}$ the dispersion along the projected major axis, and $q^{\prime}_{j}$ the apparent flattening of each Gaussian.  The Cartesian coordinates $x'$, $y'$ represent the position on the plane of the sky. The major axis of the lens galaxy is aligned with the $x'$-axis, the minor axis with the $y'$-axis.

The deprojection process depends on the assumption of the galaxies' shapes. For the commonly found elliptical galaxies with an oblate shape, the deprojection requires:
\begin{equation}
\cos^2{i} < q_{\rm min}^{\prime 2}
\end{equation}
where $i$ is the inclination angle and $q_{\rm min}^{\prime}$ is the axial ratio of the flattest Gaussian in the fit. 
The deprojected 3D luminosity density $\rho_{*}$ is  \citep[e.g.,][eq.~38]{JAM_sph}
\begin{equation}
\rho_{*}(r, \theta) = \sum_{j=1}^{N} \frac{q^{\prime}_{j}I_{0,j}}{\sqrt{2\pi}\sigma^{\prime }_{j} q_j} \exp \left[-\frac{r^2}{2\sigma^{2}_{i}} \left( \sin^2 \theta + \frac{\cos^{ 2} \theta}{q^{2}_{i}} \right) \right],
\label{eq:sb_mge_3D}
\end{equation}
where $r$ is the 3D radial distance to the galaxy centroid, $\theta$ is the polar angle \citep[see definition in][Fig.~1]{JAM_sph}, $\sigma_{j} = \sigma^{\prime }_{j}$ and $q_j = \frac{\sqrt{q^{\prime 2}_{j} - \cos{i}^2}}{\sin{i}}$ denote the dispersion and axis ratio of Gaussians after deprojection. The potential $ \psi_{\rm D,int}$ in Eqs.~\ref{eq:J1}
and \ref{eq:J2} is derived by integrating the MGE of the 3D mass density $\rho_{\rm int}$. Following the approach used to infer the light tracer \( \rho_{*} \), the 3D density profile \( \rho_{\rm int}\) is obtained by deprojecting \( \Sigma_{\rm int} \) (see Eq.~\ref{Eq:sfd final}). The surface mass density \( \Sigma_{\rm int} \) is expressed as a sum of multiple Gaussian components,
\begin{equation}
    \Sigma_{\rm int} (x^{\prime}, y^{\prime}) =  \sum_{i=1}^{M} \Sigma_{0,i} \exp \left[-\frac{1}{2\sigma^{\prime 2}_{i}} \left( x^{\prime 2} + \frac{y^{\prime 2}}{q^{\prime 2}_{i}} \right) \right].
\label{eq:smd_mge}
\end{equation}
Note we use index $i$ to denote the MGE components of the mass density, and $j$ for the luminosity components. The set of Gaussians describing the SB of lens galaxies (see Eqs.~\ref{eq:sb_mge_2D}, ~\ref{eq:smd_mge}) is not necessarily identical to the MGEs of their mass densities. Therefore, $i \neq j$ meaning that $\sigma^{\prime }_{i} \neq \sigma^{\prime }_{j}$, $q^{\prime }_{i} \neq q^{\prime }_{j}$ and $M \neq N$ unless mass follows light. The deprojected $\rho_{\rm int}(r,\theta)$ is
\begin{equation}
\rho_{\rm int} (r, \theta) = \sum_{i=1}^{N} \frac{q^{\prime}_{i}\Sigma_{0,i}}{\sqrt{2\pi}\sigma^{\prime}_{i} q_i} \exp \left[-\frac{r^2}{2\sigma^{2}_{i}} \left( \sin^2 \theta + \frac{\cos^{ 2} \theta}{q^{2}_{i}} \right) \right].
\label{eq:smd_3D}    
\end{equation}

\section{Implementation of the enfw profile}
\label{app:Implementation of the enfw profile}
\begin{figure}
  \includegraphics[width=0.97\linewidth]{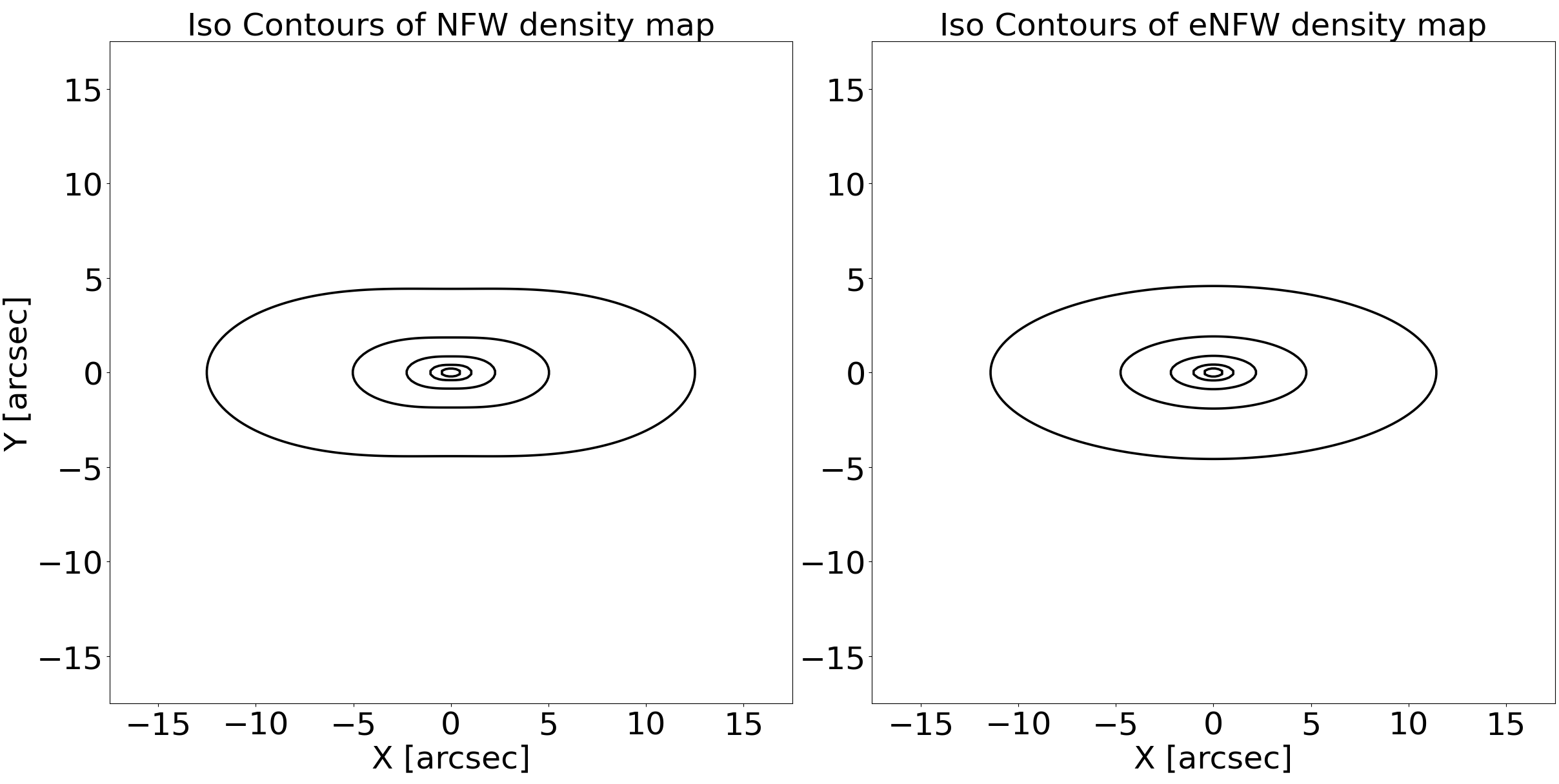}
  \caption{The mass density comparison between $\kappa_{\rm nfw}$ with $q_{\rm nfw} = 0.6$ (left panel) and $\kappa_{\rm enfw}$ with $q_{\rm enfw} = 0.4$ (right panel). The ellipticity implemented in the lensing potential leads to a dumbbell-shaped surface density (see the left panel). In contrast, applying ellipticity directly to $\kappa$ results in a more physically realistic mass distribution (see the right panel).    }
  \label{fig:NFWvsENFW_kappa}
\end{figure}

In many cases, we use the Navarro-Frenk-White (NFW) profile, derived from cosmological simulations, to model the mass density of dark matter in the lens galaxies. The classical NFW profile for lensing analyses often assumes spherical symmetry in the mass distribution, since analytical expressions for gravitational lensing properties are not available for mass distributions with ellipticity. However, observed galaxies and dark matter halos are typically not spherically symmetric but appear more elliptical when projected onto the sky. To address this challenge, one solution is to introduce ellipticity in the potential and then use Eq.~\ref{eq:poisson equation} to derive the corresponding mass density profile $\kappa_{\rm nfw} (\theta)$ \citep[e.g.,][]{2002Golse}. However, this approach can lead to unphysical mass density distributions, such as dumbbell-shaped isodensity contours, especially when the ellipticity is high ($q < 0.7$), as shown in Fig.~\ref{fig:NFWvsENFW_kappa}.
To avoid this issue, we adopt a method based on \citet{Oguri2021}, implementing a fast calculation approach that directly introduces ellipticity into $\kappa_{\rm enfw} (\theta)$. We define 

\begin{equation}
\kappa_{\rm enfw}(u) = 
\begin{cases} 
\frac{0.5~\rho_{\rm s}}{u^2 - 1} \left( 1 - \frac{1}{\sqrt{1 - u^2}} \, \text{arctanh}\left( \sqrt{1 -u^2} \right) \right), & \text{if } u < 1 \\[10pt]
\frac{0.5~\rho_{\rm s}}{u^2 - 1} \left( 1 - \frac{1}{\sqrt{u^2 - 1}} \, \text{arctan}\left( \sqrt{u^2 - 1} \right) \right), & \text{if } u > 1
\end{cases}
\label{eq:kappa_enfw}
\end{equation}

with
\begin{equation}
u = \frac{\sqrt{x^2 + y^2/q^2}} {\frac{r_{\rm s}}{\sqrt{q}}}
\end{equation}
where $r_{\rm s}$ is the scale radius and $\rho_{\rm s}$ is the characteristic density. In general, Eq.~\ref{eq:kappa_enfw} does not yield analytical expressions for lensing properties. Instead, computationally demanding numerical integration has to be performed. The idea in \citet{Oguri2021} is to decompose the Eq.~\ref{eq:kappa_enfw} into a series of basis functions, i.e., core steep ellipsoids (CSEs) which have simple analytical expressions of SL properties such as deflection angles $\boldsymbol{\alpha}_{\rm enfw}$ and the lensing potential $\psi_{\rm enfw}$.
\begin{equation}
    \frac{\kappa_{\rm enfw}}{\rho_{\rm s}}  = \sum_{i=1}^{N_{\rm enfw}}  A_{i}^{\rm enfw} \kappa_{i}^{\rm CSE} (u, s_{i}),
\end{equation}
with
\begin{equation}
    \kappa_{i}^{\rm CSE} (u, s_{i}) = \frac{1}{2(s_{i}^{2} + u^2)^{3/2}}.
    \label{eq:CSE}
\end{equation}
In \citet{Oguri2021}, they used 44 CSEs to fit $\kappa_{\rm enfw}$ (see Eq.~\ref{eq:kappa_enfw}). By minimizing
\begin{equation}
    \mathcal{L} = \exp\left[ -\frac{1}{2} \sum_j \frac{\left\{ \kappa_{\rm enfw}(u_j) - \sum_{i=1}^{N_{\rm enfw}} A_i \kappa_{\rm CSE}(u_j; s_i) \right\}^2}{\left( \kappa_{\rm enfw} \right)^2 \sigma^2} \right],
    \label{eq:likelihood_enfw}
\end{equation}
they achieved an accuracy of $\sigma = 10^{-4}$ in recovering $\kappa_{\rm gNFW}$ using CSEs, with $u_j$ spanning a wide range from $10^{-6}$ to $10^3$. The amplitude $A_i$ and core radius $s_i$ are predetermined before evaluating the lensing properties of $\kappa_{\rm gNFW}$ for any given values of $\rho_s$, $r_s$, and $q$.\footnote{Note that $\rho_s$ is omitted in Eq.~\ref{eq:likelihood_enfw} because it acts as a constant scaling factor and does not affect the decomposition process.}
The corresponding lens potential of an individual CSE is
\begin{equation}
    \psi^{\rm CSE}_{i} (x,y) = \frac{q}{2s_i}~{\rm ln}~\Psi(s_i, x,y, q) - \frac{q}{s_i}~{\rm ln}~[(1 +q)s_i],
\end{equation}
where the expression of $\Psi(s_i, x,y, q)$ does not include any complex functions. We refer readers to \citet{Oguri2021} for details. From the potential, we infer the deflection angle by calculating its gradient (see Eq.~\ref{eq:lens_equation}) and obtain an analytical expression,
\begin{equation}
    \boldsymbol{\alpha}_{\rm enfw} = \frac{r_s^2\rho_s}{\sqrt{q_0}} \sum_{i=1}^{N_{\rm enfw}}  A_{i}\boldsymbol{\nabla}  \psi^{\rm CSE}_{i} \left(\frac{\sqrt{q}}{r_s}x, \frac{\sqrt{q}}{r_s}y, s_i \right)
\end{equation}

\section{Implementation of the EPL profile}
\label{app: Implementation of the EPL profile}
We implemented the surface mass density $\kappa_{\rm epl}$ following \citet{EPL}. We define:
\begin{equation}
    \kappa_{\rm epl} = \left( \frac{3-\gamma}{2}\right) \left(\frac{b}{\sqrt{R^2 + r_{\rm soft}^2}} \right)^{\gamma - 1}
\end{equation}
with
\begin{equation}
    R = \sqrt{x^2 + y^2/q^2}
\end{equation}
where $\gamma$ represents the density slope, and $r_{\rm soft} = 0.01\arcsec$ is the softening radius introduced to prevent divergence at the central pixel. The parameter $b$ is a normalization factor, proportional to the Einstein radius $\theta_{\rm E}$, given by  
\begin{equation}
   b =  \left( \frac{2}{1+q}\right)^{\frac{1}{\gamma-1}}\theta_{\rm E}.
\end{equation}

\section{Joint modeling with ideal kinematic data across varying source grid resolutions}
\label{app: Joint modeling with ideal kinematic data across varying source grid resolutions}
\begin{figure}[H]
\includegraphics[width=0.85\linewidth]{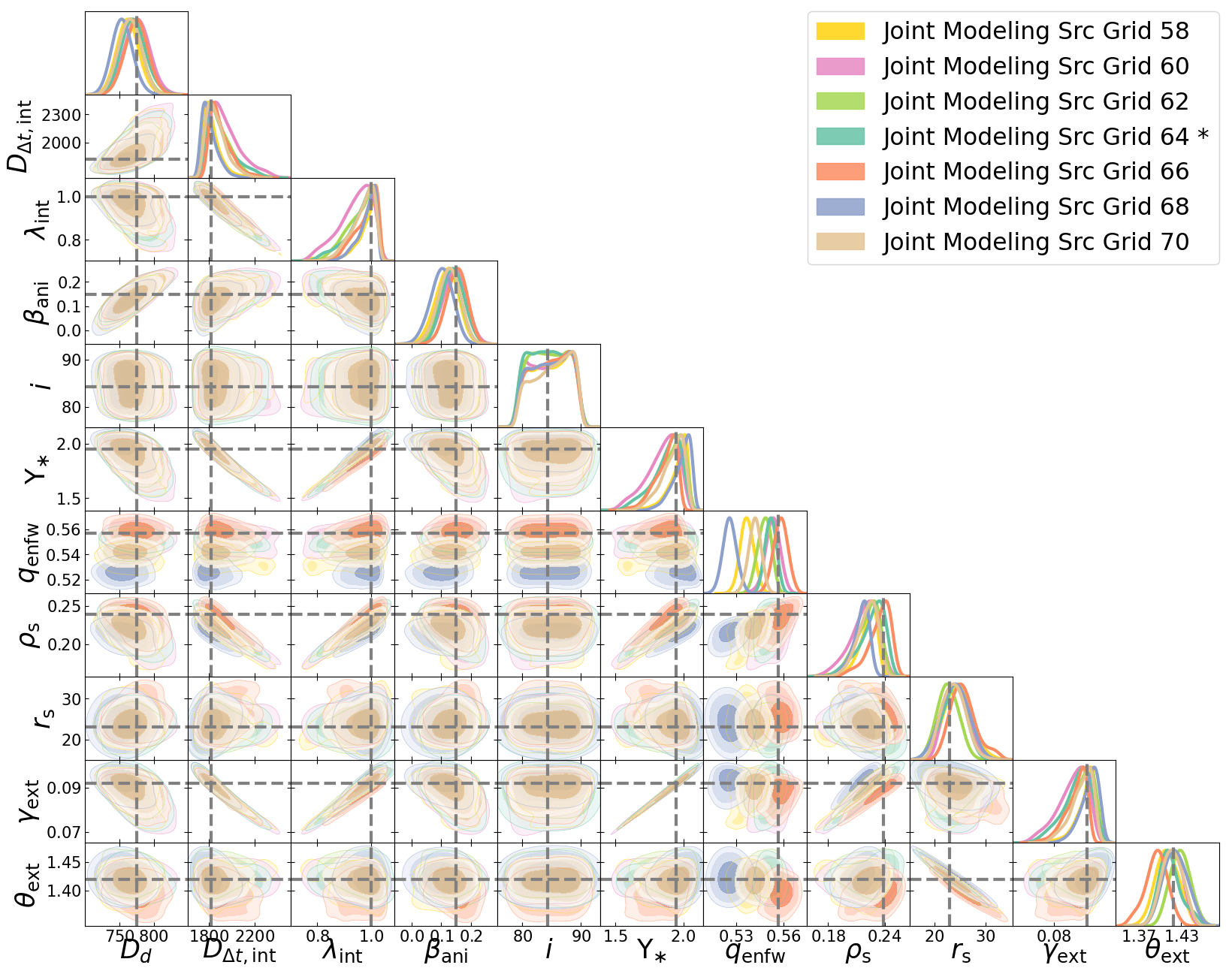}
 \caption{Equally weighted probability density distributions for all parameters in the joint modeling, given $M_{\rm BH} = 5 \times 10^{9}~{\rm M_{\odot}}$. The joint modeling is performed using different source grid resolutions (represented by different colors) to account for parameter uncertainties induced by variations in pixel size on the source plane. The simulated lensing data is generated assuming a source grid resolution of $64 \times 64$.}
 \label{fig:corner_src}
\end{figure}

To determine the resolution at which mass model parameter constraints become stable with respect to source grid resolutions, we perform joint modeling assuming $M_{\rm BH} = 5 \times 10^{9}~{\rm M_{\odot}}$. The source grid resolution varies from $58 \times 58$ to $70 \times 70$, corresponding to source pixel sizes of approximately $0.05 \pm 0.01\arcsec$ per pixel. We observe that all parameter contours stabilize when modeling with source grid resolutions beyond $\sim 60 \times 60$ (see Fig.~\ref{fig:corner_src}). Considering the computational time, we conduct joint modeling within the range of $60\times60$ to $68\times68$, excluding $58 \times 58$ and $70 \times 70$.

\begin{figure}
\includegraphics[width=1.\linewidth]{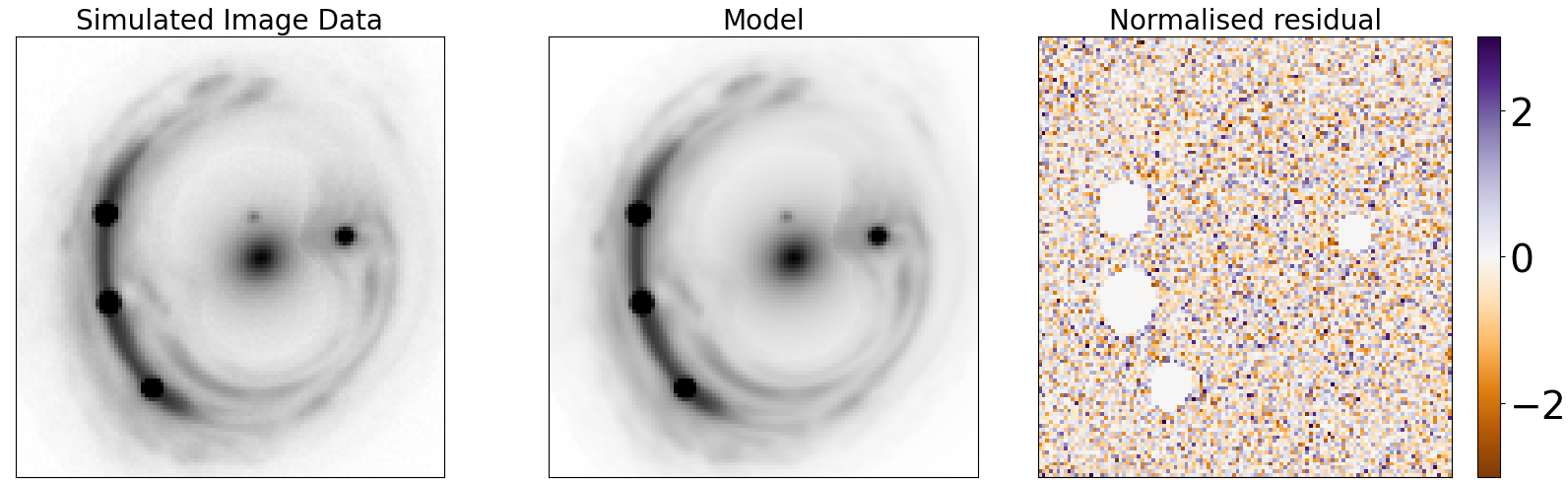}
 \caption{The best-fit composite mass model, given $M_{\rm BH} = 5 \times 10^{9}~{\rm M_{\odot}}$. The other best-fit model produces similar results. The quasar light in RXJ1131 is very strong, making it difficult to fit. To mitigate its effects, we boost the error in the quasar positions.
 }
 \label{fig:lensing_fit}
\end{figure}

\section{Joint modeling using kinematics data exclude the central bins}
\label{app: Joint modeling using kinematics data exclude the central bins}

\begin{figure}[H]
  \centering
  \includegraphics[width=0.8\linewidth]{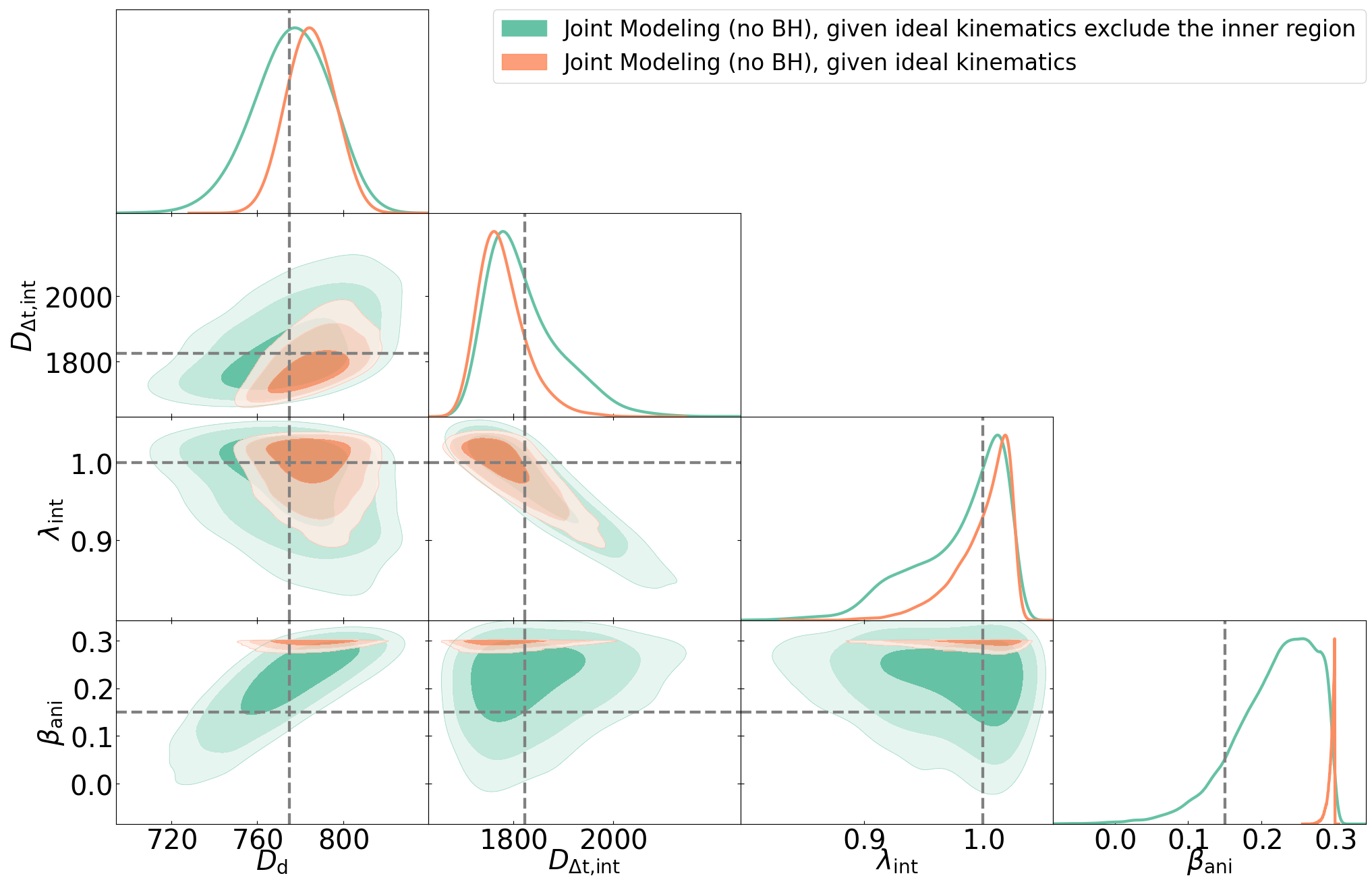}
  \caption*{(a) No black hole in the composite mass model.}
  \vspace{1em}
  
  \includegraphics[width=0.8\linewidth]{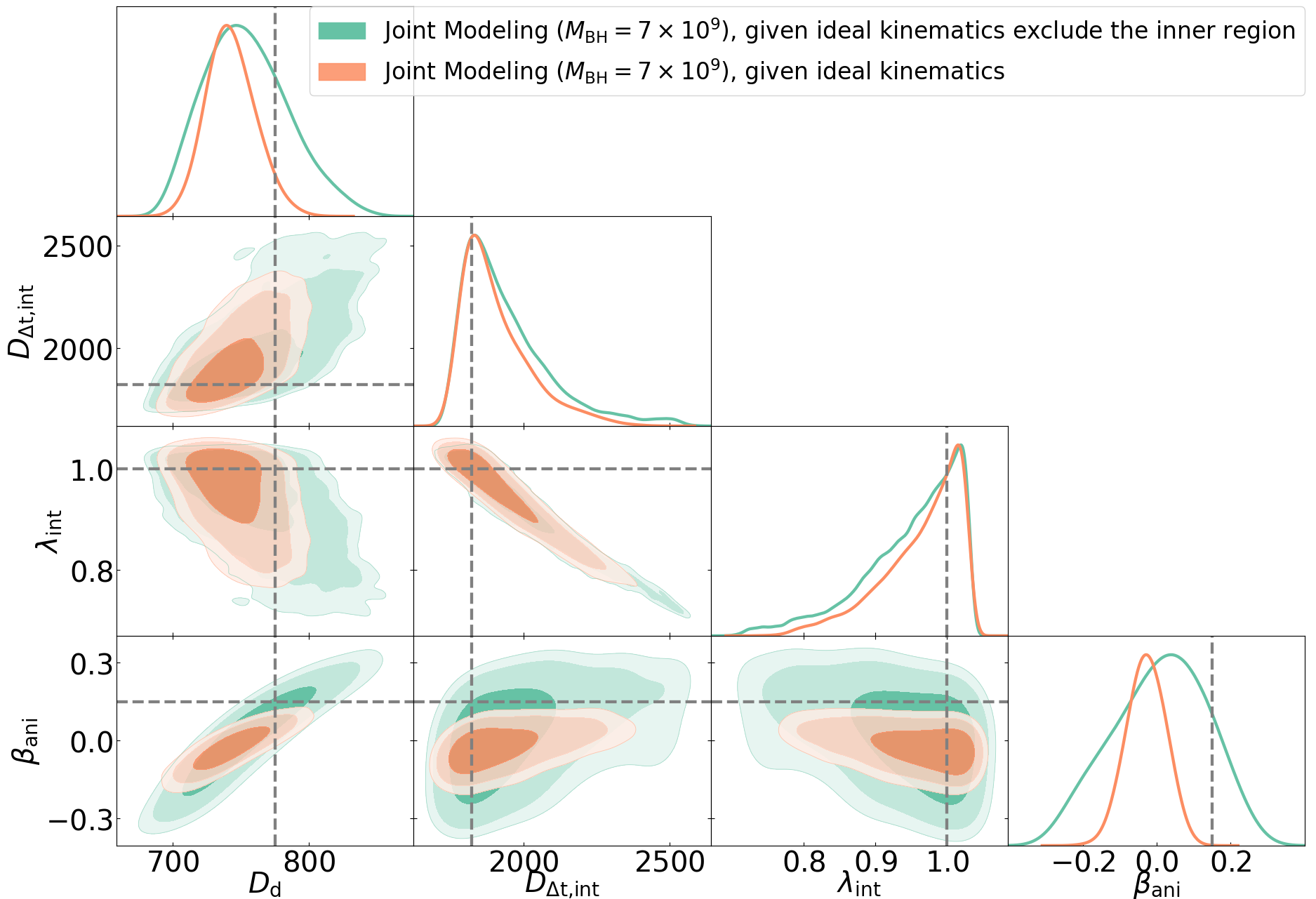}
  \caption*{(b) $M_{\rm BH} = 7 \times 10^{9}~{\rm M_{\odot}}$ in the composite mass model.}

  \caption{Comparison between joint modeling using full ideal kinematic data (orange contours) and excluding the central regions (green contours). We observe that uncertainties of parameters are enlarged, but the measurements of $D_{\rm d}$ and $\beta_{\rm ani}$ move toward the mock input values using the kinematic map excluding the central regions.}
  \label{fig:kinematics_corner_comparison}
\end{figure}
\section{The BIC weight factor \texorpdfstring{$f_{\textnormal{BIC}}^{*}$}{f_BIC*} to joint models }
\label{app:BIC weight factor for all mass models}
\begin{table}[H]
    \centering
    \caption{Comparison of models on different source resolutions, showing $\chi^2_{\text{dyn}}$ and $f^*_{\text{BIC}}$.}
    \small
    \setlength{\tabcolsep}{6pt}  
    \renewcommand{\arraystretch}{1.2}
    \begin{tabular}{llccccc}
    \toprule
    Data & Model & Source Resolution & $\chi^2_{\text{dyn}}$ & $f^*_{\text{BIC}}$& $\Delta \rm BIC$& $\sigma_{\rm BIC}$ \\ 
    \midrule
    \multicolumn{5}{l}{FoV $2'' \times 2''$} \\

    & \multirow{5}{*}{COMPOSITE} 
        & 68 & 54.11 & 0.1272& 4.154 \\
    &   & 66 & 55.08 & 0.0785 &5.120 \\
    &   & 64 & 54.61 & 0.0991 & 4.654& 0.35 \\
    &$M_{\text{BH}} = 1 \times 10^9 M_{\odot}$   & 62 & 54.21 & 0.1213 &4.250\\
    &   & 60 & 54.30 & 0.1160&4.339 \\

    \\

    \multirow{11}{*}{\shortstack[l]{Lensing \& Dynamics \\ IDEAL }} 
    & \multirow{5}{*}{$M_{\text{BH}} = 2 \times 10^9 M_{\odot}$} 
        & 68 & 50.61 & 0.7204& 0.6579 \\
    &   & 66 & 50.36 & 0.8171& 0.4060\\
    &   & 64 & 50.37 & 0.8129& 0.4163& 0.09 \\
    &   & 62 & 50.48 & 0.7712& 0.5218\\
    &   & 60 & 50.45 & 0.7805& 0.4977 \\

    \\
    & \multirow{5}{*}{$M_{\text{BH}} = 3 \times 10^9 M_{\odot}$} 
        & 68 & 50.26 & 0.8612 & 0.3015    \\
    &   & 66 & 49.96 & 1.     & 0.0   \\
    &   & 64 & 50.00 & 0.9672 & 0.0454 & 0.1  \\
    &   & 62 & 50.06 & 0.9482 & 0.0989   \\
    &   & 60 & 50.03 & 0.9567 & 0.0763    \\

    \\
    & \multirow{5}{*}{$M_{\text{BH}} = 4 \times 10^9 M_{\odot}$} 
        & 68 & 50.34 & 0.8327 & 0.3676&\\
    &   & 66 & 50.10 & 0.9325 & 0.1401&\\
    &   & 64 & 50.15 & 0.9065 & 0.1977& 0.08\\
    &   & 62 & 50.20 & 0.8866 & 0.2421&\\
    &   & 60 & 50.16 & 0.9050 & 0.2009&\\

    \\
    & \multirow{5}{*}{$M_{\text{BH}} = 5 \times 10^9 M_{\odot}$} 
        & 68 & 50.87 & 0.6350 & 0.9095     \\
    &   & 66 & 50.79 & 0.6581 & 0.8380   \\
    &   & 64 & 50.73 & 0.6790 & 0.7756 & 0.07  \\
    &   & 62 & 50.65 & 0.7080 & 0.6919   \\
    &   & 60 & 50.75 & 0.6723 & 0.7953   \\

    \\
    & \multirow{5}{*}{$M_{\text{BH}} = 6 \times 10^9 M_{\odot}$} 
        & 68 & 51.71 & 0.4162& 1.7545 \\
    &   & 66 & 51.86 & 0.3865& 1.9026\\
    &   & 64 & 51.81 & 0.3965& 1.8512&0.07\\
    &   & 62 & 51.74 & 0.4094& 1.7875\\
    &   & 60 & 51.67 & 0.4238& 1.7181\\

     \bottomrule
    \end{tabular}
   
\end{table}

\begin{table}
    \centering
    \caption*{Table F.1: continued.}
    \small
    \setlength{\tabcolsep}{6pt} 
    \renewcommand{\arraystretch}{1.2} 
    \begin{tabular}{llccccc}
    \toprule
        Data & Model & Source Resolution & $\chi^2_{\text{dyn}}$ & $f^*_{\text{BIC}}$ & $\Delta \mathrm{BIC}$ & $\sigma_{\mathrm{BIC}}$  \\ 
    \midrule
    \multicolumn{5}{l}{FoV $2'' \times 2''$} \\
    
      & \multirow{5}{*}{COMPOSITE} 

        & 68 & 52.98 & 0.2219&3.0197 \\
    &   & 66 & 53.55 & 0.1667&3.5922 \\
    &   & 64 & 53.28 & 0.1904 &3.3259 & 0.19\\
    & $M_{\text{BH}} = 7 \times 10^9 M_{\odot}$  & 62 & 53.24 & 0.1949&3.2793 \\
    &   & 60 & 53.36 & 0.1835 &3.4002\\

  \\

    \multirow{11}{*}{\shortstack[l]{Lensing \& Dynamics \\ IDEAL}} 
    & \multirow{5}{*}{$M_{\text{BH}} = 8 \times 10^9 M_{\odot}$} 
        & 68 & 54.85 & 0.0870 & 4.8943\\
    &   & 66 & 55.49 & 0.0633 & 5.5294\\
    &   & 64 & 55.11 & 0.0763 & 5.1573 &0.21 \\
    &   & 62 & 55.05 & 0.0787 & 5.0941\\
    &   & 60 & 55.13 & 0.0758 & 5.1701 \\

    \\
    & \multirow{5}{*}{$M_{\text{BH}} = 9 \times 10^9 M_{\odot}$} 
        & 68 & 56.95 & 0.0307 & 6.9940 \\
    &   & 66 & 57.94 & 0.0187 & 7.9874\\
    &   & 64 & 57.21 & 0.0269 & 7.2573 & 0.34\\
    &   & 62 & 57.22 & 0.0268 & 7.2676 \\
    &   & 60 & 57.53 & 0.0230 & 7.5741 \\

    \\
    & \multirow{5}{*}{$M_{\text{BH}} = 10 \times 10^9 M_{\odot}$} 
        & 68 & 60.93 & 0.0042 & 11.06 \\
    &   & 66 & 60.67 & 0.0047 &10.71\\
    &   & 64 & 60.73 & 0.0040 &10.98&0.23\\
    &   & 62 & 60.12 & 0.0052 &10.50\\
    &   & 60 & 60.45 & 0.0052  & 10.54\\

    \\
    & \multirow{5}{*}{EPL Model} 
        & 68 & 58.15 & 0.1338 &4.243\\
    &   & 66 & 58.12 & 0.1356 &4.217\\
    &   & 64 & 58.33 & 0.1227 &4.418& 0.93\\
    &   & 62 & 60.56 & 0.0400 &6.657\\
    &   & 60 & 58.29 & 0.1247& 4.385\\

    \bottomrule
    \end{tabular}
\end{table}

\end{appendix}

\end{document}